\documentclass[a4paper,11pt]{article}
\pdfoutput=1 

\usepackage{jheppub} 
\makeatletter
\RenewDocumentCommand{\author}{O{}m}{
    \IfBlankTF{#1}
    {\auth@toks=\expandafter{\the\auth@toks#2\ }}
    {\auth@toks=\expandafter{\the\auth@toks#2\unboldmath$^{#1}$\ }}
}
\makeatother

\NewExpandableDocumentCommand{\paperTitle}{}{%
    \texorpdfstring{\(\OO(N)\)}{O(N)} \texorpdfstring{\(\CFT[B]\)}{BCFT}: new data from conformal partial wave expansions
}
\NewExpandableDocumentCommand{\paperKeywords}{}{%
    Boundary Quantum Field Theory,
    Scale and Conformal Symmetries,
    \texorpdfstring{\(1/N\)}{1/N} Expansion,
    Global Symmetries
}

\hypersetup{
    bookmarksnumbered,   
    pdfauthor   = {Jozef Csipes, Petr Vaško},
    pdftitle    = {\paperTitle},
    pdfsubject  = {Study of the the O(N) BCFT.},
    pdfkeywords = {\paperKeywords},
    pdfdisplaydoctitle, 
}

\usepackage[T1]{fontenc} 
\usepackage{lmodern} 

\DeclareEmphSequence{\slshape,\itshape}

\usepackage[table]{xcolor} 
\usepackage{booktabs}   
\usepackage{multirow}   
\usepackage{array}      
\usepackage{tabularray} 
\usepackage{codehigh}   
\UseTblrLibrary{amsmath,booktabs}
\usepackage{enumerate}
\usepackage{adjustbox}
\usepackage{dsfont} 

\usepackage{microtype}  
\usepackage[capitalise,nameinlink]{cleveref}   
\usepackage{csquotes}   
\usepackage{enumitem}

\usepackage{standalone}
\usepackage{caption}    
\usepackage{subcaption}
\graphicspath{      
    {./figures/}
}

\usepackage{fetamont}   
\NewDocumentCommand{\githubrepository}{O{[jdujava/ONinAdS]}}{%
    \href{https://github.com/jdujava/ONinAdS}{\ttfamily #1}%
}
\NewDocumentCommand{\wolframlogo}{}{\raisebox{-0.4ex}{\hspace{0.1em}\includegraphics[height=2.2ex]{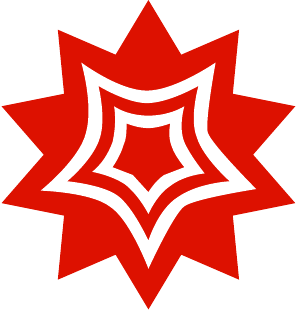}\,}}
\NewDocumentCommand{\wmathematica}{}{\wolframlogo\textffm{Wolfram Mathematica}}

\NewDocumentCommand{\todo}{O{} >{\TrimSpaces} m}{%
    \textcolor{red}{%
        \textsf{TODO\IfBlankF{#1}{(#1)}}%
        \IfBlankTF{#2}{.}{: #2}%
    }%
}
\NewDocumentCommand{\Note}{O{} >{\TrimSpaces} m}{%
    \textcolor{green!50!black}{%
        \textsf{Note\IfBlankF{#1}{(#1)}}: #2%
    }%
}

\DeclareMathOperator{\Tr}{Tr}
\DeclareMathOperator{\Res}{Res}

\NewDocumentCommand{\grayenclose}{s O{(} m O{)}}{%
    \ifmmode
        \let\colorcmd\mathcolor
        \DeclareCommandCopy{\Left}{\left}%
        \DeclareCommandCopy{\Right}{\right}%
    \else%
        \let\colorcmd\textcolor%
        \DeclareCommandCopy{\Left}{\big}%
        \DeclareCommandCopy{\Right}{\big}%
    \fi
    \IfBooleanTF{#1}{%
        \colorcmd{gray}{\Left#2}#3\colorcmd{gray}{\Right#4}%
    }{%
        \colorcmd{gray}{#2}#3\colorcmd{gray}{#4}%
    }%
}

\makeatletter
\usepackage{tikz}
\usetikzlibrary{positioning,calc}
\usetikzlibrary{arrows,arrows.meta}
\usetikzlibrary{decorations.pathmorphing,decorations.markings}
\usetikzlibrary{shapes.geometric}
\usetikzlibrary{patterns, patterns.meta}
\usetikzlibrary{fadings}

\NewDocumentCommand{\ShadowSpec}{O{1.2}}{
 \begin{tikzpicture}[x=#1cm, y=#1cm, baseline=(current bounding box.center)]
        
        \draw[thick, ->] (-4.6, 0) -- (7.8, 0) node[right] {$\widehat{\Delta}$};
        \foreach \x in {-4,-3,...,7} {
            \draw (\x, 0.1) -- (\x, -0.1);
        }
        
        \draw[gray, dashline, ultra thick] (1, -2.6) -- (1, 2.6);
        \draw[red!40, ultra thick] (1.5, -2.6) -- (1.5, 2.6);

        \foreach \start/\stop/\h in {0/2/0.8, -2/4/1.3, -4/6/1.8} {
            \draw[gray, line width=0.6pt] (\start,0) .. controls (\start, \h) and (\stop, \h) .. (\stop,0);
        }

        \foreach \start/\end/\depth in {0/3/0.8, -2/5/1.3, -4/7/1.8} {
            \draw[red!40, line width=0.6pt] (\start, 0) -- (\start, -\depth) -- (\end, -\depth) -- (\end, 0);
        }

        \def\baseY{-0.45} 
        
        \foreach \x in {-4, -3, -2, -1, 4, 6} {
             \node[anchor=base, scale=0.8] at (\x,\baseY) {$\x$};
        }

        \node[anchor=base, scale=0.7, gray, xshift=-5pt] at (1,\baseY) {$1$};
        \node[anchor=base, scale=0.7, red!80, xshift=-8pt] at (1.5,\baseY) {$\frac{3}{2}$};

        \node[anchor=base, scale=0.8, xshift=-5pt] at (0,\baseY) {$0$};
        \node[anchor=base, scale=0.8, blue, xshift=5pt] at (2,\baseY) {$2$};
        
        \foreach \x in {3, 5, 7} {
             \node[anchor=base, scale=0.8, xshift=7pt] at (\x,\baseY) {$\x$};
        }

        \foreach \x in {2, 4, 6} { \fill[blue] (\x, 0) circle (2.5pt); }
        \foreach \x in {-4, -2, 0} { \fill[black] (\x, 0) circle (2.5pt); }
        \foreach \x in {3, 5, 7} {
            \draw[red, ultra thick] (\x-0.15, -0.15) -- (\x+0.15, 0.15);
            \draw[red, ultra thick] (\x-0.15, 0.15) -- (\x+0.15, -0.15);
        }

    \end{tikzpicture}
}

\NewCommandCopy{\WD@dotaccent}{\.}
\ProvideDocumentCommand{\.}{}{\TextOrMath{\WD@dotaccent}{\mspace{1mu}}}

\RequirePackage[outline]{contour} 
\contourlength{0.08em}

\definecolor{OPEcolor}{RGB}{220,249,250}
\def\WD@dlWidth{1.2pt} 
\tikzset{
    label distance=0.8ex,
    every label/.style = {anchor=mid},
    thinline/.style  = {line width=0.85pt, draw=black, line cap=round},
    line/.style      = {line width=1pt,   draw=black},
    thickline/.style = {line width=1.5pt, draw=black},
    overline/.style  = {line width=1.5pt, double distance=1pt, draw=white, double=black},
    dashline/.style  = {line width=1pt,   draw=black, densely dashed},
    dotline/.style   = {line width=1pt,   draw=black, loosely dotted},
    doubleline/.style= {line width=0.8pt, double distance between line centers=1.4*\WD@dlWidth, double},
    doublefilled/.style = {
        pattern={Lines[angle=#1+45,distance=1.4pt,line width=0.4pt]},
        pattern color=darkgray, line width=0.8pt, draw=black
    },
    wave/.style = {line width=1pt, draw=black, decorate, decoration={snake, amplitude=1.5pt, segment length=5pt, pre length=#1}},
    correlator/.style    = {
        draw, circle, line width=0.8pt,
        inner color=black!30, outer color=black!5,
    },
    boundarypoint/.style = {draw, circle, fill=white,     line width=0.6pt, scale=.4},
    bulkpoint/.style     = {draw, circle, fill=black!80,  line width=0.6pt, scale=.35},
    vertex/.style        = {draw, circle, fill=lightgray, line width=0.8pt, scale=.5},
    vertexphi/.style     = {draw, circle, fill=black!60,  line width=0.8pt, scale=.35},
    vertexsubtle/.style  = {circle, fill=black!80, fill opacity=0.20, scale=.65},
    opevertex/.style     = {draw, circle, fill=OPEcolor,  line width=0.6pt, scale=.4},
    hide labels/.style = {every label/.append style={text opacity=0, xscale=0.5}},
}
\tikzset{ 
    dots/.style = {
        line width=1pt,
        line cap=round,
        dash pattern=on 0pt off #1,
    },
    dots/.default={4pt},
}

\NewExpandableDocumentCommand{\WD@smallerfactor}{}{0.9}
\newif\ifsmallerWD
\smallerWDfalse

\NewExpandableDocumentCommand{\WD@rDefault}{}{0.65cm}
\NewExpandableDocumentCommand{\WD@xpadNoIndices}{}{0.4ex}
\NewExpandableDocumentCommand{\WD@ypadNoIndices}{}{0.4ex}
\NewExpandableDocumentCommand{\WD@xpadWithIndices}{}{0.5em}
\NewExpandableDocumentCommand{\WD@ypadWithIndices}{}{0.6ex}
\NewDocumentCommand{\WD@SetupScaling}{m m}{
    \ifsmallerWD\def\WD@r{#2*\WD@smallerfactor*\WD@rDefault}
    \else       \def\WD@r{#2*\WD@rDefault}\fi
    \def\WD@xpad{#2*\WD@xpadNoIndices}
    \def\WD@ypad{#2*\WD@ypadNoIndices}
    \IfBooleanT{#1}{ 
        \def\WD@xpad{#2*\WD@xpadWithIndices}
        \def\WD@ypad{#2*\WD@ypadWithIndices}
    }
}

\NewDocumentCommand{\WD@ExpandBoundingBox}{O{\WD@xpad}O{\WD@ypad}}{
    \useasboundingbox ([shift={(-#1,-#2)}]current bounding box.south west)
    rectangle         ([shift={( #1, #2)}]current bounding box.north east);
}

\NewDocumentCommand{\WD@BoundaryPoints}{m}{
    \begin{scope}[overlay, \IfBooleanF{#1}{hide labels}]
        \node[boundarypoint,label={145:$    i\.  $}] at (145:\WD@r) (e1) {};
        \node[boundarypoint,label={215:$    j\.\.$}] at (215:\WD@r) (e2) {};
        \node[boundarypoint,label={ 35:$\.\.k    $}] at ( 35:\WD@r) (e3) {};
        \node[boundarypoint,label={325:$  \.l    $}] at (325:\WD@r) (e4) {};
    \end{scope}
}

\tikzfading[name=radial fade 70, inner color=transparent!70, outer color=transparent!100]
\NewDocumentCommand{\FourPt}{s O{} O{1}}{%
    \WD@SetupScaling{#1}{#3}
    \tikz[baseline={([yshift=-0.6ex]current bounding box.center)}]{
        \draw[line] (0,0) circle [radius=\WD@r];
        \WD@BoundaryPoints{#1}
        \draw[line] (e2) -- (e3);
        \draw[line] (e1) -- (e4);
        \draw[correlator] (0,0) circle [radius=5/8*\WD@r];
        \fill[fill=black!20, path fading=radial fade 70] (0,0) circle [radius=1/3*\WD@r];
        \IfBlankF{#2}{\node at (0,0) {\contour*{black!10}{#2}};}
        \WD@ExpandBoundingBox 
    }
}
\NewDocumentCommand{\tExchContrib}{s O{} O{1}}{%
    \WD@SetupScaling{#1}{#3}
    \tikz[baseline={([yshift=-0.6ex]current bounding box.center)}]{
        \draw[line] (0,0) circle [radius=\WD@r];
        \coordinate (B) at (0,-3/8*\WD@r);
        \coordinate (T) at (0, 3/8*\WD@r);
        \WD@BoundaryPoints{#1}
        \draw[line] (e2) -- (B);
        \draw[line] (e1) -- (T);
        \draw[line] (e3) -- (T);
        \draw[line] (e4) -- (B);
        \draw[correlator] (0,0) circle [x radius=1/3*\WD@r, y radius=5/8*\WD@r];
        \fill[fill=black!20, path fading=radial fade 70] (0,0) circle [x radius=1/5*\WD@r, y radius=1/3*\WD@r];
        \IfBlankF{#2}{\node at (0,0) {\contour*{black!10}{#2}};}
        \WD@ExpandBoundingBox 
    }
}

\NewDocumentCommand{\sDisc}{s O{1}}{%
    \WD@SetupScaling{#1}{#2}
    \tikz[baseline={([yshift=-0.6ex]current bounding box.center)}]{
        \draw[line] (0,0) circle [radius=\WD@r];
        \WD@BoundaryPoints{#1}
        \draw[line] (e2) to[out=  60,in=-60] (e1);
        \draw[line] (e3) to[out=-120,in=120] (e4);
        \WD@ExpandBoundingBox 
    }
}
\NewDocumentCommand{\tDisc}{s O{1}}{%
    \WD@SetupScaling{#1}{#2}
    \tikz[baseline={([yshift=-0.6ex]current bounding box.center)}]{
        \draw[line] (0,0) circle [radius=\WD@r];
        \WD@BoundaryPoints{#1}
        \draw[line] (e2) to[out= 25,in= 155] (e4);
        \draw[line] (e1) to[out=-25,in=-155] (e3);
        \WD@ExpandBoundingBox 
    }
}
\NewDocumentCommand{\uDisc}{s O{1}}{%
    \WD@SetupScaling{#1}{#2}
    \tikz[baseline={([yshift=-0.6ex]current bounding box.center)}]{
        \draw[line] (0,0) circle [radius=\WD@r];
        \WD@BoundaryPoints{#1}
        \draw[line]     (e2) -- (e3);
        \draw[overline] (e1) -- (e4);
        \WD@ExpandBoundingBox 
    }
}

\NewDocumentCommand{\cornernodes}{O{0.8}}{
    \begin{scope}[overlay]
        \node[scale=0.4] at (135:#1*\WD@r) (e1) {};
        \node[scale=0.4] at (225:#1*\WD@r) (e2) {};
        \node[scale=0.4] at ( 45:#1*\WD@r) (e3) {};
        \node[scale=0.4] at (315:#1*\WD@r) (e4) {};
    \end{scope}
}
\NewDocumentCommand{\vertexPhi}{s O{1}}{%
    \WD@SetupScaling{#1}{#2}
    \tikz[baseline={([yshift=-0.6ex]current bounding box.center)}]{
        \node[vertexphi] at (0,0) (V) {};
        \cornernodes
        \draw[line] (e1) -- (V);
        \draw[line] (e2) -- (V);
        \draw[line] (e3) -- (V);
        \draw[line] (e4) -- (V);
        \WD@ExpandBoundingBox 
    }
}
\NewDocumentCommand{\vertexPhiChannelS}{s O{1}}{%
    \WD@SetupScaling{#1}{#2}
    \tikz[baseline={([yshift=-0.6ex]current bounding box.center)}]{
        \cornernodes
        \draw[line, rounded corners=3pt] (e1) -- (0,0) -- (e2);
        \draw[line, rounded corners=3pt] (e3) -- (0,0) -- (e4);
        \node[vertexsubtle] at (0,0) (V) {};
        \WD@ExpandBoundingBox 
    }
}
\NewDocumentCommand{\vertexPhiChannelT}{s O{1}}{%
    \WD@SetupScaling{#1}{#2}
    \tikz[baseline={([yshift=-0.6ex]current bounding box.center)}]{
        \cornernodes
        \draw[line, rounded corners=3pt] (e1) -- (0,0) -- (e3);
        \draw[line, rounded corners=3pt] (e2) -- (0,0) -- (e4);
        \node[vertexsubtle] at (0,0) (V) {};
        \WD@ExpandBoundingBox 
    }
}
\NewDocumentCommand{\vertexPhiChannelU}{s O{1}}{%
    \WD@SetupScaling{#1}{#2}
    \tikz[baseline={([yshift=-0.6ex]current bounding box.center)}]{
        \cornernodes
        \draw[line]     (e2) -- (e3);
        \draw[overline] (e1) -- (e4);
        \node[vertexsubtle] at (0,0) (V) {};
        \WD@ExpandBoundingBox[0.7*\WD@xpad][0.5*\WD@ypad] 
    }
}

\NewDocumentCommand{\phiLoop}{s O{1}}{%
    \WD@SetupScaling{#1}{#2}
    \tikz[baseline={([yshift=-0.6ex]current bounding box.center)}]{
        \def\WD@rloop{0.45*\WD@r} 
        \node at (0,0) (O) {$\scriptstyle\color{gray}n\.$};
        \draw[line] (0,0) circle [radius=\WD@rloop];
        \draw[doubleline]             (\WD@rloop,0) -- (0.9*\WD@r,0);
        \draw[doubleline, rotate= 80] (\WD@rloop,0) -- (0.9*\WD@r,0);
        \draw[doubleline, rotate=-80] (\WD@rloop,0) -- (0.9*\WD@r,0);
        \node[vertex] at (  0:\WD@rloop) (v1) {};
        \node[vertex] at ( 80:\WD@rloop) (v2) {};
        \node[vertex] at (-80:\WD@rloop) (v3) {};
        \node at (  0:1.2*\WD@r) (l1) {$\sigma$};
        \node at ( 80:1.2*\WD@r) (l2) {$\sigma$};
        \node at (-80:1.2*\WD@r) (l3) {$\sigma$};
        \draw[line width=1.5pt,white] ([shift=(120:\WD@rloop)]0,0) arc [radius=\WD@rloop, start angle=120, end angle=240];
        \draw[dashline] ([shift=(135:\WD@rloop)]0,0) arc [radius=\WD@rloop, start angle=135, end angle=250];
        \draw[dots=4.24pt, gray] ([shift=(105:0.7*\WD@r)]0,0) arc [radius=0.7*\WD@r, start angle=105, end angle=260];
    }
}

\NewDocumentCommand{\SigmaFreePropag}{s O{1.2}}{%
    \WD@SetupScaling{#1}{#2}
    \tikz[baseline={([yshift=-0.6ex]current bounding box.center)}]{
        \coordinate (L) at (-3/8*\WD@r,0);
        \coordinate (R) at ( 3/8*\WD@r,0);
        \draw[doubleline] (L) -- (R);
        \node[boundarypoint] at (L) (e1) {};
        \node[boundarypoint] at (R) (e2) {};
        \WD@ExpandBoundingBox 
    }
}
\NewDocumentCommand{\SigmaFreeBubble}{s O{1.0}}{%
    \WD@SetupScaling{#1}{#2}
    \tikz[baseline={([yshift=-0.6ex]current bounding box.center)}]{
        \node[boundarypoint] at (-10/8*\WD@r,0) (LL) {};
        \node[vertex]        at ( -4/8*\WD@r,0)  (L) {};
        \node[vertex]        at (  4/8*\WD@r,0)  (R) {};
        \node[boundarypoint] at ( 10/8*\WD@r,0) (RR) {};
        \draw[doubleline] (LL) -- (L);
        \draw[doubleline] (R) -- (RR);
        \draw[line] (L) to[out= 60, in= 120] (R);
        \draw[line] (L) to[out=-60, in=-120] (R);
        \WD@ExpandBoundingBox 
    }
}
\NewDocumentCommand{\SigmaFreeBubbleTwice}{s O{1.0}}{%
    \WD@SetupScaling{#1}{#2}
    \tikz[baseline={([yshift=-0.6ex]current bounding box.center)}]{
        \node[boundarypoint] at (-10/8*\WD@r,0) (LL) {};
        \node[vertex]        at ( -4/8*\WD@r,0) (L1) {};
        \node[vertex]        at (  4/8*\WD@r,0) (R1) {};
        \node[vertex]        at ( 10/8*\WD@r,0) (L2) {};
        \node[vertex]        at ( 18/8*\WD@r,0) (R2) {};
        \node[boundarypoint] at ( 24/8*\WD@r,0) (RR) {};
        \draw[doubleline] (LL) -- (L1);
        \draw[doubleline] (R1) -- (L2);
        \draw[doubleline] (R2) -- (RR);
        \draw[line] (L1) to[out= 60, in= 120] (R1);
        \draw[line] (L1) to[out=-60, in=-120] (R1);
        \draw[line] (L2) to[out= 60, in= 120] (R2);
        \draw[line] (L2) to[out=-60, in=-120] (R2);
        \WD@ExpandBoundingBox 
    }
}
\NewDocumentCommand{\bubbleDiagram}{s O{1.4}}{%
    \WD@SetupScaling{#1}{#2}
    \tikz[baseline={([yshift=-0.6ex]current bounding box.center)}]{
        \node[boundarypoint, label={ left:$x\,$}] at (-4/8*\WD@r,0)  (L) {};
        \node[boundarypoint, label={right:$\,y$}] at ( 4/8*\WD@r,0)  (R) {};
        \draw[line] (L) to[out= 60, in= 120] (R);
        \draw[line] (L) to[out=-60, in=-120] (R);
    }
}

\NewDocumentCommand{\SigmaPropag}{s O{1.2}}{%
    \WD@SetupScaling{#1}{#2}
    \tikz[baseline={([yshift=-0.6ex]current bounding box.center)}]{
        \coordinate (L) at (-3/8*\WD@r,0);
        \coordinate (R) at ( 3/8*\WD@r,0);
        \draw[doublefilled] ([yshift=-\WD@dlWidth]L) rectangle ([yshift=+\WD@dlWidth]R);
        \node[boundarypoint] at (L) (e1) {};
        \node[boundarypoint] at (R) (e2) {};
        \WD@ExpandBoundingBox 
    }
}
\NewDocumentCommand{\vertexSigma}{s O{1}}{%
    \WD@SetupScaling{\BooleanFalse}{#2}
    \tikz[baseline={([yshift=-0.6ex]current bounding box.center)}]{
        \coordinate (L) at (-3/8*\WD@r,0);
        \coordinate (R) at ( 1/5*\WD@r,0);
        \begin{scope}[overlay, \IfBooleanF{#1}{hide labels}]
            \node[scale=0.1, inner sep=0, label={ 160:$\scriptstyle i$}] at (145:0.9*\WD@r) (e1) {};
            \node[scale=0.1, inner sep=0, label={left:$\scriptstyle j$}] at (215:0.9*\WD@r) (e2) {};
        \end{scope}
        \draw[line] (e1) -- (L);
        \draw[line] (e2) -- (L);
        \IfBooleanTF{#1} 
        {\draw[doubleline] (L) -- (R);}
        {\draw[doublefilled] ([yshift=-\WD@dlWidth]L) rectangle ([yshift=+\WD@dlWidth]R);}
        \node[vertex] at (L) (vL) {};
        \WD@ExpandBoundingBox 
    }
}
\NewDocumentCommand{\sExchFeyn}{s O{1}}{%
    \WD@SetupScaling{#1}{#2}
    \tikz[baseline={([yshift=-0.6ex]current bounding box.center)}]{
        \coordinate (L) at (-3/8*\WD@r,0);
        \coordinate (R) at ( 3/8*\WD@r,0);
        \WD@BoundaryPoints{#1}
        \draw[line] (e2) -- (L);
        \draw[line] (e1) -- (L);
        \draw[line] (e3) -- (R);
        \draw[line] (e4) -- (R);
        \draw[doublefilled] ([yshift=-\WD@dlWidth]L) rectangle ([yshift=+\WD@dlWidth]R);
        \node[vertex] at (L) (vL) {};
        \node[vertex] at (R) (vR) {};
        \WD@ExpandBoundingBox 
    }
}

\NewDocumentCommand{\bubbleChain}{s O{1}}{%
    \WD@SetupScaling{#1}{#2}
    \tikz[baseline={([yshift=-0.6ex]current bounding box.center)}]{
        \coordinate (LL) at (-14/8*\WD@r,0);
        \coordinate  (L) at ( -6/8*\WD@r,0);
        \coordinate  (R) at (  6/8*\WD@r,0);
        \coordinate (RR) at ( 14/8*\WD@r,0);
        \begin{scope}[overlay, \IfBooleanF{#1}{hide labels}]
            \node[boundarypoint] at ([xshift=-11/8*\WD@r]145:\WD@r) (e1) {};
            \node[boundarypoint] at ([xshift=-11/8*\WD@r]215:\WD@r) (e2) {};
            \node[boundarypoint] at ([xshift= 11/8*\WD@r] 35:\WD@r) (e3) {};
            \node[boundarypoint] at ([xshift= 11/8*\WD@r]325:\WD@r) (e4) {};
            \node[anchor=mid, gray] at (0,-0.01) {$\underbrace{\xmathstrut[0.05]{0}\mspace{41mu}{\color{black}\cdots}\mspace{41mu}}_{n \text{ bubbles}}$};
        \end{scope}
        \draw[line, rounded corners=3pt] (e1) -- (LL) -- (e2);
        \draw[line, rounded corners=3pt] (e3) -- (RR) -- (e4);
        \draw[line] ([xshift= 1.2]LL) .. controls +(90:0.25) and +(90:0.25) .. ([xshift=-1.2]L) .. controls +(-90:0.25) and +(-90:0.25) .. ([xshift= 1.2]LL);
        \draw[line] ([xshift=-1.2]RR) .. controls +(90:0.25) and +(90:0.25) .. ([xshift= 1.2]R) .. controls +(-90:0.25) and +(-90:0.25) .. ([xshift=-1.2]RR);
        \draw[line] ([xshift= 1.2]L)++(-60:0.22*\WD@r) .. controls +(140:0.06) and +(-90:0.05) .. ([xshift= 1.2]L) .. controls +( 90:0.06) and +(-140:0.05) .. ++( 60:0.22*\WD@r);
        \draw[dots=2pt, dash phase=0.5pt, lightgray] (L)++(  50:0.25*\WD@r) -- ++(  25:0.22*\WD@r);
        \draw[dots=2pt, dash phase=0.5pt, lightgray] (L)++( -50:0.25*\WD@r) -- ++( -25:0.22*\WD@r);
        \draw[line] ([xshift=-1.2]R)++(120:0.22*\WD@r) .. controls +(-40:0.06) and +( 90:0.05) .. ([xshift=-1.2]R) .. controls +(-90:0.06) and +( 40:0.05) .. ++(-120:0.22*\WD@r);
        \draw[dots=2pt, dash phase=0.5pt, lightgray] (R)++( 130:0.25*\WD@r) -- ++( 155:0.22*\WD@r);
        \draw[dots=2pt, dash phase=0.5pt, lightgray] (R)++(-130:0.25*\WD@r) -- ++(-155:0.22*\WD@r);
        \node[vertexsubtle] at ([xshift= 0.08]LL) (LLV) {};
        \node[vertexsubtle] at ([xshift=-0.05] L)  (LV) {};
        \node[vertexsubtle] at ([xshift= 0.05] R)  (RV) {};
        \node[vertexsubtle] at ([xshift=-0.08]RR) (RRV) {};
        \WD@ExpandBoundingBox 
    }
}
\NewDocumentCommand{\sExch}{s O{1}}{%
    \WD@SetupScaling{#1}{#2}
    \tikz[baseline={([yshift=-0.6ex]current bounding box.center)}]{
        \draw[line] (0,0) circle [radius=\WD@r];
        \coordinate (L) at (-3/8*\WD@r,0);
        \coordinate (R) at ( 3/8*\WD@r,0);
        \WD@BoundaryPoints{#1}
        \draw[line] (e2) -- (L);
        \draw[line] (e1) -- (L);
        \draw[line] (e3) -- (R);
        \draw[line] (e4) -- (R);
        \draw[doublefilled] ([yshift=-\WD@dlWidth]L) rectangle ([yshift=+\WD@dlWidth]R);
        \node[vertex] at (L) (vL) {};
        \node[vertex] at (R) (vR) {};
        \WD@ExpandBoundingBox 
    }
}
\NewDocumentCommand{\tExch}{s O{} O{1}}{%
    \WD@SetupScaling{#1}{#3}
    \tikz[baseline={([yshift=-0.6ex]current bounding box.center)}]{
        \draw[line] (0,0) circle [radius=\WD@r];
        \coordinate (B) at (0,-3/8*\WD@r);
        \coordinate (T) at (0, 3/8*\WD@r);
        \WD@BoundaryPoints{#1}
        \draw[line] (e2) -- (B);
        \draw[line] (e1) -- (T);
        \draw[line] (e3) -- (T);
        \draw[line] (e4) -- (B);
        \draw[doublefilled] ([xshift=-\WD@dlWidth]B) rectangle ([xshift=+\WD@dlWidth]T);
        \node[vertex] at (B) (vB) {};
        \node[vertex] at (T) (vT) {};
        \IfBlankF{#2}{
            \begin{scope}[overlay, inner sep=0, label distance=0.15ex] 
                \node[label={[anchor=mid]200:\(\scriptstyle{#2}\)}] at (0,0) (l) {};
            \end{scope}
        }
        \WD@ExpandBoundingBox 
    }
}
\NewDocumentCommand{\uExch}{s O{1}}{%
    \WD@SetupScaling{#1}{#2}
    \tikz[baseline={([yshift=-0.6ex]current bounding box.center)}]{
        \draw[line] (0,0) circle [radius=\WD@r];
        \coordinate (B) at (0,-3/8*\WD@r);
        \coordinate (T) at (0, 3/8*\WD@r);
        \WD@BoundaryPoints{#1}
        \draw[line]     (e2) -- (B);
        \draw[line]     (e1) -- (T);
        \draw[line]     (e3) -- (B);
        \draw[overline] (e4) -- (T);
        \draw[doublefilled] ([xshift=-\WD@dlWidth]B) rectangle ([xshift=+\WD@dlWidth]T);
        \node[vertex] at (B) (vB) {};
        \node[vertex] at (T) (vT) {};
        \WD@ExpandBoundingBox 
    }
}
\NewDocumentCommand{\sExchHarmonic}{s O{1} O{\D}}{%
    \WD@SetupScaling{#1}{#2}
    \tikz[baseline={([yshift=-0.6ex]current bounding box.center)}]{
        \draw[line] (0,0) circle [radius=\WD@r];
        \coordinate (L) at (-4/9*\WD@r,0);
        \coordinate (R) at ( 4/9*\WD@r,0);
        \WD@BoundaryPoints{#1}
        \draw[line] (e2) -- (L);
        \draw[line] (e1) -- (L);
        \draw[line] (e3) -- (R);
        \draw[line] (e4) -- (R);
        \draw[wave=1.5pt] (L) -- (R) node [above,midway]{\(\scriptstyle #3\)};
        \node[bulkpoint] at (L) (vL) {};
        \node[bulkpoint] at (R) (vR) {};
        \WD@ExpandBoundingBox[\WD@xpad][0] 
    }
}
\NewDocumentCommand{\sExchSplit}{s O{1} O{\D} O{\D*}}{%
    \WD@SetupScaling{#1}{#2}
    \tikz[baseline={([yshift=-0.6ex]current bounding box.center)}]{
        \draw[line] (0,0) circle [radius=\WD@r];
        \coordinate (L) at (-4/9*\WD@r,0);
        \coordinate (R) at ( 4/9*\WD@r,0);
        \coordinate (S) at (0,\WD@r);
        \WD@BoundaryPoints{#1}
        \draw[line] (e2) -- (L);
        \draw[line] (e1) -- (L);
        \draw[line] (e3) -- (R);
        \draw[line] (e4) -- (R);
        \draw[line] (L) -- (S) node [above left,  inner sep=-1.7pt, pos=0.5]{\contour*{white}{\(\scriptstyle\scalemath{0.8}{#3}\)}};
        \draw[line] (R) -- (S) node [above right, inner sep=-1.7pt, pos=0.5]{\contour*{white}{\(\scriptstyle\scalemath{0.8}{#4}\)}};
        \node[bulkpoint] at (L) (vL) {};
        \node[bulkpoint] at (R) (vR) {};
        \node[bulkpoint,overlay] at (S) (vS) {};
        \WD@ExpandBoundingBox[\WD@xpad][0] 
    }
}

\NewExpandableDocumentCommand{\PW@rDefault}{}{\WD@rDefault}
\NewExpandableDocumentCommand{\PW@xpadDefault}{}{0.15em}
\NewDocumentCommand{\PW@SetupScaling}{m}{
    \def\WD@r{#1*\PW@rDefault}
    \def\PW@xpad{#1*\PW@xpadDefault}
}

\NewDocumentCommand{\shadowRepCPW}{s O{1} O{\D} O{\D*}}{%
    \WD@SetupScaling{#1}{#2}
    \tikz[baseline={([yshift=-0.6ex]current bounding box.center)}]{
        \coordinate (L) at (-7/8*\WD@r,0);
        \coordinate (R) at ( 7/8*\WD@r,0);
        \coordinate (S) at (0,0);
        \begin{scope}[overlay, \IfBooleanF{#1}{hide labels}]
            \node[boundarypoint] at ([xshift=-4/8*\WD@r]145:\WD@r) (e1) {};
            \node[boundarypoint] at ([xshift=-4/8*\WD@r]215:\WD@r) (e2) {};
            \node[boundarypoint] at ([xshift= 4/8*\WD@r] 35:\WD@r) (e3) {};
            \node[boundarypoint] at ([xshift= 4/8*\WD@r]325:\WD@r) (e4) {};
        \end{scope}
        \draw[thinline] (e2) -- (L);
        \draw[thinline] (e1) -- (L);
        \draw[thinline] (e3) -- (R);
        \draw[thinline] (e4) -- (R);
        \draw[thinline] (L) -- (S) node [xshift= 1/13*\WD@r, above, inner sep=1pt, midway]{\(\scriptstyle #3\)};
        \draw[thinline] (R) -- (S) node [xshift=-1/13*\WD@r, above, inner sep=1pt, midway]{\(\scriptstyle #4\)};
        \node[opevertex,scale=2] at (L) (vL) {};
        \node[opevertex,scale=2] at (R) (vR) {};
        \node[bulkpoint] at (S) (vS) {};
        \WD@ExpandBoundingBox[3*\WD@xpad] 
    }
}

\NewDocumentCommand{\sPartWave}{s O{\D} O{J} O{1}}{%
    \PW@SetupScaling{#4}
    \tikz[baseline={([yshift=-0.6ex]current bounding box.center)}, color=black]{
        \node[opevertex] at (-1/2*\WD@r,0) (vL) {};
        \node[opevertex] at ( 1/2*\WD@r,0) (vR) {};
        \draw[thinline] (215:\WD@r) -- (vL);
        \draw[thinline] (145:\WD@r) -- (vL);
        \draw[thinline] ( 35:\WD@r) -- (vR);
        \draw[thinline] (325:\WD@r) -- (vR);
        \draw[thickline] (vL) -- (vR);
        \begin{scope}[overlay, inner sep=0, label distance=0.1ex] 
            \node[label={[above]\(\scriptstyle\IfBooleanF{#1}{#2\IfBlankF{#3}{,\.#3}}\)}] at (0,0) (l) {};
        \end{scope}
        \WD@ExpandBoundingBox[\PW@xpad][0] 
    }
}
\NewDocumentCommand{\tPartWave}{s O{\Dpr} O{J'} O{1}}{%
    \PW@SetupScaling{#4}
    \tikz[baseline={([yshift=-0.6ex]current bounding box.center)}, color=black]{
        \node[opevertex] at (0,-1/3*\WD@r) (vB) {};
        \node[opevertex] at (0, 1/3*\WD@r) (vT) {};
        \draw[thinline] (215:\WD@r) -- (vB);
        \draw[thinline] (145:\WD@r) -- (vT);
        \draw[thinline] ( 35:\WD@r) -- (vT);
        \draw[thinline] (325:\WD@r) -- (vB);
        \draw[thickline] (vB) -- (vT);
        \begin{scope}[overlay, inner sep=0, label distance=0.15ex] 
            \node[label={[anchor=mid west]right:\(\scriptstyle\IfBooleanF{#1}{#2\IfBlankF{#3}{\mspace{-2mu},\.#3}}\)}] at (0,0) (l) {};
        \end{scope}
        \WD@ExpandBoundingBox[\PW@xpad][0] 
    }
}
\usepackage{mathtools}
\usepackage{amsmath,amsfonts,amssymb,bbold,bm}
\usepackage{physics}
\usepackage{scalerel}   

\usepackage{mathfixs}   
\ProvideMathFix{autobold}
\ProvideMathFix{greekcaps=it}
\ProvideMathFix{frac,rfrac,vfrac,vfracskippre=4mu}

\NewCommandCopy{\dotaccent}{\.}
\RenewDocumentCommand{\.}{}{\TextOrMath{\dotaccent}{\mspace{1mu}}}
\NewCommandCopy{\acuteaccent}{\'}
\renewcommand*{\'}{\TextOrMath{\acuteaccent}{\mspace{-1mu}}}

\usepackage{siunitx}
\NewDocumentCommand{\eqend}{}{\,.}
\NewDocumentCommand{\eqcomma}{}{\,,}

\NewDocumentCommand{\eqdim}{O{} m}{
    \IfBlankTF{#1}
    {\mathrel{\overset{\mathcolor{black!70}{d\.=\.#2}}{\scalebox{2.8}[1]{\(=\)}}}}
    {\mathrel{\overset{\mathcolor{black!70}{#2}}{\scalebox{#1}[1]{\(=\)}}}}
}

\NewCommandCopy{\olddet}{\det}
\RenewDocumentCommand{\det}{}{\olddet\nolimits}

\NewCommandCopy{\transpose}{\intercal}

\NewDocumentCommand{\exchange}{O{\quad}}{\xleftrightarrow{#1}}

\NewDocumentCommand{\E}{s}{\IfBooleanTF{#1}{\mathrm{e}}{\mathinner{\mathrm{e}}}}

\NewDocumentCommand{\bigO} {l m}{\fbraces#1{\lparen}{\rparen}                   {O} {#2}}
\NewDocumentCommand{\biggO}{l m}{\fbraces#1{\lparen}{\rparen}{\raisemath{-0.1ex}{O}}{#2}}

\NewDocumentCommand{\R}{}{\mathbb{R}}
\NewDocumentCommand{\Z}{}{\mathbb{Z}}

\NewCommandCopy{\leftOrig}{\left}
\NewCommandCopy{\rightOrig}{\right}
\NewCommandCopy{\middleOrig}{\middle}
\RenewDocumentCommand{\left}{}{\mathopen{}\mathclose\bgroup\leftOrig}
\RenewDocumentCommand{\right}{}{\aftergroup\egroup\rightOrig}
\RenewDocumentCommand{\middle}{sm}{\IfBooleanTF{#1}{\.\middleOrig#2\.}{\mathrel{}\middleOrig#2\mathrel{}}}
\NewDocumentCommand{\innermiddle}{m}{\mathinner{}\middleOrig#1\mathinner{}}

\NewDocumentCommand{\Ext}{e{^}}{
    \scalerel*{\bigwedge}{\xmathstrut[-0.1]{0.1}} 
    \IfValueT{#1}{^{\mspace{-2mu}#1}} 
}
\NewDocumentCommand{\Odot}{e{^}}{
    \scalerel*{\bigodot}{\xmathstrut[-0.1]{0.1}} 
    \IfValueT{#1}{^{\mspace{-0mu}#1}} 
}

\DeclareDocumentCommand{\raisemath}{m}{\mathpalette{\raisemathAux{#1}}}
\DeclareDocumentCommand{\raisemathAux}{mmm}{\raisebox{#1}{\(#2#3\)}}
\DeclareDocumentCommand{\scalemath}{m}{\mathpalette{\scalemathAux{#1}}}
\DeclareDocumentCommand{\scalemathAux}{mmm}{\scalebox{#1}{\(#2#3\)}}

\NewDocumentCommand{\sbullet}{O{0.5}}{%
    \mathbin{\ThisStyle{\vcenter{\hbox{\scalebox{#1}{\(\SavedStyle\bullet\)}}}}}
}

\NewDocumentCommand{\idot}{s}{\mathcolor{darkgray}{\IfBooleanTF{#1}{\.\sbullet[0.4]\.}{\sbullet[0.4]}}}

\NewDocumentCommand{\ind}{}{\mathcolor{lightgray}{\bullet}}

\NewDocumentCommand{\argument}{s O{} O{1}}{%
    \def\squaresize{1.0}
    \IfBooleanT{#1}{\def\squaresize{0.7}}
    \IfBlankF{#2}{\def\squaresize{#2}}
    \scalebox{\squaresize}{%
        \begin{tikzpicture}[baseline=-#3*0.6ex]
            \node(char)[draw, shape=rectangle, dash=on 1.2pt off 1.05pt phase 0.5pt, dash expand off,
                inner ysep=2pt, inner xsep=2pt, minimum size=0.6em, rounded corners=2pt] {};
        \end{tikzpicture}%
    }%
}

\tikzset{
    leadsto/.style={-{Stealth[length=0.6em,open,round]},decorate,decoration={snake,amplitude=0.20ex,segment length=0.5em,pre length=0.2em,post length=0.6em}},
    toleads/.style={{Stealth[length=0.6em,open,round]}-,decorate,decoration={snake,amplitude=0.20ex,segment length=0.5em,pre length=0.6em,post length=0.2em}},
    correspondsto/.style={{Stealth[length=0.6em,open,round]}-{Stealth[length=0.6em,open,round]},decorate,decoration={snake,amplitude=0.20ex,segment length=0.5em,pre length=0.7em,post length=0.7em}},
}
\NewDocumentCommand{\longleadsto}{s O{} O{}}{%
    \ensuremath{\mathrel{
            \tikz[baseline=-0.5ex, inner sep=0ex, font=\scriptsize]{
                \node[minimum width=2.15em, inner xsep=0.6em, align=center] (a){\hphantom{#2}\\[0ex]\hphantom{#3}};
                \IfBlankF{#2}{\node[inner sep=0.3ex, above=0.3ex, xshift=-0.1em] at (a){#2};}
                \IfBlankF{#3}{\node[inner sep=0.3ex, below=0.3ex, xshift=-0.1em] at (a){#3};}
                \IfBooleanTF{#1}
                {\draw[line width=0.6pt, toleads] (a.west) -- (a.east);}
                {\draw[line width=0.6pt, leadsto] (a.west) -- (a.east);}
            }
        }}%
}
\NewDocumentCommand{\correspondsto}{O{}O{}}{%
    \ensuremath{\mathrel{
            \tikz[baseline=-0.5ex, inner sep=0ex, font=\scriptsize]{
                \node[minimum width=3.48em, inner xsep=0.8em, align=center] (a){\hphantom{#1}\\[0ex]\hphantom{#2}};
                \IfBlankF{#1}{\node[inner sep=0.3ex, above=0.3ex] at (a){#1};}
                \IfBlankF{#2}{\node[inner sep=0.3ex, below=0.3ex] at (a){#2};}
                \draw[line width=0.6pt, correspondsto] (a.west) -- (a.east);
            }
        }}%
}

\NewDocumentCommand{\bigslant}{O{0.2}O{1.7}mm}{
    \left.\mspace{-1mu}\raisemath{#1em}{#3}
    \mspace{-#2mu} \middleOrig/ \mspace{-\fpeval{#2+1}mu}
    \raisemath{-#1em}{#4} \mspace{-\fpeval{5*#1}mu} \right.
}

\NewDocumentCommand{\dAlembertian}{s}{\mathord{\raisemath{-0.05ex}{\square}}}

\DeclareMathOperator{\OO}{\mathsf{O}} 

\NewDocumentCommand{\EAdS}{}  {\ensuremath{\mathsf{EAdS}}}

\NewDocumentCommand{\QFT}{O{}}{\ensuremath{\mathsf{#1QFT}}}
\NewDocumentCommand{\CFT}{O{}}{\ensuremath{\mathsf{#1CFT}}}
\NewDocumentCommand{\AdSCFT}{}{\textsf{AdS/CFT}}

\DeclareDocumentCommand{\correlator}{ l m o }{
    \braces#1{\langle}{\rangle\IfValueT{#3}{_{\mspace{-2mu}#3}}}{#2}
}

\NewDocumentCommand{\OperSymbol}{e{_^}}{ 
    \mathsf{O\xmathstrut[-0.1]{-1}}\IfValueT{#1}{_{\mspace{-2mu}#1}}\IfValueT{#2}{^{#2}}
}
\NewDocumentCommand{\Oper}{s O{} e{_^}}{ 
    \IfBooleanTF{#1}{\mspace{2mu}\widehat{\mspace{-2mu}\OperSymbol}}{\OperSymbol}%
    \IfBlankF{#2}{^{#2}}\IfValueT{#3}{_{\mspace{-2mu}#3}}\IfValueT{#4}{^{#4}}
}
\NewExpandableDocumentCommand{\Opr}{}{\Oper^{\prime}} 
\NewDocumentCommand{\Ostar}{O{}}{\OperSymbol_{\star}\IfBlankF{#1}{^{(#1)}}} 
\NewDocumentCommand{\Oprstar}{O{}}{\OperSymbol^{\prime}_{\star}\IfBlankF{#1}{^{(#1)}}} 
\NewDocumentCommand{\Obullet}{O{}}{\OperSymbol_{\bullet}\IfBlankF{#1}{^{(#1)}}} 
\NewDocumentCommand{\Ophi}{O{}}{\OperSymbol_{\mspace{-1.5mu}\phi^{#1}}} 
\NewDocumentCommand{\Osigma}{O{}}{\widehat{\sigma}}

\NewDocumentCommand{\SpecNoArgs}{O{}}{\operatorname{\mathsf{Spec}}\IfBlankF{#1}{_{#1}}} 
\NewDocumentCommand{\Spec}{O{s} O{\D} O{\mathcolor{gray}{0}} m}{ 
    \SpecNoArgs[\IfBlankF{#1}{#1\!}]
    \left\lbrack
    \IfBlankF{#2}{
        \begin{+matrix}[cells={c},rowsep=-0.5pt,colsep=2pt]
        #2 \\ #3
        \end{+matrix}
        \mspace{-2mu} \mathcolor{black!60}{\innermiddle|}
    }
    \smallerWDtrue #4\smallerWDfalse
    \right\rbrack
}
\NewDocumentCommand{\ope}  {l m}{\fbraces#1{\lbrack}{\rbrack}{\operatorname{\mathsf{ope}}}{#2}} 
\NewDocumentCommand{\opesq}{l m}{\fbraces#1{\lbrack}{\rbrack}{\operatorname{\mathsf{ope^{2}}}}{#2}} 

\NewDocumentCommand{\Bubble}{s}{\opbraces{\widetilde{B}\IfBooleanT{#1}{^{\raisemath{0.4ex}{\.\prime\!}}}}}

\NewDocumentCommand{\G}{e{^} O{}}{%
    \IfBlankTF{#2}%
    {\opbraces{\mathop{\Gamma\xmathstrut{-0.1}}\IfValueT{#1}{^{#1}}}}%
    {\mathop{\Gamma\xmathstrut{-0.1}}\nolimits\IfValueT{#1}{^{#1}}_{\!#2}}%
}
\RenewDocumentCommand{\digamma}{e{^} O{}}{%
    \IfBlankTF{#2}%
    {\opbraces{\psi\IfValueT{#1}{^{#1}}}}%
    {\psi\IfValueT{#1}{^{#1}}_{\mspace{-2mu}#2}}%
}

\NewDocumentCommand{\widetildesmashAux}{O{0ex}m}{ 
    \mathrlap{\smash{\raisemath{#1}{\widetilde{\phantom{#2}}}}}#2
}
\NewDocumentCommand{\widetildesmash}{O{}m}{
    \IfBlankTF{#1}{%
        \mathchoice%
        {\widetildesmashAux[0ex]{#2}}%
        {\widetildesmashAux[-0.04ex]{#2}}%
        {\widetildesmashAux[-0.07ex]{#2}}%
        {\widetildesmashAux[-0.15ex]{#2}}%
    }{%
        \widetildesmashAux[#1]{#2}%
    }%
}
\NewDocumentCommand{\D}   {s O{}}{\IfBooleanTF{#1}{\widetildesmash{\Delta}}{\Delta}\IfBlankF{#2}{^{\!#2}}}
\NewDocumentCommand{\Dpr} {s}{\IfBooleanTF{#1}{\widetildesmash{\Delta}}{\Delta}^{\mspace{-2mu}\prime}}
\NewDocumentCommand{\Dstar}{s O{}}{\IfBooleanTF{#1}{\widetildesmash{\Delta}}{\Delta}_{\star}\IfBlankF{#2}{^{\!#2}}}
\NewDocumentCommand{\Dbullet}{s O{}}{\IfBooleanTF{#1}{\widetildesmash{\Delta}}{\Delta}_{\bullet}\IfBlankF{#2}{^{\!#2}}}
\NewDocumentCommand{\Dprstar}{s}{\IfBooleanTF{#1}{\widetildesmash{\Delta}}{\Delta}^{\mspace{-2mu}\prime}_{\star}}
\NewDocumentCommand{\Dphi}{s}{\IfBooleanTF{#1}{\widetildesmash{\Delta}}{\Delta}_{\phi}}

\NewDocumentCommand{\MFT}{s}{\ensuremath{\IfBooleanTF{#1}{\text{MFT}}{(\text{MFT})}}}
\NewDocumentCommand{\OPEcoef}{O{}}{f\IfBlankF{#1}{^{(#1)}}}
\NewDocumentCommand{\OPEfunc}{O{}}{C\IfBlankF{#1}{^{(#1)}}}
\NewDocumentCommand{\OPEsq}{O{}}{C\IfBlankF{#1}{^{(#1)}}}
\NewDocumentCommand{\anomdim}{O{1}}{\gamma\IfBlankF{#1}{^{(#1)}}}

\NewDocumentCommand{\Cnorm}{O{\smash{\Dphi}}}{\mathfrak{C}_{\mspace{-1mu}#1}}
\NewDocumentCommand{\Gbulkbulk}{O{\Dphi}}{G^{bb}_{\mspace{-2mu}#1}}
\NewDocumentCommand{\Gbulkbdry}{O{\Dphi}}{G^{b\partial}_{\mspace{-2mu}#1}}
\NewDocumentCommand{\Gbdrybdry}{O{\Dphi}}{G^{\partial\partial}_{\mspace{-2mu}#1}}

\NewDocumentCommand{\HypGeoNoArgs}{s O{2} O{1}}{
    \mathop{{}_{#2}\bm{\IfBooleanTF{#1}{\widetilde{\mathsf{F}}}{\mathsf{F}}}_{#3}}
}
\NewDocumentCommand{\HypGeo}{s O{2} O{1} m m O{1}}{
    \IfBooleanTF{#1}{\HypGeoNoArgs*[#2][#3]}{\HypGeoNoArgs[#2][#3]}
    \left\lbrack
    \begin{+matrix}[cells={c},rowsep=0.5pt,colsep=2pt]
    #4 \\ #5 \\
    \end{+matrix}
    \mathcolor{black!60}{\innermiddle|} #6
    \mspace{2mu}\right\rbrack
}
\NewDocumentCommand{\anomdimFfunction}{O{d} m m}{
    \operatorname{\mathsf{F}}^{\smash{\raisemath{0.4ex}{\mathcolor{darkgray}{(#1)}}}}_{\!#2,\.#3}
}

\NewDocumentCommand{\Repre}    {s}{\ensuremath{\mathtt{\IfBooleanTF{#1}{R }{(R) }}}}
\NewDocumentCommand{\Singlet}  {s}{\ensuremath{\mathtt{\IfBooleanTF{#1}{S }{(S) }}}}
\NewDocumentCommand{\AntiSym}  {s}{\ensuremath{\mathtt{\IfBooleanTF{#1}{AS}{(AS)}}}}
\NewDocumentCommand{\SymTrless}{s}{\ensuremath{\mathtt{\IfBooleanTF{#1}{ST}{(ST)}}}}
\NewDocumentCommand{\NonSinglet}{s}{\SymTrless/\AntiSym}

\NewDocumentCommand{\ARepre}{O{\Repre}}{\mathcal{A}_{#1}}

\NewDocumentCommand{\FourPtNonSinglet}{O{1}}{\FourPt[\(\substack{\SymTrless*\\\AntiSym*}\)][#1]}

\NewDocumentCommand{\Proj}{O{} o}{%
    \IfValueTF{#2}%
    {\fbraces{\lparen}{\rparen}{\mathcal{P}_{#1}}{#2}}%
    {\mathcal{P}_{#1}}%
}

\NewDocumentCommand{\Knorm}{O{\D*,J}}{\operatorname{\mathnormal{K}}_{\xmathstrut[-1]{0.33}\mspace{-4mu}#1}}
\NewDocumentCommand{\Snorm}{O{\D*,J}}{\operatorname{\mathnormal{S}}_{\xmathstrut[-1]{0.33}\mspace{-2mu}#1}}
\NewDocumentCommand{\ConfBlock}{s O{\D,J} O{s}}{
    \operatorname{\mathnormal{G}}
    \IfBooleanTF{#1}
    {_{#2}^{(#3)}}
    {_{\xmathstrut[-1]{0.33}\mspace{-2mu}#2}^{\raisemath{0.5ex}{(#3)}}}
}
\NewDocumentCommand{\CPWnorm}{O{\D,J}}{n_{#1}}
\NewDocumentCommand{\sixjsymbol}{s}{%
    \IfBooleanTF{#1}
    {\ensuremath{\mathsf{6j}\text{--Symbol}}}
    {\ensuremath{\mathsf{6j}\text{--symbol}}}%
}
\NewDocumentCommand{\CrK}{O{\D,J}O{\Dpr,J'}O{20mu}}{\operatorname{\mathsf{CrK}}\IfBlankTF{#1}{^{s\leftarrow t}}{^{\mspace{#3}s\,\longleftarrow\,t}_{\bra{#1}\ket{#2}}}}
\makeatother

\title{\paperTitle}

\author[a]{Jozef Csipes}\emailAdd{csipesjozef@gmail.com}
\author[b,c]{and Petr Vaško}\emailAdd{petr.vasko@matfyz.cuni.cz}

\affiliation[a]{Institute of Theoretical Physics,          Charles University, V Holešovičkách 2, 180\,00 Prague, Czech Republic}
\affiliation[b]{Institute of Particle and Nuclear Physics, Charles University, V Holešovičkách 2, 180\,00 Prague, Czech Republic}
\affiliation[c]{Institute of Physics of the Czech Academy of Sciences, Na Slovance 1999/2, 182\,00 Prague, Czech Republic}

\abstract{The \( \OO(N) \) \CFT[B] is analyzed in the large \(N\) expansion in a generic bulk dimension \(2<d+1<4\). Focus is on boundary conditions corresponding to the \emph{ordinary} transition, however techniques used can be generalized to special or extraordinary transitions. We study the system of bulk \(2\)-pt functions \(\langle \phi\phi\rangle\), \(\langle \phi^2 \phi^2\rangle\) and \(\langle \sigma\sigma \rangle\), where \(\sigma\) is the Hubbard--Stratonovich field. They are expanded in bulk/boundary conformal partial waves. The coefficient functions of this expansion -- the \emph{spectral functions} -- encode \CFT[B] data and determine bulk/boundary conformal block expansions. We prove conjectures about boundary spectral functions in Dujava et al.~\cite{Dujava:2025php} and reproduce the bulk expansion of \(\langle \phi\phi \rangle\) in Giombi et al.~\cite{Giombi:2020rmc}. The bulk expansion of \(\langle\sigma\sigma\rangle\) is new and allows to extract the leading large \(N\) expression for an unknown OPE coefficient \(C_{\sigma\sigma\sigma^3}\) (the computation outputs as a byproduct also a mixing coefficient, which matches an existing result by Derkachov et al.~\cite{Derkachov:1997gc}, providing further support for the main result). Besides these derived data infinitely many constraints are produced, though due to operator mixing growing in complexity as degeneracy increases, they cannot be disentangled without further input.
}


\arxivnumber{}

\begin{document}
\pdfbookmark[2]{Title Page}{titlepage} 
\maketitle
\flushbottom

\section{Introduction}
\label{sec:intro}
The history of the critical \(\OO(N)\) model reaches far back in time and that is true even for the variant with a boundary, which is of interest to us in this work. Without any doubt it is a fundamental theory in the realm of theoretical physics, importantly still a subject of current research. Our goal isn't to give a detailed review (that was already done by authors cited below), instead we wish to move straight to the point: \textit{(i)} derive complete bulk/boundary conformal block decompositions of the two simplest bulk-bulk \(2\)-pt functions \(\left\langle \phi\phi \right\rangle\), \(\left\langle \sigma\sigma \right\rangle\) (to be defined below) and \textit{(ii)} extract from them as much \CFT[B] data as possible.

Nevertheless, let us point readers interested in an introduction to the critical \(\OO(N)\) model with boundary (\(\OO(N)\) \CFT[B]) to existing literature. Its history began in the statistical mechanics context by the work~\cite{Mills1971} in 1971 followed by two similar ones in spirit~\cite{BinderHohenberg1972,BinderHohenberg1974}. Shortly afterwards, in 1975, first field-theoretic papers~\cite{LubenskyRubin1975a,LubenskyRubin1975b,BrayMoore1977} started to appear and implement the idea of renormalization group~\cite{Wilson1971a,Wilson1971b,WilsonFisher1972,Wilson1975}. These papers recognized the importance of boundary universality classes and coined the terms \emph{ordinary, special, extraordinary} transition (we treat just the \emph{ordinary} one). Further progress came at the beginning of 80's with~\cite{DiehlDietrich1980,DiehlDietrich1981}, on the computational front \(\varepsilon\)-expansions were pushed to higher orders, but perhaps more significantly these authors systematized various scattered ideas to a unified field-theoretic treatment of boundary critical phenomena (the concept of a boundary operator expansion was introduced). This effort culminated in the 1986 review~\cite{Diehl1986}, which became and still is a standard reference in the field. A systematic large \(N\) approximation was applied to the \(\OO(N)\) \CFT[B] in a series of papers~\cite{OhnoOkabe1983a,OhnoOkabe1983b,OhnoOkabe1983c,OhnoOkabe1984} starting in 1983.

In parallel, boundary critical behavior specifically in two (bulk) dimensions was connected to \CFT{} by Cardy. His groundbreaking 1984 paper~\cite{Cardy:1984bb} demonstrated an isomorphism between boundary conditions and boundary states that he introduced and constructed. It opened the door to exact solutions of \(2\)D \CFT[B]s and clearly a large number of vital follow up work was published. However, specialized 2D techniques are beyond the scope of this paper, thus we mention only his essential contribution~\cite{Cardy:1989ir} that in some sense completed developments in this era and field of research. What was accomplished by Cardy and his followers in two dimensions was ported to higher dimensions in the first half of the 90's by~\cite{McAvity:1993ue,McAvity:1995zd}, establishing a self-contained \CFT[B] framework (obviously without the power leading to exact solutions as conformal algebra in higher dimensions is finite dimensional). This progress was synthesized in Diehl's 1997 review~\cite{Diehl:1996kd}, updating his previous one~\cite{Diehl1986}.

The modern era of \CFT[B] stays on the shoulders of revived conformal bootstrap methods~\cite{Rattazzi:2008pe,El-Showk:2012cjh}, which were adapted to theories with a boundary (in particular the \(\OO(N)\) model) in~\cite{Liendo:2012hy}. Since this seminal paper, the number of publications studying critical theories in the presence of boundaries/defects grows rapidly. Roughly (with unavoidable overlaps and omissions, in particular we exclude Monte Carlo, Hamiltonian truncations and Functional renormalization group that we are not familiar with as well as specialized 2D techniques), they can be categorized into the following classes:
\begin{itemize}
\item bootstrap methods:
  \begin{itemize}
  \item analytic: unfortunately, we can't give proper credit to all the people contributing in this area; the list~\cite{Billo:2016cpy,Hogervorst:2017kbj,Rastelli:2017ecj,Lauria:2017wav,Lemos:2017vnx,Mazac:2018biw,Bissi:2018mcq,Liendo:2019jpu,Bissi:2022mrs} represents solely references we came across while working on this project
  \item numeric: these studies~\cite{Padayasi:2021sik,Hu:2025yrs} concern the extraordinary/normal boundary universality class of the \(\OO(N)\) \CFT[B] and were designed to illuminate the existence and properties of the newly discovered ``extraordinary-log'' transition~\cite{Metlitski:2020cqy}
  \end{itemize}
\item perturbations in a parameter (either large \(N\) or \(\varepsilon\)-expansions), often upgraded compared to previous standard treatments on half space by Weyl mapping the critical theory to hyperbolic space (\EAdS) to mitigate the effects of boundary as much as possible: only a limited selection we managed to familiarize ourselves with contains~\cite{Giombi:2020rmc,Diehl:2020rfx,Cuomo:2021kfm,Diatlyk:2024ngd,Giombi:2025pxx,Drukker:2026nvv,Sun:2026mib,Diatlyk:2026eta}
\item fuzzy sphere approach~\cite{Zhu:2022gjc}: an incomplete list of contributions treating the \(\OO(N)\) model consist of~\cite{Lauchli:2025fii,Dey:2026cso} while a selection of papers focusing on \CFT[B] includes~\cite{Zhou:2024dbt,Dedushenko:2024nwi,Feng:2026iii}
\end{itemize}

Perhaps the last major development partial to the \(\OO(N)\) model came with~\cite{Metlitski:2020cqy}, where  a standard belief regarding the extraordinary transition was revised. It was realized that the decay of boundary correlations for the \(3\)D \(\OO(N)\) \CFT[B] tuned to the extraordinary boundary universality class is not power-law (as was suggested by previous studies, especially \(\varepsilon\)-expansions) for all \(N\geq 2\), instead they decay logarithmically in a certain range \(2\leq N<N_c\). Thus a novel ``extraordinary-log'' boundary universality class was introduced and a chase after the critical value \(N_c\) started~\cite{Padayasi:2021sik,Toldin:2021kun}.

The \(\OO(N)\) model will remain a cornerstone for next generations of theoretical physicists to come. While a complete solution remains elusive, the pursuit itself is deeply inspiring. It is remarkable that the first direct experimental test of conformal invariance in critical phenomena, as proposed recently in~\cite{Podo:2026hfh}, is based on a \(2\)-pt function in the presence of a boundary. We hope that our study of \(2\)-pt functions within the \(\OO(N)\) \CFT[B] can contribute at least a tiny bit along this fascinating journey.

\subsection{Summary of results: \texorpdfstring{\CFT[B]}{BCFT} data}
\label{sec:summary-results}
Below, we point readers who don't wish to go through the whole text to the main results achieved in this paper. Note that all computations were done for the \emph{ordinary} transition. The expansion coefficients calculated in this text are also available in the supplementary \wmathematica{} notebook\footnote{The attached file \texttt{ONBCFTdata.m} is essentially a text file that can be read by any software, including standard text editors.} including some that did not fit into this text. Possible extensions of this work are discussed in~\Cref{sec:outlook}.

\paragraph{Boundary data.}
\begin{enumerate}
\item\label{item:1} In~\eqref{eq:spec-bdy-sigma-even-dim}, we prove conjectures made in~\cite{Dujava:2025php} about the boundary spectral function associated with the bulk-bulk \(2\)-pt function \(\left\langle \delta\sigma\delta\sigma \right\rangle\) that are valid for even \(d\) (so that the bulk dimension of the theory \( \CFT[B]_{d+1} \) is odd).
\item\label{item:2} In~\eqref{eq:sigma-BOE} we define an infinite tower of operators, \(\widehat{\sigma}_k,\,k\in\mathbb{Z}_{\geq 0}\), appearing in the BOE of \(\sigma\) and in~\eqref{eq:sigma-BOE-coeff} their squared BOE coefficients \(( b_{\sigma \widehat{\sigma}_k} )^2\) are provided for any \(d\). Their asymptotics for \(k\to\infty\) is verified to obey (complex) Tauberian theorems, as it should. We didn't find the coefficients for the entire tower elsewhere in the literature, typically only first few are presented by Taylor expanding the position space correlator around the boundary limit. Correctness of the BOE coefficients is verified by explicit boundary conformal block expansions, see~\Cref{fig:boundary-cb-exp-convergence-2d} and~\Cref{fig:boundary-cb-exp-convergence}.
\end{enumerate}

\paragraph{Bulk data.}
\begin{enumerate}
\item\label{item:3} In~\Cref{sec:bulk-exp} bulk spectral functions associated with \(\left\langle \phi\phi \right\rangle\) and \(\left\langle \sigma\sigma \right\rangle\) were computed and from those coefficients of bulk conformal block decompositions derived. We don't link here to concrete expressions as they are not very illuminating. However, let as point out that their correctness was verified by explicitly plotting the sequence of partial bulk block sums and checking that they converge to the exact correlators, as displayed in~\Cref{fig:bulk-cb-exp-convergence-2d} and~\Cref{fig:bulk-cb-exp-convergence}.
\item\label{item:4} In~\eqref{eq:bulk-primaries}, a unique infinite tower of bulk primaries contributing both to the \(\phi\times\phi\) and \(\sigma\times\sigma\) OPE defined in~\eqref{eq:phi-sig-bulk-ope} was identified. Degeneracy of operators in this tower with equal scaling dimensions was worked out for low scaling dimensions needed.
\item\label{item:5} The above identification was verified by various \emph{crosschecks}. Note that for OPE coefficients we offer slightly more than just a test as we fixed also their signs (in most of the literature OPE coefficients were computed from bulk \(4\)-pt functions sensitive just to their squares, contrary to \CFT[B] capable of fixing also signs). All checks regard the leading non-trivial large \(N\) order (emphasized as a superscript enclosed in parenthesis for each quantity of interest) and references used for crosschecking are included at relevant places where the links bellow point to.
  \begin{itemize}
  \item[--] Bulk-bulk \(2\)-pt function \(\left\langle \phi\phi \right\rangle\):
    \begin{itemize}[label=\textbullet]
    \item The OPE coefficient \(C_{\phi\phi\sigma}\) was extracted in~\eqref{eq:ope-coeff-phi-phi-sigma} and successfully matched against a known result.
    \item Prediction for the OPE coefficient \(C_{\phi\phi\sigma^2}\) in~\eqref{eq:ope-coeff-phi-phi-sigma2} was verified to agree with existing literature.
    \end{itemize}
  \item[--] Bulk-bulk \(2\)-pt function \(\left\langle \sigma\sigma \right\rangle\):
    \begin{itemize}[label=\textbullet]
    \item The OPE coefficient \(C_{\sigma\sigma\sigma}\) was derived in~\eqref{eq:ope-coeff-sigma-sigma-sigma} and checked to conform to previous studies.
    \item Result for the OPE coefficient \(C_{\sigma\sigma\sigma^2}\) given in~\eqref{eq:ope-coeff-sigma-sigma-sigma2} was found to coincide with existing calculations.
    \item Anomalous dimension of \(\sigma^2\) is determined in~\eqref{eq:k2-anom-dim-predict} and observed to comply with known results.
    \end{itemize}
  \end{itemize}
  \item\label{item:6} Finally, we were able to derive also a new \emph{prediction} that to our knowledge didn't appear elsewhere.
    \begin{itemize}
    \item The yet unknown OPE coefficient \(C_{\sigma\sigma\sigma^3}\) was computed in~\eqref{eq:OPE-sig-sig-sig3}.
    \item As a byproduct of the analysis leading to its derivation, also a mixing coefficient, defined in~\eqref{eq:eigenstate-k3-mix-SigBoxSig}, linking the two primaries \(\sigma^3\) and \([\sigma\sigma]_{1,0}\) (a double-twist operator defined in~\eqref{eq:sigma-box-sigma-explicit}) was given in~\eqref{eq:mix-coeff-sol}. Its expression exactly matches a previous computation.
    \end{itemize}
    \item\label{item:7} The remaining infinite sequence of bulk conformal block coefficients (composed of a product of an OPE coefficient with a \(1\)-pt function) imposes severe constraints on \CFT[B] data. However, due to ever growing operator degeneracy implying more involved operator mixing, we were unable to harness it immediately for extraction of either the OPE coefficients or the \(1\)-pt functions. An example of such a constraint is shown in~\eqref{eq:phi-k3-eqn}.
\end{enumerate}

\subsection{Conventions}
\label{sec:conventions}
Before closing the introduction, let us compile here our conventions (which can be safely skipped, except~\eqref{eq:cross-ratios}, by expert readers familiar with standard \CFT[B] notation and different large \(N\) counting schemes), in order to facilitate the reading of this paper.
\paragraph{Coordinates and cross ratios.}
\begin{itemize}
\item A bulk point has coordinate \(x= \left(  \mathbf{x}, x_{\perp}\right)\), where \(x_{\perp}\) is the distance from the bulk point to the boundary in flat half plane.
\item
In a \CFT[B] there is a single real-valued cross ratio for bulk-bulk \(2\)-pt functions, whose definition is not unique. We will work with the following two definitions
\begin{align}
\label{eq:cross-ratios}
&\xi\left(x^1,x^2\right)=\frac{\left(x^1-x^2\right)^2}{\left(2x^1_{\perp}\right)\left(2x^2_{\perp}\right)}\in [0,\infty], && \rho=\frac{\xi}{1+\xi}\in [0,1]\eqend
\end{align}
It is worth recalling two important limits: \textit{(i)} the bulk/OPE limit, when the two points collide in the bulk away from the boundary corresponds to \(\xi\to 0\) or \(\rho\to 0\); \textit{(ii)} the boundary/BOE limit when either of the points approaches the boundary while separated in the bulk corresponds to \(\xi\to\infty\) or \(\rho\to 1\).
\end{itemize}

\paragraph{Bulk/boundary objects.}
\begin{itemize}
\item Bulk/boundary conformal blocks, partial waves, spectral functions are denoted by the superscript \(B/\partial\): \(\mathsf{G}^{B/\partial},\;\mathsf{CPW}^{B/\partial},\;\mathsf{Spec}^{B/\partial}\).
\item Bulk/boundary operators or scaling dimensions are denoted as: \(\mathsf{O}/\widehat{\mathsf{O}}\), \(\Delta/\widehat{\Delta}\).
\end{itemize}

\paragraph{Correlators.}
\begin{itemize}
\item Mostly, we consider correlators in \(\CFT[B]_{d+1}\) and by the Weyl mapping it is convenient to define them on the hyperbolic space \(\mathsf{H}_{d+1}\). So the symbol \(\langle \ind \rangle\) defines a correlator on \(\mathsf{H}_{d+1}\).
\item When explicitly needed, we define correlators on flat half plane \(\mathrm{FHP} \coloneq \R^d\times \R_{\geq 0}\) as \(\langle \ind \rangle_{\mathrm{FHP}}\). They are related to those on the hyperbolic space by standard Weyl factors to be defined below.
\item Occasionally, we write correlators on the flat plane without boundary \(\mathrm{FP}\coloneq \R^{d+1}\) as \(\langle \ind \rangle_{\mathrm{FP}}\).
\item All correlators are understood as the complete ones, including disconnected parts. However, since we study the ordinary transition, \(1\)-pt functions of the \(\phi\)-field vanish. Thus, the only case where this distinction matters concerns the \(\sigma\)-field, \(\sigma=\langle\sigma\rangle\widehat{\mathds{1}}+\delta\sigma\), such that \(\langle \delta\sigma \rangle=0 \) and consequently \(\langle\sigma\sigma\rangle\coloneq\langle\sigma\rangle^2+\langle\delta\sigma\delta\sigma\rangle\).
\end{itemize}

\paragraph{Large \(N\) counting.}
\begin{itemize}
\item We employ normalization of the Hubbard--Stratonovich field $\sigma$, so that the interaction vertex \(\sigma\left( \phi^{\ind} \right)^2\) scales as \(\sim N^{-\tfrac{1}{2}}\).
\item When we need to show an explicit order of a quantity in the large \(N\) expansion, we do so by a superscript enclosed in parenthesis. For instance, with the above normalization, some important leading order observables scale as: OPE coefficient \(C_{\phi\phi\sigma}\sim \langle \phi\phi\delta\sigma \rangle_{\mathrm{FP}}^{\left(-1/2\right)}\sim N^{-\tfrac{1}{2}}\), \(1\)-pt function \(\langle \sigma \rangle^{\left(1/2\right)}\sim N^{\tfrac{1}{2}}\), connected correlator \(\langle \delta\sigma\delta\sigma\rangle^{(0)}\sim N^0\) or \(\langle\phi\phi\rangle^{(0)}\sim N^{0}\).
\end{itemize}

 \section{Preliminaries}
\label{sec:preliminaries}

We begin with a preparatory section that serves two purposes: \textit{(i)} to introduce the $\mathsf{O}(N)$ model and establish our conventions, including the essential elements of \(\CFT[B]\); \textit{(ii)} to define the spectral transformations that will be central to extraction of the data.

\subsection{The \texorpdfstring{\(\OO(N)\)}{O(N)}  model}

Let us start by recalling certain aspects of the large \(N\) limit of the \(\mathsf{O} (N)\) model which will be fundamental to our work.
We are interested in its interacting fixed point (the Wilson-Fisher fixed point) in the presence of a boundary.
More thorough treatment of this model on flat space can be found in specialized reviews \cite{Moshe:2003xn, Fei:2014yja}, or for the case of \(\CFT[B]\) \cite{Giombi:2020rmc}.
Recently, the behavior of this model on \(\mathsf{AdS}\) space has also been examined in \cite{Carmi:2018qzm, Dujava:2025php}.

The \(\mathsf{O} (N)\) vector model describes \(N\) interacting scalar fields with the Euclidean bulk action
\begin{align}
    \label{eq:ON-action-non-conformal}
	\mathcal{S} [\phi^{i} ] &= \int_{\mathcal{M}}  \mathrm{d}^{d+1} x \, \left[ \frac{1}{2} \left( \partial \phi^{i}  \right)^{2} + \frac{1}{2}\left(  m^{2} + \xi R\right) \phi^{i} \phi_{i}  +   \frac{\lambda }{4 N} \left( \phi^{i} \phi_{i}   \right)^{2}    \right]\eqcomma
\end{align}
where we have included the potential coupling to the curvature \(\frac{1}{2} \xi  R \phi^{2} \) which is important if the underlying manifold \(\mathcal{M} \) is not flat.
Here, \(R\) is the Ricci scalar.

If, however, the underlying manifold \(\mathcal{M}\) is flat, the theory has a Gaussian fixed point in the UV, which corresponds to the free massless theory (\(m^{2} = \lambda = 0\)).
In dimensions \(2 < d+1 < 4\) the mass and the  \(\phi^{4} \) interaction term are relevant and drive the RG flow to the interacting Wilson-Fisher fixed point (with a specific scheme-dependent value of \(m^{2} / \lambda \)).
To better understand the large \(N\) behavior of this theory, one can introduce the auxiliary Hubbard-Stratonovich field \(\sigma\) and consider the action
\begin{align}
    \label{eq:HS-action-non-conformal}
	\mathcal{S}_{\text{HS}} [\phi^{i} , \sigma ] &= \int_{\mathcal{M}} \mathrm{d}^{d+1} x \,
		 \left[ \frac{1}{2} \left( \partial \phi^{i}  \right)^{2}  + \frac{1}{2}(m^{2} + \xi  R) \phi^{i} \phi_{i}  - \frac{1}{4 \lambda }\sigma^{2} + \frac{1}{2\sqrt{N} } \sigma \phi^{i} \phi_{i}   \right]\eqcomma
\end{align}
which is equivalent to \eqref{eq:ON-action-non-conformal} as can be seen by the equation of motion for \(\sigma\) which just sets \(\sigma = \frac{\lambda }{\sqrt{N} } \phi^{2}\).
Here, the critical theory can be obtained by performing the limits \(m^{2} \to 0\), \(\lambda \to \infty\), and \(\xi \to \frac{1}{4} \frac{d-1}{d}\), resulting in the simplified action
\begin{align}
    \label{eq:HS-action-critical}
	\mathcal{S}_{\text{HS}} [\phi^{i}, \sigma  ] &= \int_{\mathcal{M}}\mathrm{d}^{d+1} x  \, \left[ \frac{1}{2} \left( \partial \phi^{i}  \right)^{2} + \frac{d-1}{8d}R \phi^{i}\phi_{i}    + \frac{1}{2\sqrt{N} } \sigma  \phi^{i} \phi_{i}   \right] \eqend
\end{align}
The benefit being that this action is quadratic which allows us to integrate out the \(\phi^{i}\) resulting in an effective action for \(\sigma \).
The partition function for this effective action becomes
\begin{align}
    \label{eq:Z-conformal-general}
    \mathcal{Z} = \int \mathcal{D} \sigma \, \exp \left[ - \frac{N}{2} \Tr \log \left( - \nabla^{2} + \frac{d-1}{4d}R + \frac{\sigma }{\sqrt{N} } \right)  \right] \eqend
\end{align}

For our purposes, the space manifold \(\mathcal{M}\) will be a half-space \(\mathbb{R}^{d+1}_{\ge 0} \) with a flat boundary at \(x^{d+1} \equiv x_{\perp} = 0\).
Considering the action \eqref{eq:HS-action-critical} on it leads to a \(\CFT[B]\) which has several available conformal boundary conditions that are connected by boundary RG flows (see~\cite[Fig. 2]{Giombi:2020rmc} for a useful summary).
Starting with the \textit{special transition} in the UV, one can add a relevant boundary interaction term \(c \hat{\phi}^{2} \) (with positive \(c\)) which drives the RG flow to the \textit{ordinary transition} while adding the same operator with a negative \(c\) leads to an RG flow to the \textit{extraordinary transition}.
On the other hand, if one starts in the ordinary transition, a deformation by the operator \(h \cdot \hat{\phi} \) (where \(h\) is a vector of constants) leads again to the extraordinary transition.
Our interest will be in the ordinary transition which can intuitively be understood—if one considers the model as a lattice of spins—as a transition connecting a phase where both bulk and boundary are ordered with a phase where they are both disordered.

A different perspective on the boundary critical theory is provided by \(\mathsf{AdS}/\CFT\) techniques, from which it is well known that one can map the half-space \(\mathbb{R}^{d+1}_{\ge 0}\) onto a hyperboloid (the Euclidean Anti-de Sitter space) \(\mathsf{H}_{d+1}\) by a Weyl transformation.
This map is particularly clear if one considers the metric for \(\mathsf{H}_{d+1}\) in the Poincaré coordinates,
\begin{align}
    \label{eq:AdS-Poincare-metric}
    &\mathrm{d} s^{2}_{\mathsf{H}_{d+1}} = \frac{1}{x_{\perp}^{2}  } \left( \mathrm{d} x_{\perp}^{2}  + \mathrm{d} \mathbf{x}^{2} \right)\eqcomma \quad x_{\perp} \ge 0 \eqcomma
\end{align}
which is clearly Weyl equivalent to the flat metric on a half-space.
We always consider the hyperboloid to have a unit radius.
This provides a correspondence between \(\CFT[B]\)s and \(\CFT\)s on the hyperboloid, meaning that the correlators in \(\CFT[B]\) can be translated into correlators on \(\mathsf{H}_{d+1}\) by simply performing a Weyl transformation.
The Ricci scalar on \(\mathsf{H}_{d+1}\) is a constant \(R = -d(d+1)\) which can readily be substituted into \eqref{eq:Z-conformal-general} and then one can employ the saddle point approximation to find the large \(N\) description of the theory.
We can split the operator \(\sigma\) as
\begin{align}
    \label{eq:sigma-split}
    \sigma(x) = \langle \sigma \rangle + \delta \sigma(x)
\end{align}
where \(\langle \sigma \rangle\) is a constant when considered on the hyperbolic space (the saddle point) and \(\langle \delta \sigma \rangle = 0\).
Plugging this back into the Hubbard-Stratonovich action \eqref{eq:HS-action-critical} makes it clear that the saddle point \(\langle \sigma \rangle\) only modifies the effective mass of the \(\phi \) field. Expanding the effective action in \eqref{eq:Z-conformal-general} around the saddle point and only keeping the leading terms in large \(N\) gives the connected part of effective propagator for the \(\sigma\) field
\begin{align}
    \label{eq:sigma-prop-from-bubble}
    \langle \delta \sigma \delta \sigma \rangle &= - 4 B^{-1} (x,y)
\end{align}
where \(B(x, y ) = \frac{\langle \phi^{2}(x) \phi^{2}(y) \rangle}{N} = \frac{2}{N} \langle  \phi^{i}\phi^{j} \rangle \langle  \phi_{i} \phi_{j} \rangle\) is the bubble function and by \(B^{-1}\) we mean its inverse with respect to convolutions on \(\mathsf{H}_{d+1} \).

In the large \(N\) limit we can solve the saddle point equation and find \(\langle \sigma \rangle\) that extremizes the effective action for \(\sigma\) (see \cite{Giombi:2020rmc} for detailed derivation). In particular, the solution for the ordinary transition reads
\begin{align}
    \label{eq:sigma-saddle-point}
    \langle \sigma  \rangle &= \langle \sigma \rangle^{\left(1 / 2\right)} + \mathcal{O} (1) =  \sqrt{N}\frac{(d-1) (d-3)}{4} + \mathcal{O} (1) \eqend
\end{align}
The saddle point of \(\sigma\) contributes to the effective mass of the \(\phi\) fields and their propagator is just a free propagator on \(\mathsf{H}_{d+1}\) with this mass. This can be translated by standard \AdSCFT{} dictionary to correspond to boundary scaling dimension \(\widehat{\Delta}_{\widehat{\phi}}=d-1\)
\begin{align}
    \label{eq:phi-prop}
    \langle \phi^{i} \phi^{j} \rangle &= \langle \phi^{i} \phi^{j} \rangle^{\left( 0 \right)} + \mathcal{O} \left( \tfrac{1}{N} \right) =\delta^{i j} \mathcal{N}_\phi \rho^{-\frac{d-1}{2}} \left( 1- \rho  \right)^{d-1}+ \mathcal{O} \left( \tfrac{1}{N} \right) \eqend
\end{align}
Here we have used the \(\CFT[B]\) cross-ratio \(\rho\) from \eqref{eq:cross-ratios} and the normalization factor is given by
\begin{align}
\label{eq:phi-norm}
&\mathcal{N}_\phi=\mathcal{N}_\phi^{\mathrm{FP}} 2^{1-d},\quad \text{with} \quad \mathcal{N}_\phi^{\mathrm{FP}}=  \frac{\Gamma \left( \tfrac{d-1}{2} \right)}{4\pi^{\frac{d+1}{2}}} \eqend
\end{align}
Then the bulk OPE limit \(\rho\to 0\) yields the \(2\)-pt function of the canonically normalized free masless scalar on \(\R^{d+1}\)
\begin{align}
\label{eq:phi-prop-ope-lim}
  \left(\frac{1}{x_\perp}\right)^{\frac{d-1}{2}}\left(\frac{1}{y_\perp}\right)^{\frac{d-1}{2}}\langle\phi^i\phi^j\rangle=\langle\phi^i\phi^j\rangle_{\mathrm{FHP}}\xrightarrow[\rho\to 0]{} \langle\phi^i\phi^j\rangle_\mathrm{FP}=\delta^{ij} \frac{\mathcal{N}_\phi^{\mathrm{FP}}}{\left[\left( x-y \right)^2\right]^{\frac{d-1}{2}}} \eqcomma
\end{align}
and we introduced the normalization \(\mathcal{N}_{\sigma}^{\mathrm{FP}}\) in the OPE limit since it will play a role when we will be computing \CFT[] data from bulk conformal block expansions.

The connected \(\sigma\)-propagator for the ordinary transition can be found by inverting the bubble function as in \eqref{eq:sigma-prop-from-bubble} which yields (see \cite{Giombi:2020rmc} for details)
\begin{align}
    \label{eq:sigma-prop-connected}
    \langle \delta \sigma \delta \sigma \rangle &= \mathcal{N}_\sigma \left( \frac{\rho }{1-\rho } \right) ^{-d-1}  \HypGeo[2][1]{d-1,\, d+1}{2d-2}[- \frac{1-\rho }{\rho }] +\mathcal{O} \left( \tfrac{1}{N} \right)\eqcomma
\end{align}
where the normalization takes the form
\begin{align}
\label{eq:sigma-prop-norm}
\mathcal{N}_\sigma=\mathcal{N}_\sigma^{\mathrm{FP}} \frac{\Gamma \left( d-1 \right)\Gamma \left( d+1 \right)}{16\Gamma \left( 2d-2 \right)},\quad \text{with} \quad \mathcal{N}_\sigma^{\mathrm{FP}} =\frac{16\left( 3-d \right) \Gamma \left( d-1 \right)}{\Gamma \left( \frac{3-d}{2} \right) \Gamma \left( \frac{d-1}{2} \right)^{3} }\eqend
\end{align}
The OPE limit then reduces~\eqref{eq:sigma-prop-connected} to (in the Weyl factors we use \(\Delta_\sigma=2+\mathcal{O} \left( \tfrac{1}{N} \right)\))
\begin{align}
\label{eq:sigma-prop-ope-lim}
  \left(\frac{1}{x_\perp}\right)^2\left(\frac{1}{y_\perp}\right)^2\langle\delta\sigma\delta\sigma\rangle=\langle\delta\sigma\delta\sigma\rangle_{\mathrm{FHP}}\xrightarrow[\rho\to 0]{} \langle\delta\sigma\delta\sigma\rangle_\mathrm{FP}=\frac{\mathcal{N}_\sigma^{\mathrm{FP}}}{\left[\left( x-y \right)^2\right]^2} \eqcomma
\end{align}
and we want to stress that our \(2\)-pt functions on flat space are not unit normalized, which will have to be taken into account when using OPE data from the literature.

\begin{figure}[h!]
\centering
\includegraphics[width=\linewidth]{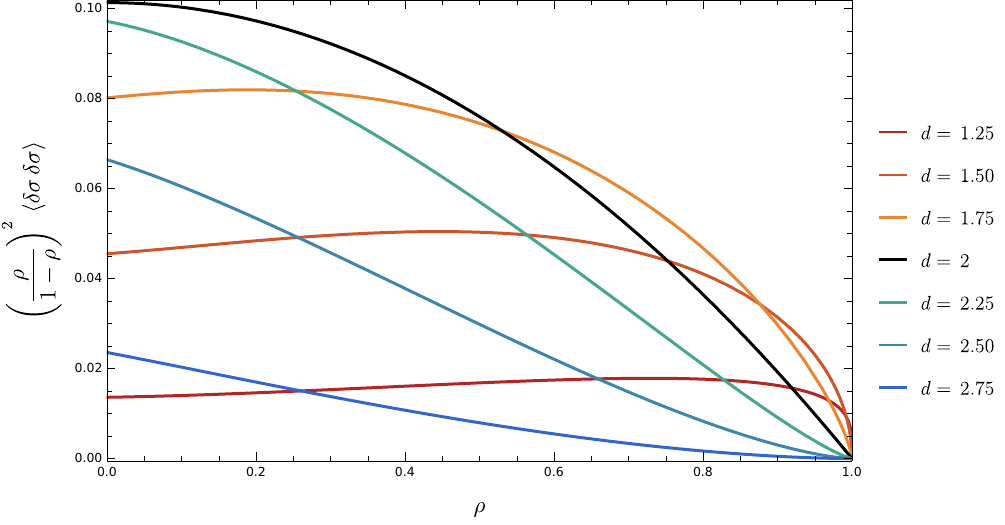}
\caption{Modified connected \(\sigma\)-propagators \(\left( \frac{\rho}{1-\rho} \right)^2 \left\langle \delta\sigma\delta\sigma \right\rangle\) across different dimensions in the interval \(1<d<3\).}
\label{fig:sig-conn-propag}
\end{figure}

For later convenience, it is useful to have a basic picture of these functions across different dimensions. Their OPE bulk limit grows like a power \(\left\langle \delta\sigma\delta\sigma \right\rangle\underset{\rho\to 0}{\sim} \frac{1}{\rho^2}\), while their boundary limit vanishes as \(\left\langle \delta\sigma\delta\sigma \right\rangle\underset{\rho\to 1}{\sim} \left( 1-\rho\right)^{d+1}\). In~\Cref{fig:sig-conn-propag}, we plot a modified function (that will be considered for bulk conformal block expansions in~\Cref{sec:bulk-exp}) with the bulk singularity removed, \(\left( \frac{\rho}{1-\rho} \right)^2 \left\langle \delta\sigma\delta\sigma \right\rangle\).

\subsection{\texorpdfstring{\CFT[B]}{BCFT} kinematics}
\label{sec:bcft-kinematics}

We provide a short review of the \CFT[B] kinematics to fix notation and to present the main tools used to extract the $\CFT$ data. A more thorough treatment can be found in \cite{Liendo:2012hy,McAvity:1995zd, Hogervorst:2017kbj, Billo:2016cpy}.

The presence of the boundary breaks the conformal group of the theory to a subgroup of transformations that stabilize the boundary.
In effect, bulk operators gain a non-zero expectation value whose form is entirely fixed by symmetries except for the overall normalization,
\begin{align}
    \label{eq:1-pt-function-general}
    \langle \mathsf{O} (x) \rangle_{\text{FHP}}  &= \frac{b_{\mathsf{O} \widehat{\mathds{1}}} }{x_{\perp}^{\Delta_{\mathsf{O}} } } \eqend
\end{align}
The coefficient \(b\) is a boundary operator expansion (BOE) coefficient\footnote{The conventions for the BOE coefficient vary in literature. Many authors include an extra factor of \(2\) when defining the 1-pt function so that it is \(\langle  \mathsf{O}(x) \rangle_{\text{FHP}} = b_{\mathsf{O} \hat{\mathds{1}}} / \left( 2 x_{\perp}\right)^{\Delta_{\mathsf{O}}}\). We chose not to do this since we work mostly on \(\mathsf{H}_{d+1}\) and with our conventions the BOE coefficients have the nice interpretation that they are equivalent to the value of the 1-pt function. The downside of this is that we will have to include some extra factors of \(2\) when defining our expansion into conformal blocks in \eqref{eq:bulk-CB-decomposition-general}.}, which appears in the two point function of a bulk and a boundary scalar.
Such a 2-pt function is again completely fixed by conformal invariance except the normalization
\begin{align}
    \label{eq:2-pt-function-bulk-boundary}
    \langle \mathsf{O} (x) \widehat{\mathsf{O}}(\mathbf{y}) \rangle_{\text{FHP}} &=  \frac{b_{\mathsf{O} \widehat{\mathsf{O}}} }{ x_{\perp}^{\Delta -\widehat{\Delta}} \left( x_{\perp}^{2} + \left( \mathbf{x} -\mathbf{y}  \right)^{2} \right)^{\widehat{\Delta }}} \eqend
\end{align}
Clearly, the case of the 1-pt function is just a special case of this formula for \(\widehat{\mathsf{O}} = \widehat{\mathds{1}}\).
The simplest correlator not fixed by the conformal invariance is the 2-pt function of two bulk operators.
Conformal invariance alone dictates that the 2-pt function of two operators of bulk scaling dimensions \(\Delta_{1}\) and \(\Delta_{2}\) can be an arbitrary function of the cross-ratio \(\rho\)
\begin{align}
    \label{eq:2-pt-general}
    \langle \mathsf{O}_{1} \mathsf{O}_{2} \rangle &= \left( \frac{\rho }{1-\rho} \right)^{-\frac{\Delta_{1}+\Delta_{2}}{2}} \mathcal{F}_{\mathsf{O}_{1} \mathsf{O}_{2}}  (\rho) \eqend
\end{align}
Here, the prefactor is a convention and we stress that this is a correlator on the hyperboloid \(\mathsf{H}_{d+1}\), not on a half-plane.
To obtain the correlator on the flat half-plane one would have to perform the Weyl transformation which results in multiplying the above expression by a factor~\(\frac{1}{ x_{\perp}^{\Delta_{1}} y_{\perp}^{\Delta_{2}}}\).

Bulk operators in \CFT[B] allow for an OPE of the standard form (schematically)
\begin{align}
    \label{eq:bulk-OPE}
    \mathsf{O}_{1}(x) \mathsf{O}_{2}(y) &= \sum_{k} \frac{C_{\mathsf{O}_{1} \mathsf{O}_{2} \mathsf{O}_{k}}}{C_{\mathsf{O}_{k}\mathsf{O}_{k}}} \big[  \mathsf{O}_{k}(y) + \text{descendants} \big] \eqcomma
\end{align}
where the sum runs over the operators appearing in the bulk OPE of \(\mathsf{O}_{1}\) and \(\mathsf{O}_{2}\), \(C_{\mathsf{O}_{1} \mathsf{O}_{2} \mathsf{O}_{k}}\) is the corresponding OPE coefficient, and \(C_{\mathsf{O}_{k}\mathsf{O}_{k}}\) is the optional normalization factor appearing in the 2-pt function \(\langle \mathsf{O}_{k}\mathsf{O}_{k} \rangle_{\text{FP}}\), which is only needed if one works with not canonically normalized operators.
This allows for the 2-pt function to be written as a sum of bulk conformal blocks
\begin{align}
    \label{eq:bulk-CB-decomposition-general}
    \mathcal{F}_{\mathsf{O}_{1}\mathsf{O}_{2} }(\rho ) &= \sum_{k} 2^{-\Delta_{1} -\Delta_{2} + \Delta_{k}} \frac{C_{\mathsf{O}_{1} \mathsf{O}_{2} \mathsf{O}_{k} }}{C_{\mathsf{O}_{k}\mathsf{O}_{k}}} b_{\mathsf{O}_{k} \widehat{\mathds{1}}} \mathsf{G}^{B}_{\Delta_{k} }  (\rho) \eqcomma
\end{align}
where the first term is the contribution of the bulk identity \(\mathds{1}\) in the OPE and
\begin{align}
    \label{eq:bulk-CB}
    \mathsf{G}_{\Delta}^{B}(\rho ) &= \rho^{\frac{\Delta}{2}} (1-\rho )^{\frac{\Delta_{1} -\Delta_{2} }{2}} \HypGeo{ \frac{\Delta }{2} + \frac{\Delta_{1}-\Delta_{2}}{2}, \, \frac{\Delta }{2} + \frac{1-d}{2} + \frac{\Delta_{1}-\Delta_{2}}{2}}{\Delta  + \frac{1-d}{2}}[\rho ]\eqend
\end{align}
The precise form of the coefficients appearing in front of the conformal block can be deduced by performing the OPE limit \(\rho \to 0\) and comparing the coefficients with the standard flat space OPE as we show in \Cref{app:rel-cb-coeff-OPE-data}.

On the other hand, in \CFT[B] one can bring a bulk operator close to the boundary and expand it as a sum of boundary operators.
This is the boundary operator expansion (BOE) and can be schematically written as
\begin{align}
    \label{eq:BOE-general}
    \mathsf{O}(x) &= \sum_{k}  b_{\mathsf{O} \widehat{\mathsf{O}}_{k}} \left[ \widehat{\mathsf{O}}_{k}  \left( \mathbf{x}  \right)  + \text{descendants} \right]\eqend
\end{align}
Performing this operation on both operators in a 2-pt function leads to a decomposition into boundary conformal blocks
\begin{align}
    \label{eq:boundary-CB-decomposition-general}
    \langle \mathsf{O}_{1} \mathsf{O}_{2} \rangle &= \sum_{k} 2^{-2 \widehat{\Delta}_{k}} b_{\mathsf{O}_{1} \widehat{\mathsf{O}}_{k} }      b_{\mathsf{O}_{2} \widehat{\mathsf{O}}_{k}} \mathsf{G}^{\partial}_{\widehat{\Delta}_{k} } (\rho ) \eqcomma
\end{align}
where
\begin{align}
    \label{eq:boundary-CB}
    \mathsf{G}^{\partial}_{\widehat{\Delta}} (\rho ) &= (1-\rho )^{\widehat{\Delta }} \HypGeo{\widehat{\Delta }, \, \widehat{\Delta }+ \frac{1-d}{2}}{2 \widehat{\Delta}+1-d}[1-\rho ]\eqend
\end{align}
The boundary CB happens to coincide with the bulk-to-bulk propagator on \(\mathsf{H}_{d+1}\).

The two channels of the CB expansion \eqref{eq:bulk-CB-decomposition-general} and \eqref{eq:boundary-CB-decomposition-general} must reproduce the same function, which is just the statement of the crossing equation, and it puts constraints on the \CFT[B] data allowing for bootstrap solutions \cite{Liendo:2012hy}.

\subsection{Spectral representations}
\label{sec:spectral-rep}
In order to extract the \CFT[B] data from 2-pt functions, it is imperative to have a reliable technique for finding the conformal block decomposition of these functions.
One such technique is provided by the spectral representations \cite{Hogervorst:2017kbj} which first decompose the 2-pt functions into a complete basis of the so-called conformal partial waves (CPWs) and then by the deformation of a contour one obtains a sum of conformal blocks.

\paragraph{Boundary channel.} For the boundary channel one defines the CPWs as a combination of a boundary CB and a shadow CB
\begin{align}
    \label{eq:boundary-CPW-def}
    \begin{split}
        \mathsf{CPW}_{\widehat{\Delta}}^{\partial} \left( \rho  \right) &=  \frac{1}{2} \left[ \mathcal{Q}^{\partial}_{\widehat{\Delta }}  \mathsf{G}^{\partial }_{\widehat{\Delta }} \left( \rho  \right)  + \left( \widehat{\Delta } \to d-\widehat{\Delta} \right)  \right] \\
                                                                        &= \HypGeo{\widehat{\Delta}, \, d- \widehat{\Delta}}{ \frac{d+1}{2}}[- \frac{\rho }{1-\rho}]\eqcomma
    \end{split}
\end{align}
where we defined
\begin{align}
    \label{eq:Q-boundary-def}
    \mathcal{Q}_{\widehat{\Delta }}^{\partial } &= \frac{2 \Gamma \left( \frac{d+1}{2} \right) \Gamma  \left( d - 2\widehat{\Delta} \right)  }{ \Gamma \left(d - \widehat{\Delta} \right) \Gamma \left( \frac{d+1}{2} - \widehat{\Delta} \right)} \eqend
\end{align}
If the scaling dimension \(\widehat{\Delta}\) is chosen to lie on the principal series, \(\widehat{\Delta} \in \frac{d}{2} + i \mathbb{R}\), then the CPWs form a complete orthogonal basis of eigenfunctions of the conformal Casimir operator.
This allows to decompose an arbitrary 2-pt function on \(\mathsf{H}_{d+1}\)—which is just a function of the cross-ratio \(\rho\)—into CPWs as
\begin{align}
    \label{eq:boundary-spec-rep}
    \langle \mathsf{O}_{1} \mathsf{O}_{2} \rangle \left( \rho  \right) &=  \frac{1}{2 \pi i} \int_{\frac{d}{2} + i \mathbb{R}} \frac{\mathrm{d} \widehat{\Delta}}{\mathcal{N}^{\partial}_{\widehat{\Delta}}} \mathsf{Spec}_{\mathsf{O}_{1} \mathsf{O}_{2}}^{\partial} \left(\widehat{\Delta}\right) \mathsf{CPW}_{\widehat{\Delta}}^{\partial } \left( \rho  \right)\eqcomma && \text{where } \mathcal{N}_{\widehat{\Delta }}^{\partial } = \frac{\lvert \mathcal{Q}_{\widehat{\Delta }}^{\partial } \rvert^{2}  }{2}  \eqend
\end{align}
The CPWs are analogous to the \enquote{harmonic functions} that appear in the \(\mathsf{AdS}/\CFT\) literature \cite{Penedones:2010ue, Cornalba:2007fs, Penedones:2007ns, Carmi:2018qzm}.
The spectral representation can also be viewed as a special case of the Jacobi transform \cite{Hogervorst:2017kbj, Koornwinder1984}.
The orthogonality of the CPWs allows the spectral representation to be inverted and the \(\mathsf{Spec}_{\mathsf{O}_{1}\mathsf{O}_{2}}^{\partial } \left( \widehat{ \Delta } \right)\) to be expressed as
\begin{align}
    \label{eq:boundary-inv-formula}
    \mathsf{Spec}_{\mathsf{O}_{1}\mathsf{O}_{2}}^{\partial} \left( {\widehat{\Delta}} \right) &= \int_{0}^{1} \mathrm{d} \rho \, \frac{\rho^{\frac{d-1}{2}}}{\left(1-\rho\right)^{d+1}} \langle \mathsf{O}_{1} \mathsf{O}_{2} \rangle \left( \rho  \right) \, \mathsf{CPW}_{\widehat{\Delta}}^{\partial}  \left( \rho \right)  \eqend
\end{align}
This is the Euclidean inversion formula for the boundary channel.
Once we have the decomposition into CPWs, it is easy to find the decomposition into conformal blocks by closing the integration contour in \eqref{eq:boundary-spec-rep} which results in
\begin{align}
    \label{eq:boundary-CB-decomposition}
    \langle \mathsf{O}_{1} \mathsf{O}_{2} \rangle \left( \rho  \right) &= \sum_{\widehat{\Delta }_{n} \in \{\text{poles}\}}-2 \Res_{\widehat{\Delta }  = \widehat{\Delta }_{n}} \left[ \frac{\mathsf{Spec}_{\mathsf{O}_{1}\mathsf{O}_{2}}^{\partial }\left( \widehat{\Delta } \right)}{\mathcal{Q}_{d-\widehat{\Delta }}^{\partial}} \mathsf{G}^{\partial}_{\widehat{\Delta}}\left( \rho  \right) \right].
\end{align}

\paragraph{Bulk channel.} A very similar decomposition exists for the bulk channel as well. Here one defines the bulk CPW as
\begin{align}
    \label{eq:bulk-CPW-def}
    \begin{split}
        \mathsf{CPW}_{\Delta}^{B} \left( \rho  \right) &=  \frac{1}{2} \left[ \mathcal{Q}^{B}_{\Delta }  \mathsf{G}^{B }_{\Delta} \left( \rho  \right)  + \left( \Delta  \to d+1-\Delta \right)  \right] \\
                                                       &= \left( \frac{1-\rho }{\rho } \right)^{ \frac{\Delta_{1}-\Delta_{2}}{2}} \HypGeo{ \frac{\Delta_{1}-\Delta_{2}}{2} + \frac{\Delta }{2}, \, \frac{\Delta_{1} - \Delta_{2}}{2} + \frac{d+1-\Delta }{2}}{1 + \Delta_{1} - \Delta_{2}}[\frac{\rho -1}{\rho }] 
      \end{split}
\end{align}
where
\begin{align}
    \mathcal{Q}_{\Delta }^{B} &= \frac{2 \Gamma \left( 1 + \Delta_{1} - \Delta_{2} \right) \Gamma \left( \frac{d+1}{2} - \Delta  \right)}{\Gamma \left( \frac{\Delta_{1} - \Delta_{2}}{2} + \frac{d+1 - \Delta }{2} \right) \Gamma \left( \frac{\Delta_{1} - \Delta_{2}}{2} +1 - \frac{\Delta }{2} \right)} \eqend
\end{align}
Notice that the above formulas simplify substantially if the external scaling dimensions \(\Delta_{1}\) and \(\Delta_{2}\) are the same, which will be the most important scenario for our purposes.
The corresponding spectral representation and inversion formula is
\begin{align}
    \label{eq:bulk-spec-rep}
    \mathcal{F}_{\mathsf{O}_{1}\mathsf{O}_{2}} \left( \rho \right) &= \frac{1}{2 \pi  i}\int_{ \frac{d+1}{2} + i \mathbb{R}} \frac{\mathrm{d} \Delta }{2 \mathcal{N}_{\Delta }^{B}} \mathsf{Spec}_{\mathsf{O}_{1}\mathsf{O}_{2}}^{B} \left( \Delta  \right) \, \mathsf{CPW}^{B}_{\Delta } \left( \rho  \right) \\
    \label{eq:bulk-inv-formula}
    \mathsf{Spec}_{\mathsf{O}_{1}\mathsf{O}_{2}}^{B}\left( \Delta  \right) &= \int_{0}^{1} \mathrm{d} \rho \, \frac{1}{\rho^{ \frac{d+3}{2}}} \mathcal{F}_{\mathsf{O}_{1}\mathsf{O}_{2}} \left( \rho  \right) \, \mathsf{CPW}_{\Delta }^{B} \left( \rho  \right) \eqend
\end{align}
Here we again defined \(\mathcal{N}_{\Delta }^{B} = \frac{\lvert \mathcal{Q}_{\Delta }^{B} \rvert }{2}\).
The integration contour runs over the bulk principal series \(\Delta \in \frac{d+1}{2} + i\mathbb{R}\).
Notice that we work with the factorized 2-pt function \(\mathcal{F}_{\mathsf{O}_{1} \mathsf{O}_{2}}\) which relates to \(\langle  \mathsf{O}_{1} \mathsf{O}_{2} \rangle\) as in \eqref{eq:2-pt-general}.
To decompose this function into CBs, one can again close the contour in \eqref{eq:bulk-spec-rep} and reduce the integral to a sum of residues,
\begin{align}
    \label{eq:bulk-CB-decomposition}
    \mathcal{F}_{\mathsf{O}_{1} \mathsf{O}_{2}}\left( \rho \right) &= \sum_{\Delta_{n} \in \{ \text{poles} \}} - \Res_{\Delta = \Delta_{n}} \left[ \frac{\mathsf{Spec}_{\mathsf{O}_{1}\mathsf{O}_{2}}^{B} \left( \Delta  \right)}{\mathcal{Q}^{B}_{d+1 - \Delta }} \mathsf{G}^{B}_{\Delta } \left( \rho  \right) \right] \eqend
\end{align}
Notice that if the \(\mathsf{Spec}_{\mathsf{O}_{1}\mathsf{O}_{2}}^{B}\) function has only simple poles to the right of the principal series, then the sum above will only contain conformal blocks multiplied by a factor of~\( \Res_{\Delta = \Delta_{n}} \mathsf{Spec}_{\mathsf{O}_{1}\mathsf{O}_{2}}^{B} \left( \Delta  \right) / \mathcal{Q}_{d+1 - \Delta }^{B}\).
But if \(\mathsf{Spec}_{\mathsf{O}_{1}\mathsf{O}_{2}}^{B}\) contains higher order poles, then \eqref{eq:bulk-CB-decomposition} will contain derivatives of CBs as well. This subtlety will play an important role in our extraction of BCFT data in~\Cref{sec:extraction-bcft-data}.

 \section{Boundary channel expansion}
\label{sec:boundary-channel-expansion}

As the first application of spectral representations, we will compute the decomposition of the \(\langle  \phi \phi  \rangle\) and \(\langle  \sigma \sigma  \rangle\) \(2\)-pt functions into boundary conformal blocks.
The \(\langle  \phi \phi  \rangle\) correlator is simple, indeed~\eqref{eq:phi-prop} is directly proportional to a conformal block of dimension \(\widehat{\Delta} = d-1\) and we fix normalization of the BOE coefficient as
\begin{align}
\label{eq:phi-boundary-cb-exp}
\left\langle \phi\phi \right\rangle=\underbrace{2^{-2(d-1)} \left( b_{\phi \widehat{\phi}} \right)^{2}}_{\mathcal{N}_{\phi}} \mathsf{G}_{d-1}^{\partial}\eqcomma
\end{align}
which means \( \left( b_{\phi \widehat{\phi}} \right)^{2} = \mathcal{N}_{\phi }^{\text{FP}}2^{d-1}\).
For the \(\langle  \sigma \sigma  \rangle\) correlator we can separate the disconnected contribution from \(\langle  \delta \sigma  \delta \sigma  \rangle \),
\begin{align}
\label{eq:sigma-boundary-cb-exp}
\left\langle \sigma\sigma \right\rangle\coloneq \left\langle \sigma \right\rangle^2+ \left\langle \delta\sigma\delta\sigma \right\rangle=\left\langle \sigma \right\rangle^2 \mathsf{G}_{\widehat{\mathds{1}}}^{\partial}+ \sum_{k\geq 0} 2^{-2(d+1+2k)} \left( b_{\sigma \widehat{\sigma}_k} \right)^2 \mathsf{G}_{d+1+2k}^{\partial} \eqend
\end{align}
Notice that the disconnected part \(\langle  \sigma  \rangle^{2}\) corresponds to the identity block.
The main difficulty will thus lie in decomposing the connected part \(\langle  \delta \sigma \delta \sigma  \rangle\).

We know from \eqref{eq:sigma-prop-from-bubble} that \(\langle  \delta \sigma \delta \sigma  \rangle\) is proportional to the inverse of the bubble function.
This can be exploited by using the remarkable property of the boundary CPWs which are orthogonal under convolutions in the hyperbolic space.
\subsection{Boundary spectral functions}
\label{sec:bdry-spec-fn}
Orthogonality implies that the boundary spectral representation of a convolution of two functions \(f\) and \(g\) on \(\mathsf{H}_{d+1}\) is proportional to the product of their spectral representations.
To be precise, the exact relation is
\begin{align}
    \label{eq:boundary-spec-convolution}
    \mathsf{Spec}_{f \star g}^{\partial } \left( \widehat{\Delta } \right) &= \frac{\mathsf{Spec}_{f}^{\partial } \left( \widehat{\Delta } \right) \mathsf{Spec}_{g}^{\partial }\left( \widehat{\Delta } \right)}{\mathsf{Spec}_{\delta }^{\partial } \left( \widehat{\Delta } \right)}
\end{align}
where \(\mathsf{Spec}_{\delta }^{\partial } \left( \widehat{\Delta } \right)\) is the spectral representation of the \(\delta\)-function on \(\mathsf{H}_{d+1}\) which in our conventions is
\begin{align}
    \label{eq:boundary-spec-delta}
    \mathsf{Spec}_{\delta }^{\partial } \left( \widehat{\Delta } \right) &= \frac{\Gamma \left( \frac{d+1}{2} \right)}{\left( 4 \pi \right)^{ \frac{d+1}{2}} } \eqend
\end{align}
Equation \eqref{eq:boundary-spec-convolution} allows us to express the spectral representation of the inverse of bubble function as
\begin{align}
    \label{eq:B-inverse-from-boundary-spec}
    \mathsf{Spec}_{B^{-1}}^{\partial } \left( \widehat{\Delta } \right) &= \frac{\mathsf{Spec}_{\delta }^{\partial } \left( \widehat{\Delta } \right)^{2}}{\mathsf{Spec}_{B}^{\partial } \left( \widehat{\Delta } \right)} \eqend
\end{align}

The usefulness of this trick stems from the fact that computation of the spectral representation of \(B\) is much simpler than that of \(\langle  \delta \sigma \delta \sigma  \rangle\).
The bubble function is
\begin{align}
    B \left( \rho  \right) &= \frac{2}{N} \langle  \phi^{i}\phi^{j} \rangle \langle  \phi_{i}\phi_{j} \rangle = 2 \mathcal{N}_{\phi }^{2} \rho^{-d+1} \left( 1- \rho  \right)^{2d-2}\eqcomma
\end{align}
and the inversion formula \eqref{eq:boundary-inv-formula} tells us to integrate this function against the CPW, which is in this case convenient to view as a combination of a CB and its shadow.
The inversion integral can be resolved by the standard integral identity
\begin{align}
    \label{eq:integral-formula-2F1}
    \int_{0}^{1} \mathrm{d}x \, x^{\rho  - 1} \left( 1-x \right)^{\sigma  -1} \HypGeo{\alpha , \, \beta }{\gamma}[x] &= \frac{\Gamma \left( \rho  \right) \Gamma \left( \sigma  \right)}{\Gamma \left( \rho  + \sigma  \right)} \HypGeo[3][2]{\alpha , \, \beta ,\,\rho}{\gamma ,\,\rho +\sigma }[1]\eqcomma
\end{align}
which expresses the spectral representation as a \(\HypGeoNoArgs[3][2]\)-function,
\begin{align}
    \label{eq:boundary-spec-3F2}
    \mathsf{Spec}_{B}^{\partial} \left( \widehat{\Delta} \right) &= 2 \mathcal{N}_{\phi }^{2} \frac{\mathcal{Q}_{\widehat{\Delta}}^{\partial}}{2} \frac{\Gamma \left( d-2+\widehat{\Delta } \right)\Gamma \left( \frac{3-d}{2} \right)}{\Gamma \left( \frac{d-1}{2} +\widehat{\Delta } \right)} \HypGeo[3][2]{\widehat{\Delta } + \frac{1-d}{2},\, d-2+\widehat{\Delta },\, \widehat{\Delta}}{2 \widehat{\Delta }+1-d ,\, \frac{d-1}{2} + \widehat{\Delta}}[1] + \left( \widehat{\Delta} \to d-\widehat{\Delta}\right)\eqend
\end{align}
The hypergeometric function here diverges in all dimensions \(d > 2\) and thus has to be defined by analytic continuation.
Fortunately, the arguments of the hypergeometric function satisfy the special requirements for the application of Watson's theorem which allows to simplify this in terms of pure \(\Gamma\)-functions.
The resulting spectral function is valid even in dimensions greater than \(2\) and can easily be inverted to find the spectral representation of \(\langle  \delta \sigma \delta \sigma  \rangle\) as explained in \eqref{eq:B-inverse-from-boundary-spec}.
The result is (boundary spectral function for this correlator was very recently also computed in~\cite{Diatlyk:2026eta})
\begin{align}
    \label{eq:spec-bdy-sigma}
    \begin{split}
        \mathsf{Spec}_{\langle  \delta \sigma \delta \sigma  \rangle}^{\partial } \left( \widehat{\Delta } \right) &= - \frac{2^{4-2d}\Gamma \left( d \right)}{\pi^{4}} \left[ \Gamma \left( \widehat{\Delta } \right)\Gamma \left( 1-\frac{\widehat{\Delta }}{2} \right) \Gamma \left( 2-d+\frac{\widehat{\Delta }}{2} \right) \times \text{shadow} \right] \\
            &\times \frac{ \sin \left( d \pi  \right) \left( \sin \left( \pi \widehat{\Delta } \right) - \sin \left( \pi \left( d-\widehat{\Delta } \right) \right) \right)}{\csc \left( \frac{\pi  \widehat{\Delta }}{2} \right) \csc \left( \frac{\pi }{2} \left( d+\Delta  \right) \right) - \csc \left( \frac{\pi }{2} \left( d-\Delta  \right) \right)\csc \left( \frac{\pi }{2}\left( 2d-\Delta  \right) \right)} \eqend
    \end{split}
\end{align}
Here the \enquote{shadow} term stands for the same terms that appear in the parenthesis in front of them but with a substitution \(\widehat{\Delta} \to d-\widehat{\Delta}\).
This formula fails in integer dimensions due to the expression on the second line which is of the form \(0 / 0\). However, the limit can be taken and yields \(0\) in odd dimensions, while in even dimensions it is
\begin{align}
    \label{eq:spec-bdy-sigma-even-dim}
    \mathsf{Spec}_{\langle  \delta \sigma \delta \sigma  \rangle}^{\partial } \left( \widehat{\Delta } \right) \overset{\mathcolor{black!70}{d \,\,\text{even}}}{\scalebox{2.8}[1]{\(=\)}} -\frac{2^{5-2d}\Gamma \left( d \right) (-1)^{\frac{d}{2}}}{\pi^{4}} \sin^{4} \left( \frac{\pi \widehat{\Delta }}{2} \right) \left[ \Gamma \left( \widehat{\Delta } \right)\Gamma \left( 1-\frac{\widehat{\Delta }}{2} \right) \Gamma \left( 2-d+\frac{\widehat{\Delta }}{2} \right) \times \text{shadow} \right]\eqend
\end{align}
This formula can be shown to agree with a conjecture made in~\cite[(4.21)]{Dujava:2025php} for the spectral representation of the bubble function based on bootstrap arguments (recall~\eqref{eq:B-inverse-from-boundary-spec} \(\mathsf{Spec}_{\left\langle \delta\sigma\delta\sigma \right\rangle}^{\partial}\sim 1/\widetilde{B}\) and the identification \(\widehat{\Delta}_{\mathcolor{black!70}{\text{here}}}=\Delta_{\mathcolor{black!70}{\text{there}}}\) should be made)
\begin{align}
\label{eq:conjecture}
    \Bubble(\D)\eval_{\Dphi=d-1} \mathrel{\overset{\mathcolor{black!70}{d\text{ even}}}{\scalebox{3.5}[1]{\(=\)}}}
    -\frac{\cot({\frac{\pi}{2}}\D)}{2^{2d-1} \pi^{\frac{d}{2}-1} \G(\frac{d}{2})}
    \left(
        \,
        \prod_{\substack{a\.=\.4-d\\a\text{ even}}}^{2(d-2)} (\D-a)
        \middle/
        \prod_{\substack{b\.=\.1\\b \text{ odd}}}^{d-1} (\D-b)
    \right)
    \eqcomma
\end{align}
up to a normalization dependent only on the dimension \(d\), which establishes one of the important results accomplished in this work.

The most interesting even dimension for us is of course \(d = 2\). Here \(\mathsf{Spec}_{\langle  \delta \sigma \delta \sigma  \rangle}^{\partial }\) simplifies just to
\begin{align}
    \label{eq:spec-bdy-d2}
    \mathsf{Spec}_{\langle  \delta \sigma \delta \sigma  \rangle}^{\partial } \left( \widehat{\Delta } \right) &\eqdim{2} \frac{1}{\pi} \left( \widehat{\Delta} -1 \right) \tan \left( \frac{\pi \widehat{\Delta }}{2} \right)\eqend
\end{align}
In \Cref{app:bdry-spec-bootstrap} we show that this simple result can be derived by an alternative bootstrap approach.

Main conclusions of the above analysis can be summarized in terms of the BOE of \(\phi\) and \(\sigma\).
\subsection{BOE of \texorpdfstring{\(\phi\)}{ɸ}}
\label{sec:boe-phi}
From~\eqref{eq:phi-boundary-cb-exp} we know that the BOE of \(\phi\) at leading large \(N\) contains only a single boundary primary operator (we slightly abuse notation by stripping indices in the vector \(\OO(N)\) irrep and ignoring the \(x\)-dependence of operators)
\begin{align}
\label{eq:phi-BOE}
\phi_{(d-1)/2} \underset{\mathrm{BOE}}{\sim}  \left[\widehat{\phi}_{d-1} + \mathrm{desc.}\right] \eqcomma
\end{align}
such that in our normalization of the \(\phi\) operator, the BOE coefficient is \( \left(b_{\phi \widehat{\phi}}\right)^{2}= \mathcal{N}_{\phi}^{\text{FP}} 2^{d-1}\). It is clear that the scaling dimension of the boundary operator \(\widehat{\Delta}=d-1\) valid for ordinary transition is far from the free Dirichlet value \(\widehat{\Delta}=\tfrac{d+1}{2}\) -- a hallmark of a strongly interacting theory -- as long as \(d\) is at finite distance below the critical dimension \(d=3\), where the Wilson--Fischer fixed point approaches the Gaussian one and becomes free.

The above BOE obviously reproduces the boundary conformal block expansion with a single block
\begin{align}
\label{eq:phi-prop-BOE}
\left\langle \phi\phi \right\rangle= \mathcal{N}_{\phi}\underbrace{\left\langle  \left[\widehat{\phi}_{d-1} + \mathrm{desc.}\right]\left[\widehat{\phi}_{d-1} + \mathrm{desc.}\right]\right\rangle}_{\mathsf{G}_{d-1}^{\partial}=\left[\frac{4 x_{\perp} y_{\perp}}{\left( \mathbf{x}-\mathbf{y} \right)^{2} }\right]^{d-1}+\ldots}=\eqref{eq:phi-boundary-cb-exp} \eqend
\end{align}
We were explicit and included contributions of descendants. From now on they will be suppressed and every boundary primary operator will be assumed to contribute with its whole conformal multiplet to the BOE.
\subsection{BOE of \texorpdfstring{\(\sigma\)}{σ}}
\label{sec:boe-sigma}
The BOE of \(\sigma\) contains an infinite tower of boundary primaries (in addition to the identity responsible for a non-vanishing \(1\)-pt function) with scaling dimensions \(\widehat{\Delta}_k=d+1+2k,\;k\in\Z_{\geq 0}\). The first \(k=0\) operator in this tower is the protected displacement operator \(\widehat{D}\) with an exact scaling dimension \(\widehat{\Delta}_{\widehat{D}}=d+1\) (the remaining scaling dimensions are corrected at subleading orders in the large \(N\) expansion)
\begin{align}
\label{eq:sigma-BOE}
\sigma_2\underset{\mathrm{BOE}}{\sim} \left\langle \sigma \right\rangle \widehat{\mathds{1}} \oplus \underbrace{b_{\sigma \widehat{D}} \widehat{D}_{d+1} \oplus \sum_{k\geq 1} b_{\sigma \widehat{\sigma}_k} \widehat{\sigma}_{d+1+2k}}_{\delta\sigma}.
\end{align}
We omitted position dependent factors dictated by dilatation symmetry for clarity. The first BOE coefficient \(\left\langle \sigma \right\rangle\) was defined in~\eqref{eq:sigma-saddle-point} and the rest can be read off the boundary conformal block expansion~\eqref{eq:sigma-boundary-cb-exp}
\begin{align}
\label{eq:sigma-BOE-coeff}
\begin{split}
    &\left( b_{\sigma \widehat{\sigma }_{k}} \right)^{2} = \mathcal{N}_{\sigma} \frac{2^{-d+8} \Gamma (2 d-2) \Gamma \left(-d-k+\frac{3}{2}\right) \Gamma \left(-\frac{d}{2}-k+\frac{1}{2}\right)}{\pi  (d-3) \Gamma \left(\frac{d}{2}\right) \Gamma (d+1) \Gamma (k+1) \Gamma \left(\frac{1}{2} (d+4 k+2)\right)} \times \\
    &\times \frac{\sin \left(\frac{\pi  d}{2}\right) \sin (\pi  d) \Gamma \left(\frac{d}{2}+k+1\right) \Gamma (d+2 k+1)^2 \Gamma \left(-\frac{d}{2}+k+\frac{5}{2}\right)}{\sqrt{\pi } 2^{d+2 k} \sec \left(\frac{\pi  d}{2}\right) \Gamma \left(\frac{d}{2}+k+1\right)-\sec (\pi  (d+k)) \Gamma \left(-\frac{d}{2}-k+\frac{1}{2}\right) \Gamma (d+2
   k+1)}\eqend
\end{split}
\end{align}
This expression materializes one of the important results worked out in this paper, so for convenience we recall the definition~\eqref{eq:sigma-prop-norm} of \(\mathcal{N}_{\sigma}\).

It is worth noticing that (in normalization conventions defined in~\eqref{eq:sigma-boundary-cb-exp}) the BOE coefficient of the displacement operator corresponding to \(k=0\) in the above formula simplifies substantially, \(b_{\sigma \widehat{D}}^{2}=2^{2(d+1)}\mathcal{N}_{\sigma}\). This is not obvious, nevertheless true after applying identities for \(\Gamma\)-functions. Moreover, it obeys a Ward identity~\cite{McAvity:1995zd,Billo:2016cpy} linking it to the \(1\)-pt function \(\left\langle \sigma \right\rangle\). Readers interested in its verification including some subtle normalizations can find it in~\Cref{app:Ward-id}.

Since~\eqref{eq:sigma-BOE-coeff} is rather complicated for a generic dimension, let us focus on the most interesting case \(d=2\) of a \(\CFT[B]_{3}\), where the BOE coefficients simplify to
\begin{align}
\label{eq:sigma-BOE-coeff-d2}
\left( b_{\sigma \widehat{\sigma}_k} \right)^2 \eqdim{2} 64\underbrace{\mathcal{N}_{\sigma}}_{\frac{2}{\pi^2}}\left(k+1\right)^{2},\; k\in\Z_{\geq 0} \eqend
\end{align}
This is most easily derived by directly working in \(d=2\), where the \(\sigma\)-propagator, its boundary spectral function and thus also the BOE coefficients simplify dramatically. We verified that it follows also from~\eqref{eq:sigma-BOE-coeff}, which however requires substantial use of special function identities.

Asymptotics of this result for \(k\to\infty\) are in agreement with general conclusions obtained using (complex) Tauberian theorems in~\cite{Pappadopulo:2012jk,Qiao:2017xif,Rychkov:2015lca,Mukhametzhanov:2018zja,Lauria:2017wav}. The full coefficient multiplying the boundary conformal block, being equal to \(2^{-2(d+1+2k)} \left( b_{\sigma \widehat{\sigma }_{k}} \right)^{2}\), has a leading exponential fall off coming from spacing of the boundary primaries' scaling dimensions with a step of \(2\), while the power of the subleading term \(k^2\) is dictated by the scaling dimension \(\Delta_{\sigma}=2\) of the operator whose BOE is being considered.

\begin{figure}[h!]
\centering
\includegraphics[width=\linewidth]{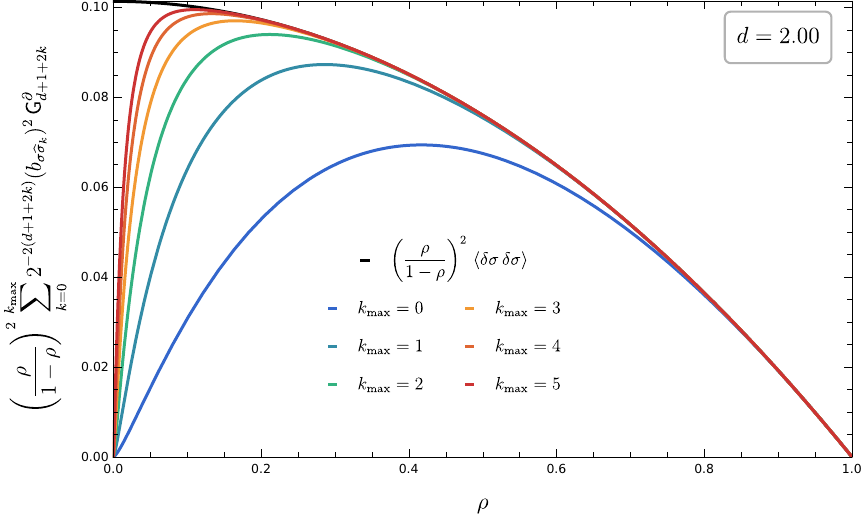}
\caption{Convergence of boundary conformal block expansion for the modified correlator \(\left( \tfrac{\rho}{1-\rho} \right)^2 \left\langle \delta\sigma\delta\sigma \right\rangle\) in the case \(d=2\) of a \(\CFT[B]_3\). For \(k_{\mathrm{max}}=0\), there is a single boundary block in the sum corresponding to the protected displacement operator \(\widehat{D}\).}
\label{fig:boundary-cb-exp-convergence-2d}
\end{figure}

\begin{figure}[h!]
\begin{subfigure}{0.49\textwidth}
\includegraphics[width=\linewidth]{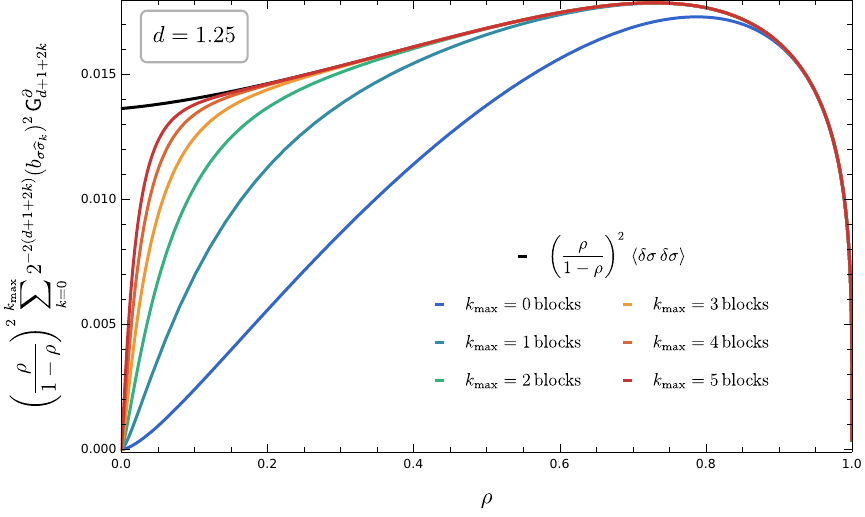}
\end{subfigure}
\hfill
\begin{subfigure}{0.49\textwidth}
\includegraphics[width=\linewidth]{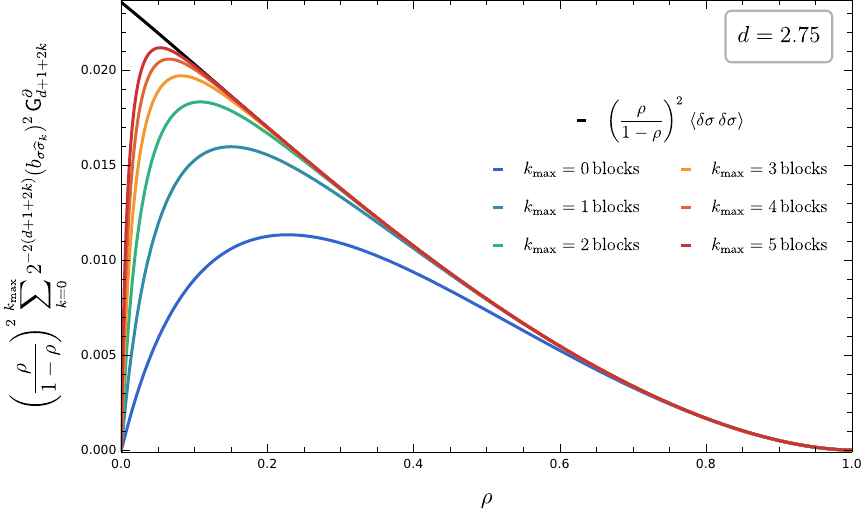}
\end{subfigure}

\caption{Convergence of boundary conformal block expansion for the modified correlator \(\left( \tfrac{\rho}{1-\rho} \right)^2 \left\langle \delta\sigma\delta\sigma \right\rangle\) in \(d=1.25\) and \(d=2.75\). For \(k_{\mathrm{max}}=0\), there is a single boundary block in the sum corresponding to the protected displacement operator \(\widehat{D}\).}
\label{fig:boundary-cb-exp-convergence}
\end{figure}

\paragraph{Convergence.} Let us end this section by discussing convergence of the boundary expansion~\eqref{eq:sigma-boundary-cb-exp} (more precisely of the connected correlator only). It is best visible by plotting a sequence of partial sums of boundary blocks -- \Cref{fig:boundary-cb-exp-convergence-2d} for the case \(d=2\) of a \(\CFT[B]_3\),~\Cref{fig:boundary-cb-exp-convergence} showing the lower/upper end of the interval \(1<d<3\), \(d=1.25\)/\(d=2.75\) -- and observing how they approach the connected correlator.
As in~\Cref{fig:sig-conn-propag}, bulk singularity determined by the bulk identity is removed, thus we plot the modified correlator \(\left( \tfrac{\rho}{1-\rho} \right)^2 \left\langle \delta\sigma\delta\sigma \right\rangle\) (used as a starting point for bulk expansions in~\eqref{eq:sigma-corr-mod} of the next section). As is clear from the figures, convergence is pretty fast across different dimensions. It is uniform on any interval \(\rho\in( \varepsilon,\,1],\;\varepsilon>0\) that doesn't include the bulk OPE point \(\rho=0\). This analysis also verifies correctness of BOE coefficients given in~\eqref{eq:sigma-BOE-coeff}.


\section{Bulk channel expansion}
\label{sec:bulk-exp}
The purpose of this section is to derive bulk partial wave expansions of the modified correlators (prefactors cancel leading singularities from bulk identity operators in the respective OPEs)
\begin{align}
\left(\frac{\rho}{1-\rho}\right)^{\Delta_{\phi}}  \langle \phi\phi \rangle \label{eq:phi-corr-mod} \\
\left(\frac{\rho}{1-\rho}\right)^{\Delta_{\sigma}}   \langle \sigma\sigma \rangle \label{eq:sigma-corr-mod}
\end{align}
To establish what expressions need to be decomposed into partial waves, large \(N\) analysis of the above functions has to be discussed first.

\subsection{Large \(N\) expansion of modified bulk-bulk \(2\)-pt functions}
\label{sec:largeN-exp-bulk}
Large \(N\) expansion for \(\phi\) is completely straightforward, while that of \(\sigma\) is slightly more involved.
\paragraph{Large \(N\) expansion for \(\phi\).} All terms start at order \(N^0\), like \(\Delta_{\phi}=\frac{d-1}{2}+ \mathcal{O} \left( \frac{1}{N} \right)\), and subleading pieces can be dropped. Recalling~\eqref{eq:phi-prop}, one gets
\begin{align}
\label{eq:phi-corr-mod-lead}
\left(\frac{\rho}{1-\rho}\right)^{\Delta_{\phi}}  \langle \phi\phi \rangle = \left(\frac{\rho}{1-\rho}\right)^{\frac{d-1}{2}}  \langle \phi\phi \rangle^{(0)} = \mathcal{N}_{\phi} \left( 1-\rho \right)^{\frac{d-1}{2}} \eqend
\end{align}
\paragraph{Large \(N\) expansion for \(\sigma\).}
Individual terms in~\eqref{eq:sigma-corr-mod} have the following large \(N\) expansions
\begin{align}
\label{eq:sigma-terms-largeN}
  &\Delta_{\sigma}=2+\gamma_{\sigma}^{(-1)}+ \mathcal{O} \left( \frac{1}{N^{2}} \right),&& \left\langle \delta\sigma\delta\sigma \right\rangle= \left\langle \delta\sigma\delta\sigma \right\rangle^{(0)}+ \mathcal{O} \left( \frac{1}{N} \right), \notag \\
  & \left\langle \sigma \right\rangle= \left\langle \sigma \right\rangle^{(1/2)}+ \left\langle \sigma \right\rangle^{(-1/2)}+ \mathcal{O} \left( N^{- \frac{3}{2}} \right) \eqend
\end{align}
Note that a subleading \(1\)-pt function \(\left\langle \sigma \right\rangle^{(-1/2)}\) had to be introduced, which is  an unknown constant (function of space dimension \(d\), really) to the best of our knowledge. The anomalous dimension of \(\sigma\) is known~\cite{Vasiliev:1981dg} and will be used as input in our computations
\begin{align}
\label{eq:anom-dim-sigma}
\gamma_{\sigma}^{(-1)}= \frac{2^d (d-1) d \cos \left(\frac{\pi  d}{2}\right) \Gamma \left(\frac{d}{2}\right)}{\pi ^{3/2} \Gamma \left(\frac{d+3}{2}\right)} \eqend
\end{align}
Plugging these expansions to~\eqref{eq:sigma-corr-mod} splits it into two parts. The leading one of order \(N\)
\begin{align}
\label{eq:sigma-corr-mod-lead}
\left( \frac{\rho}{1-\rho} \right)^2 \left[ \left\langle \sigma \right\rangle ^{(1/2)}\right]^2 \eqcomma
\end{align}
and a subleading term of order \(N^0\)
\begin{align}
\label{eq:sigma-corr-mod-sublead}
\left( \frac{\rho}{1-\rho} \right)^2 \left[ \left\langle \delta\sigma\delta\sigma \right\rangle ^{(0)}+2 \left\langle \sigma \right\rangle^{(1/2)} \left\langle \sigma \right\rangle ^{(-1/2)}+ \left( \left\langle \sigma \right\rangle ^{(1/2)}\right)^{2}\gamma_{\sigma}^{(-1)}\log \frac{\rho}{1-\rho}\right] \eqend
\end{align}
We will compute bulk spectral functions via conformal partial wave expansions for both these terms in the next section.

\subsection{Bulk partial wave decomposition}
\label{sec:bulk-PW-decomp}

As was already explained in \Cref{sec:preliminaries}, the bulk conformal partial waves form a complete basis for 2-pt functions, and the inversion formula \eqref{eq:bulk-inv-formula}—producing the corresponding spectral function—requires integrating the 2-pt function against the conformal partial wave (with a specific weight).
In this section we explain how to compute these integrals and list the resulting spectral functions for~\eqref{eq:phi-corr-mod-lead},~\eqref{eq:sigma-corr-mod-lead} and~\eqref{eq:sigma-corr-mod-sublead}.


\subsubsection{Derivation of bulk spectral functions}
\label{sec:bulk-Spec}

We will use the inversion formula \eqref{eq:bulk-inv-formula} applied to the functions \eqref{eq:phi-corr-mod-lead}, \eqref{eq:sigma-corr-mod-lead} and \eqref{eq:sigma-corr-mod-sublead} to obtain the spectral representations and then use the contour trick to express the functions as sums of conformal blocks as explained in section \Cref{sec:bcft-kinematics}.

\paragraph{Spectral representation for \texorpdfstring{\(\langle\phi \phi\rangle\)}{<ϕϕ>}.}

The leading part \eqref{eq:phi-corr-mod-lead} of the \(\langle  \phi \phi  \rangle\) correlator is just a simple power \( \mathcal{N}_{\phi} \left( 1-\rho  \right)^{ \frac{d-1}{2}}\) which facilitates the computation of the inversion integral \eqref{eq:bulk-inv-formula}.
It can be performed in a straightforward way by applying the integral identity \eqref{eq:integral-formula-2F1}. The resulting spectral function is
\begin{align}
    \label{eq:bulk-spec-phi}
    \mathsf{Spec}^{B}_{\eqref{eq:phi-corr-mod-lead}} \left( \Delta  \right) &= \mathcal{N}_{\phi} \frac{\pi  \cos \left(\frac{\pi  d}{2}\right) \Gamma
   \left(\frac{d+1}{2}\right)^2 \csc \left(\frac{\pi  \Delta }{2}\right) \csc\left( \frac{\pi}{2}  (d+1-\Delta )\right)}{\Gamma \left(\frac{\Delta }{2}\right) \Gamma \left(\frac{\Delta}{2}+1\right) \Gamma
   \left( \frac{d+1-\Delta }{2}\right) \Gamma \left( \frac{d+1-\Delta }{2} + 1 \right)}\eqend
\end{align}
Notice that the spectral function is shadow-symmetric as it must be (this means that it is invariant under \(\Delta \to d+1-\Delta\)) and it has simple poles which are located at \(\Delta = 2k\) for \(k \in \mathbb{Z}_{\geq 1}\) (and the corresponding shadow pairs) coming from the \(\csc\) functions in the numerator.

\paragraph{Spectral representation of the leading part of \texorpdfstring{\(\langle  \sigma \sigma \rangle \)}{<σσ>}.}
The leading part of \(\langle  \sigma \sigma  \rangle\) is just a constant \(\langle  \sigma  \rangle^{2}\) which, however, needs to be multiplied by the conventional prefactor \( \left( \frac{\rho }{1-\rho } \right)^{2}\). In order to find its spectral representation, let us actually consider a general scaling dimension \(\Delta_{\sigma }\) and set it equal to 2 later. Thus we will focus on the function
\begin{align}
    \label{eq:f-bulk-function-leading-sigma}
    f \left( \rho  \right) &= \left( \frac{\rho }{1-\rho } \right)^{\Delta_{\sigma}}\eqcomma
\end{align}
(the constant \(\langle  \sigma  \rangle^{2}\) will be added at the end).
The inversion integral is again straightforward to carry out using the identity \eqref{eq:integral-formula-2F1} and yields the result
\begin{align}
    \label{eq:spec-sigma-leading-general}
    \mathsf{Spec}_{f}^{B} \left( \Delta  \right) &= \frac{\pi  \sin \left(\pi  \Delta _{\sigma }\right) \csc \left(\frac{\pi }{2}  \left(\Delta -2 \Delta _{\sigma }\right)\right) \Gamma \left(1-\Delta _{\sigma}\right){}^2 \csc \left(\frac{\pi }{2}  \left(-d+2 \Delta _{\sigma }+\Delta +1\right)\right)}{\Gamma \left(1-\frac{\Delta }{2}\right) \Gamma \left( \frac{-d+\Delta +1}{2} \right) \Gamma \left(\frac{\Delta }{2}-\Delta _{\sigma }+1\right) \Gamma \left( \frac{d-\Delta -2 \Delta _{\sigma }+3}{2} \right)} \eqend
\end{align}
Notice that this function is divergent in the limit \(\Delta_{\sigma } \to 2\) due to the \(\Gamma  \left( 1-\Delta_{\sigma } \right)^{2}\) factor. However, the function can be expanded in a Laurent series around \(\Delta_{\sigma } = 2\) which reveals that the divergent part is
\begin{align}
    \frac{\left( \Delta -2 \right) \left( d-1-\Delta \right)}{4 \left( \Delta_{\sigma } - 2 \right)}.
\end{align}
Since this is a polynomial in \(\Delta\), it can be subtracted from the original spectral function \eqref{eq:spec-sigma-leading-general} resulting in the regular part where the limit can be taken. The resulting regularized spectral function for \(\Delta_{\sigma } = 2\) is
\begin{align}
    \begin{split}
        \mathsf{Spec}_{\eqref{eq:sigma-corr-mod-lead}}^{B} \left( \Delta  \right) = \left[ \langle  \sigma  \rangle^{\left( 1 / 2 \right)} \right]^{2} &\frac{\left( \Delta -2 \right) \left( d-1 - \Delta  \right)}{4} \bigg[\pi \left( \cot \frac{\pi \Delta }{2} + \cot \frac{\pi \left( d+1-\Delta  \right)}{2} \right) + \\
                                                                                          &-2 + 2 \gamma_{E} +  \psi \left( \frac{\Delta }{2} -1 \right) + \psi \left( \frac{d+1-\Delta }{2} -1 \right) \bigg]\eqcomma
    \end{split} \label{eq:bulk-spec-sigma-lead}
\end{align}
where \(\gamma_{E}\) is the Euler's constant and \( \psi \) is the digamma function. It can be checked that this spectral function correctly reproduces the original function \eqref{eq:sigma-corr-mod-lead}. Clearly, the same derivation can be applied to the term in the subleading part of \(\langle  \sigma \sigma  \rangle\) which is proportional to \(2 \langle  \sigma  \rangle^{(1 / 2)} \langle  \sigma  \rangle^{( - 1 / 2 )}\) (see the second term in \eqref{eq:sigma-corr-mod-sublead}) since it is also just a constant. The spectral representation contains simple poles at \(\Delta = 2k\) with \(k \in \mathbb{Z}_{\geq 2}\) (so the first pole is at \(\Delta = 4\)).

\paragraph{Spectral representation of the logarithmic part of \(\langle \sigma \sigma  \rangle\).} Let us now focus on the logarithmic part of \eqref{eq:sigma-corr-mod-sublead} which contains the function
\begin{align}
    g \left( \rho  \right) &= \left( \frac{\rho }{1-\rho } \right)^{2} \log \frac{\rho }{1-\rho } = \frac{\partial }{\partial \Delta_{\sigma }} \left( \frac{\rho }{1-\rho } \right)^{\Delta_{\sigma}}\Bigg|_{\Delta_{\sigma} = 2} \eqend
\end{align}
Since it is just the derivative of the function \(f\) from \eqref{eq:f-bulk-function-leading-sigma}, we can find its spectral representation by taking the derivative of \eqref{eq:spec-sigma-leading-general} with respect to \(\Delta_{\sigma}\), regularizing the result in the same way as before (subtracting the part that is divergent in \(\Delta_{\sigma}\)) and then setting \(\Delta_{\sigma }= 2\). The resulting spectral representation is
\begin{align}
    \label{eq:spec-sigma-log-term}
    \mathsf{Spec}_{g}^{B}\left( \Delta \right) &= b \tilde{b} + a \tilde{c} + \tilde{a}c
\end{align}
where we have denoted
\begin{align}
    a &= \frac{\Delta -2}{2} \label{eq:spec-log-term-a}\\
    b &= \frac{\Delta -2}{2} \left( -1 + \gamma_{E} + \psi \left( \frac{\Delta }{2}-1 \right) + \pi  \cot \frac{\pi \Delta }{2} \right) \label{eq:spec-log-term-b}\\
    \begin{split}
            c &= \frac{1}{4} (\Delta -2) \bigg[-\psi ^{(1)}\left(\frac{\Delta }{2}-1\right)+2 \pi  \left((\gamma_{E} -1) \cot \left(\frac{\pi  \Delta }{2}\right)+\pi  \csc ^2\left(\frac{\pi
       \Delta }{2}\right)\right)+ \\
          & +\psi \left(\frac{\Delta }{2}-1\right) \left(2 \pi  \cot \left(\frac{\pi  \Delta }{2}\right)+\psi \left(\frac{\Delta
      }{2}-1\right)+2 \gamma_{E} -2\right)-\pi ^2+(\gamma_{E} -2) \gamma_{E} +2\bigg] \label{eq:spec-log-term-c}
    \end{split}
\end{align}
and the variables with a tilde denote the bulk shadow transformed versions of \(a\), \(b\), \(c\) (this means substituting \(\Delta \to d+1-\Delta\)) and \(\psi^{(1)}\) is the derivative of the digamma function, \(\psi^{(n)}(z) = \frac{\mathrm{d}^{n}}{\mathrm{d}z^{n}}\psi (z)\). This spectral function contains only double poles at locations \(\Delta = 2k\) for \(k \in \mathbb{Z}_{\geq 2}\). It is important, since as we remarked below \eqref{eq:bulk-CB-decomposition}, the presence of double poles in the spectral representation implies the appearance of derivatives of conformal blocks in the corresponding conformal block expansion. In the end, these terms will be responsible for the shift in the scaling dimensions corresponding to the anomalous dimensions of the exchanged operators.

\paragraph{Spectral representation of the connected part \(\langle  \delta \sigma \delta \sigma  \rangle\).} Now we come to the most difficult part---the spectral representation of \(\langle  \delta \sigma \delta \sigma  \rangle\). The inversion integral here is a bit more involved, since the correlator~\eqref{eq:sigma-prop-connected} contains a hypergeometric function, but can be solved by applying the Barnes integral representation,
\begin{align}
  \label{eq:barnes-representation}
  \HypGeo{a , b}{c}[z] &= \frac{\Gamma \left( c \right)}{\Gamma \left( a \right)\Gamma \left( b \right)} \frac{1}{2 \pi  i} \int_{-i \infty}^{i \infty} \mathrm{d}u \, \frac{\Gamma \left( a+u \right)\Gamma \left( b+u \right)\Gamma \left( -u \right)}{\Gamma \left( c+u \right)} \left( -z \right)^{u} \eqend
\end{align}
Substituting this representation for the hypergeometric function appearing in \(\langle  \delta \sigma \delta \sigma  \rangle\) and swapping the order of integration, we reduce the inversion integral to the form that is again solvable by the formula \eqref{eq:integral-formula-2F1} (the remaining hypergeometric function comes from the \(\mathsf{CPW}\)). The result of this integral simplifies to an expression that contains pure \(\Gamma \)-functions and the remaining integration in \(u\), which we artificially introduced in~\eqref{eq:barnes-representation}, can be carried out by closing the contour in the right half plane and summing up the residues. Since the resulting spectral function is significantly more complicated here, we present it in the following form:
\begin{align}
  \label{eq:bulk-spec-delta-sigma}
  \mathsf{Spec}_{\langle  \delta \sigma \delta \sigma  \rangle}^{B} \left( \Delta  \right) &= \mathcal{A} + \mathcal{B} + \tilde{\mathcal{B}} \eqcomma
\end{align}
where \(\tilde{\mathcal{B}}\) means the shadow of \(\mathcal{B}\) (which here means substituting \(\Delta \to d+1- \Delta\)) and \(\mathcal{A}\) is shadow symmetric. The expressions for \(\mathcal{A}\) and \(\mathcal{B}\) are
\begin{align}
  \begin{split} \label{eq:bulk-spec-delta-sigma-a}
    \mathcal{A} &= - \mathcal{N}_{\sigma } \frac{\pi  \Gamma \left( d \right)^{2}  \left[ \sin \frac{\pi \Delta }{2} \csc \frac{\pi }{2} \left( 2d +\Delta  \right) \csc \pi \left( \frac{d+1}{2} - \Delta  \right) + \text{shadow} \right]}{ \left[ \Gamma \left( 1- \frac{\Delta }{2} \right) \Gamma \left( d+\frac{\Delta }{2} \right) \times \text{shadow} \right]} \\
                &\hspace{2cm} \times \HypGeo[4][3]{d ,\, d ,\, d-1 ,\, d+1}{2d-2 ,\, d+ \frac{\Delta }{2} ,\, d + \frac{d+1-\Delta }{2}}[1]\eqcomma
  \end{split} \\
        \mathcal{B} &= \mathcal{N}_{\sigma } \frac{2^{-3+2d} \sqrt{\pi } \Gamma \left( d-\frac{1}{2} \right)\Gamma \left( - \frac{\Delta }{2} \right)\Gamma \left( 1-\frac{\Delta }{2} \right) \Gamma \left( 2- \frac{\Delta }{2} \right) \csc \pi \left( d+\frac{\Delta }{2} \right) \csc \pi \left( \frac{d+1}{2}-\Delta  \right) \sin \frac{\pi \Delta }{2} }{\Gamma \left( 1+d \right) \Gamma \left( \frac{3}{2} + \frac{d}{2} - \Delta  \right) \Gamma \left( 2-d-\frac{\Delta }{2} \right) \Gamma \left( d-1-\frac{\Delta }{2} \right)\Gamma \left( \frac{1}{2} - \frac{d}{2} + \frac{\Delta }{2} \right)}\nonumber \\
                    &\hspace{2cm} \times \HypGeo[4][3]{1- \frac{\Delta }{2} ,\, 1 - \frac{\Delta }{2} ,\, 2- \frac{\Delta }{2} ,\, - \frac{\Delta }{2}}{\frac{3}{2} + \frac{d}{2} - \Delta ,\, 2-d-\frac{\Delta }{2} ,\, d-1 - \frac{\Delta }{2}}[1] \eqend \label{eq:bulk-spec-delta-sigma-b}
\end{align}
The spectral function contains a simple pole at \(\Delta = 2\) which corresponds to the operator \(\sigma\) itself and double poles at locations \(\Delta = 2k\) for \(k \in \mathbb{Z}_{\geq 2}\) (so starting at dimension \(4\)).
The presence of the double poles again introduces derivatives of conformal blocks in the resulting conformal block expansion.

With this, we have all the ingredients to write down the subleading part of the propagator \eqref{eq:sigma-corr-mod-sublead},
\begin{align}
\label{eq:bulk-spec-sigma-sublead}
  \mathsf{Spec}^B_{\eqref{eq:sigma-corr-mod-sublead}}= \mathsf{Spec}_{\langle  \delta \sigma \delta \sigma  \rangle}^{B} + 2 \langle  \sigma  \rangle^{(1 / 2)} \langle  \sigma  \rangle^{(- 1 / 2 )} \frac{\mathsf{Spec}_{\eqref{eq:sigma-corr-mod-lead}}^{B} }{\left[ \langle  \sigma  \rangle^{(1 / 2 )} \right]^{2}} + \left[ \langle  \sigma  \rangle^{(1 / 2 )} \right]^{2} \gamma_{\sigma }^{(-1)} \mathsf{Spec}_{g}^{B} \eqcomma
\end{align}
which we will utilize in the next section.

\subsection{Bulk conformal block decomposition}
\label{sec:bulk-CB-decomp}

Knowing bulk conformal partial wave expansions of~\eqref{eq:phi-corr-mod-lead},~\eqref{eq:sigma-corr-mod-lead} and~\eqref{eq:sigma-corr-mod-sublead} allows to easily obtain their corresponding bulk conformal block decompositions by closing the contour in the right half \(\Delta\)-plane. Coefficients of this decomposition were derived in~\eqref{eq:bulk-CB-decomposition} to be related to residues of spectral functions. Applying this master formula to~\eqref{eq:bulk-spec-phi},~\eqref{eq:bulk-spec-sigma-lead} and~\eqref{eq:bulk-spec-sigma-sublead} fixes coefficients of bulk conformal block expansions of~\eqref{eq:phi-corr-mod-lead},~\eqref{eq:sigma-corr-mod-lead} and~\eqref{eq:sigma-corr-mod-sublead}. These will be presented below and constitute the main achievement of this section.

\subsubsection{\texorpdfstring{\(\left\langle \phi\phi \right\rangle\)}{<ϕϕ>} bulk conformal block expansion}
\label{sec:phi-bulk-cb-exp}

Closing the contour as described above generates bulk conformal block expansion of~\eqref{eq:phi-corr-mod-lead}
\begin{align}
  \label{eq:bulk-cb-exp-phi}
  \left(\frac{\rho}{1-\rho}\right)^{\frac{d-1}{2}}  \langle \phi\phi \rangle^{(0)}&= \mathcal{N}_{\phi}\left[ \mathsf{G}^B_{\mathds{1}} + \sum_{k\geq 1}C[\langle \phi\phi \rangle]^{(0)}_k\mathsf{G}^B_{2k} \right] \eqend
\end{align}
Coefficients are derived by substituting the spectral function~\eqref{eq:bulk-spec-phi} to the general residue formula~\eqref{eq:bulk-CB-decomposition}, giving
\begin{align}
\label{eq:coeff-bulk-phi-explicit}
C[\langle \phi\phi \rangle]^{(0)}_k= \frac{\Gamma \left(\frac{d+1}{2}\right)^2 \Gamma \left(k-\frac{d+1}{2}\right)}{k! \Gamma \left(\frac{d+1}{2}-k\right)^2 \Gamma \left(2 k-\frac{d+1}{2}\right)} \eqend
\end{align}
They were already computed in~\cite[(A.8)]{Giombi:2020rmc} and both results match (\(\left( d+1 \right)_{\text{\textcolor{gray}{us}}}=d_{\text{\cite{Giombi:2020rmc}}}\) and \(k_{\text{\textcolor{gray}{us}}}= \left( n+1 \right)_{\text{\cite{Giombi:2020rmc}}}\)), even though basic special function identities are needed to show it.

\subsubsection{Leading \texorpdfstring{\(\left\langle \sigma\sigma \right\rangle\)}{<σσ>} bulk conformal block expansion}
\label{sec:sigma-lead-bulk-cb-exp}

Spectral function~\eqref{eq:bulk-spec-sigma-lead} determines by the residue theorem~\eqref{eq:bulk-CB-decomposition} bulk conformal block expansion of~\eqref{eq:sigma-corr-mod-lead}
\begin{align}
\label{eq:bulk-cb-exp-sigma-lead}
\left( \frac{\rho}{1-\rho} \right)^2 \left[ \left\langle \sigma \right\rangle ^{(1/2)}\right]^2 =  \sum_{k\geq 2} C \left[ \left\langle \sigma\sigma \right\rangle \right]^{(1)}_k \mathsf{G}^B_{2k} \eqend
\end{align}
Let us point out that there is no contribution of identity, neither of an operator with scaling dimension \(\Delta=2\). Vanishing of this coefficient will play a role below. This is clear as the OPE expansion \(\rho\to 0\) of the left hand side starts like \(\rho^2 \left( 1+\cdots \right)\), while that of a bulk conformal block as \(\mathsf{G}^B_{\Delta}=\rho^{ \frac{\Delta}{2}} \left( 1+\cdots \right)\).

The coefficients are easily computed as
\begin{align}
\label{eq:sigma-lead-bulk-cb-coeff}
C \left[ \left\langle \sigma\sigma \right\rangle \right]^{(1)}_k = \left[ \left\langle \sigma \right\rangle ^{(1/2)}\right]^2  \frac{(k-1) \Gamma (k) \Gamma \left(\frac{3-d}{2}+k\right)}{\Gamma \left(2 k-\frac{d+1}{2}\right)},\;k\in\Z_{\geq 2} \eqcomma
\end{align}
with \(\left\langle \sigma \right\rangle^{(1/2)}\) defined in~\eqref{eq:sigma-saddle-point}.

\subsubsection{Subleading \texorpdfstring{\(\left\langle \sigma\sigma \right\rangle\)}{<σσ>} bulk conformal block expansion}
\label{sec:sigma-sublead-bulk-cb-exp}
The same strategy as for the leading term leads to conformal block expansion of~\eqref{eq:sigma-corr-mod-sublead}. There is one crucial difference, however. Constituent spectral functions contain double poles, therefore produce terms with derivatives (with respect to the scaling dimension) of conformal blocks. Concretely, the last logarithmic term in~\eqref{eq:sigma-corr-mod-sublead} has an associated spectral function with double poles for all space dimensions \(d\). In contrast, the connected \(2\)-pt function \(\left\langle \delta\sigma\delta\sigma \right\rangle^{(0)}\) in that equation has an associated spectral function with double poles in all space dimensions \(d\), importantly except for \(d=2\), where they become simple poles and no longer contribute to derivatives of blocks (and thus anomalous dimensions of exchanged operators).

We get for the decomposition of the subleading \(N^0\) term
\begin{align}
\label{eq:bulk-cb-exp-sigma-sublead}
&\left( \frac{\rho}{1-\rho} \right)^2 \left[ \left\langle \delta\sigma\delta\sigma \right\rangle ^{(0)}+2 \left\langle \sigma \right\rangle^{(1/2)} \left\langle \sigma \right\rangle ^{(-1/2)}+ \left( \left\langle \sigma \right\rangle ^{(1/2)}\right)^{2}\gamma_{\sigma}^{(-1)}\log \frac{\rho}{1-\rho}\right] \notag \\
 &= \mathcal{N}_{\sigma}^{\mathrm{FP}} \left\{\mathsf{G}^B_{\mathds{1}}-\frac{(d-1)(d-2)}{16} \mathsf{G}_2^B\right\}+\sum_{k\geq 2}\left\{C\left[\langle \sigma\sigma \rangle\right]^{(0)}_k\mathsf{G}^B_{2k} + D\left[\langle \sigma\sigma \rangle\right]^{(0)}_k\partial_{\Delta} \mathsf{G}^B_{2k}\right\} \eqcomma
\end{align}
where by \(\partial_{\Delta} \mathsf{G}_{2k}^B\) we really mean \(\left( \partial_{\Delta} \mathsf{G}_{\Delta}^B \right)\bigg\rvert_{\Delta=2k}\). The first two blocks get contributions only from the connected correlator and we recall the definition of \(\mathcal{N}_{\sigma}^{\mathrm{FP}}\) in~\eqref{eq:sigma-prop-norm}. All three terms on the left hand side contribute to the non-derivative tower of blocks with \(k\geq 2\). Finally, the tower of differentiated blocks gets contributions from the connected correlator and the logarithmic term in a generic dimension \(d\), while in \(d=2\) only the logarithmic term contributes.

The expansion coefficients \(C,\,D\) are rather complicated functions expressed in terms of (derivatives of) \(\HypGeoNoArgs[4][3] \left[ \ldots \vert 1 \right]\) hypergeometric functions with arguments dependent only on the space dimension \(d\) and scaling dimensions \(\Delta_k=2k,\;k\geq 2\) of bulk primaries contributing to the \(\sigma\times\sigma\) OPE (and implicitly also on the scaling dimension of \(\sigma\) itself set to \(\Delta_{\sigma}=2\)). Thus we do not provide them here, but they can be accessed in the supplementary \wmathematica{} notebook together with a number of other expansion coefficients.

\begin{figure}[h!]
\centerline{\includegraphics[]{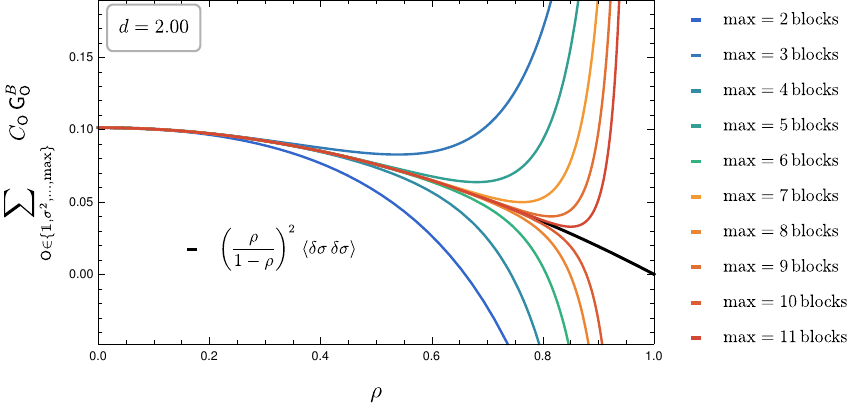}}
\caption[]{Bulk conformal block expansion of the modified connected correlator \(\left( \frac{\rho}{1-\rho} \right)^2 \left\langle \delta\sigma\delta\sigma \right\rangle\) in \(\CFT[B]_3\). The lowest block corresponds to the bulk identity and the next one to an operator of dimension \(\Delta=4\) (which is \(\sigma^2\) as we prove in the next section). Notice that the boundary asymptotics of the sequence of partial sums oscillates due to sign flips between subsequent coefficients in the expansion.}
\label{fig:bulk-cb-exp-convergence-2d}
\end{figure}

\begin{figure}[h!]
\centering
\begin{subfigure}{0.49\textwidth}
\includegraphics[width=\linewidth]{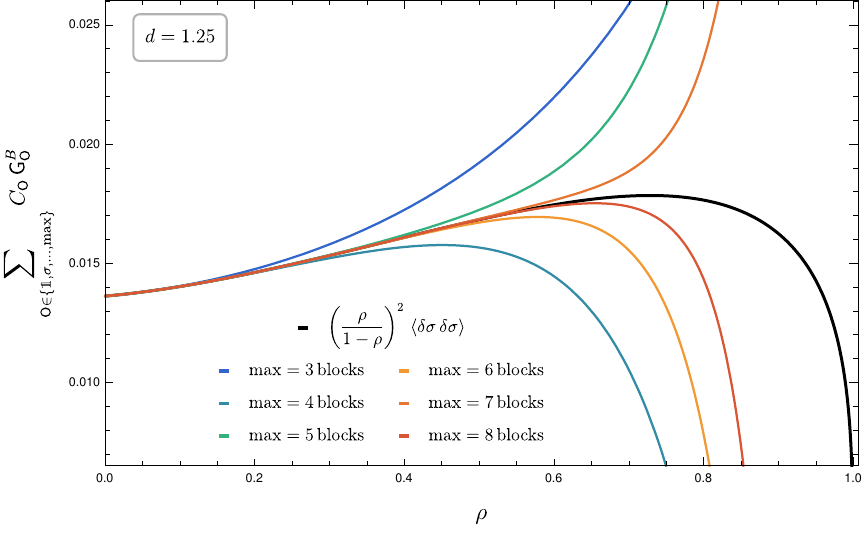}
\end{subfigure}
\hfill
\begin{subfigure}{0.49\textwidth}
\includegraphics[width=\linewidth]{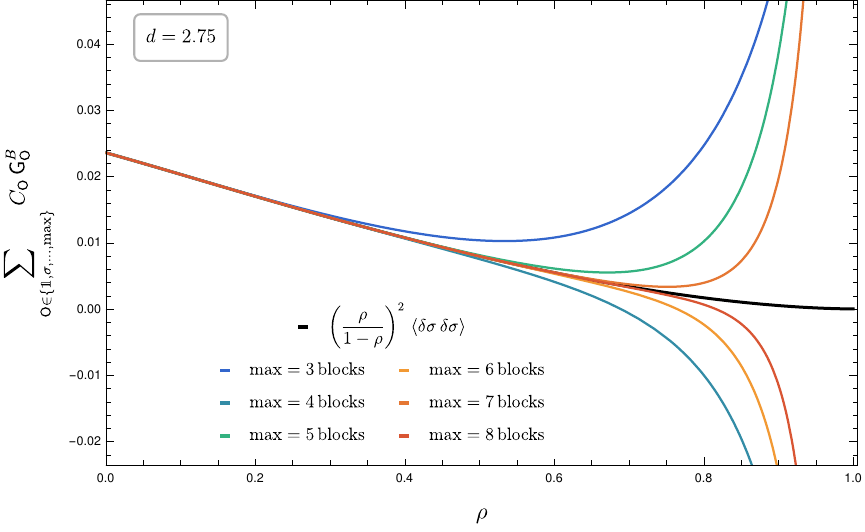}
\end{subfigure}
\caption{Bulk conformal block expansion of the modified connected correlator \(\left( \frac{\rho}{1-\rho} \right)^2 \left\langle \delta\sigma\delta\sigma \right\rangle\) in \(d=1.25\) (slightly above the lower critical dimension) and in \(d=2.75\) (slightly below the upper critical dimension). The lowest block corresponds to the bulk identity and the next one to an operator of dimension \(\Delta=2\) (which is \(\sigma\) as we prove in the next section). Notice that the boundary asymptotics of the sequence of partial sums oscillates due to sign flips between subsequent coefficients in the expansion.}
\label{fig:bulk-cb-exp-convergence}
\end{figure}

\paragraph{Convergence.} Even more than for the boundary expansion, it is instructive to analyze convergence of the bulk conformal block expansion. This is a place where boundary conditions especially matter and the \emph{ordinary transition} we are focusing on will exhibit a peculiar behavior. \Cref{fig:bulk-cb-exp-convergence-2d} shows bulk expansion for the case \(d=2\) of a \(\CFT[B]_3\) while~\Cref{fig:bulk-cb-exp-convergence} displays the same for \(d=1.25\)/\(d=2.75\) (that is the region near the lower/upper critical dimension). The key feature of these plots is that subsequent coefficients of the expansion change sign as can be seen from the oscillatory behavior. This prevents the ordinary transition from being investigated by the (rigorous, \texttt{SDPB} like) numerical bootstrap algorithms as they require positivity of expansion coefficients. For this reason the special~\cite{Liendo:2012hy} and extraordinary/normal~\cite{Padayasi:2021sik} transitions are more amenable to numerical bootstrap.

As a side remark, let us point out that the coefficient of the \(\sigma\)-block \(\mathsf{G}_2^B\) in~\eqref{eq:bulk-cb-exp-sigma-sublead} vanishes in \(d=2\), therefore the gap between bulk identity and the lowest contributing operator gets abruptly lowered as we move away from \(\CFT[B]_3\).

In general, convergence of the bulk expansion is slower compared to the boundary case, but it is still uniform on any interval \(\rho\in[0,1-\varepsilon)\) with the BOE limit point \(\rho=1\) excluded.

Similar convergence plots were checked to reproduce the remaining ``trivial'' parts of the left hand side in~\eqref{eq:bulk-cb-exp-sigma-sublead} and all of them together verify correctness of the expansion coefficients \(C,\,D\) on the right hand side of that equation.

\subsection{Bulk OPE}
\label{sec:bulk-ope}
From~\eqref{eq:bulk-cb-exp-phi},~\eqref{eq:bulk-cb-exp-sigma-lead} and~\eqref{eq:bulk-cb-exp-sigma-sublead} one can already infer the structure of the \(\phi\times\phi\) and \(\sigma\times\sigma\) OPE
\begin{align}
\label{eq:phi-sig-bulk-ope}
  &\phi\times\phi \sim \mathcal{N}_{\phi}^{\mathrm{FP}}\mathds{1} +\bigoplus_{k\geq1}C_{\phi\phi\Oper_k}\Oper_{\Delta_k=2k+\gamma_k^{(-1)}+\cdots} \\
  &\sigma\times\sigma\sim \mathcal{N}_{\sigma}^{\mathrm{FP}}\mathds{1}+ \bigoplus_{k\geq1}C_{\sigma\sigma\Oper_k}\Oper_{\Delta_k=2k+\gamma_k^{(-1)}+\cdots} \eqend
\end{align}
Note that the set of operators appearing in both OPEs is the same. For the \(\left\langle \phi\phi \right\rangle\) correlator anomalous dimensions \(\gamma_k^{(-1)}\) don't matter at all, but they do for \(\left\langle  \sigma\sigma\right\rangle\) as confirmed in~\eqref{eq:bulk-cb-exp-sigma-sublead} by the appearance of differentiated blocks.

We postpone detailed analysis of the OPE to the next section, where we identify the operators. It will turn out that at (leading) dimension \(\Delta=2,\,4\) there is a single primary, however starting from \(\Delta=6\) degeneracy kicks in (at \(\Delta=6\) are two primaries, at \(\Delta=8\) three and so on).

 \section{Extraction of \texorpdfstring{\CFT[B]}{BCFT} data}
\label{sec:extraction-bcft-data}

To determine \CFT[B] data from bulk conformal block expansions~\eqref{eq:bulk-cb-exp-phi},~\eqref{eq:bulk-cb-exp-sigma-lead} and~\eqref{eq:bulk-cb-exp-sigma-sublead} requires significantly more effort compared to the boundary channel. The reason is that coefficients of the bulk expansion are composite objects consisting of the product of a \(1\)-pt function and a bulk OPE coefficient, whose orders in the large \(N\) expansion mix.

To match \CFT[B] data with formulas~\eqref{eq:bulk-cb-exp-phi},~\eqref{eq:bulk-cb-exp-sigma-lead} and~\eqref{eq:bulk-cb-exp-sigma-sublead}, one should consider an exact bulk conformal block expansion of complete \(2\)-pt functions (which is easy for \(\phi\), but more difficult for \(\sigma\)) and expand it in large \(N\). In our conventions~\cref{sec:conventions} we work to order \(N^0\) and thus drop all terms that are \(\mathcal{O} \left( \tfrac{1}{N} \right)\).
\subsection{\texorpdfstring{\(\left\langle \phi\phi \right\rangle\)}{<ϕϕ>} bulk expansion}
The bulk-bulk \(2\)-pt function of the fundamental field \(\phi\) (whose vector index we suppress) has the following bulk conformal block expansion
\begin{align}
\label{eq:phi-bulk-cb-exact}
&\left( \frac{\rho}{1-\rho} \right)^{\Delta_{\phi}} \left\langle \phi\phi \right\rangle= \mathcal{N}_{\phi} \left\{ \mathsf{G}_{\mathds{1}}^B +\sum_{k\geq 1}A \left[ \left\langle \phi\phi \right\rangle \right]_k \mathsf{G}^B_{\Delta_k}  \right\},&&A \left[ \left\langle \phi\phi \right\rangle \right]_k=2^{\Delta_k} \frac{C_{\phi\phi\Oper_k}}{C_{\Oper_k\Oper_k}} \left\langle \Oper_k \right\rangle\eqend
\end{align}
We should explain the relation between the coefficient of the block \(A_k\) and \CFT[B] data. Generally, for \(\left\langle \Oper\Oper \right\rangle\supset A_k \mathsf{G}_{\Oper_k}^B\), the relation is given by
\begin{align}
\label{eq:rel-block-coeff-ope-data}
A_k=2^{-\Delta_{\Oper}-\Delta_{\Oper}+\Delta_{\Oper_k}} \frac{C_{\Oper\Oper\Oper_k}}{C_{\Oper_k\Oper_k}}\left\langle \Oper_k \right\rangle\eqcomma
\end{align}
where the norm of the \(2\)-pt function \(C_{\Oper_k\Oper_k}\) appears since we don't work with unit normalized operators and powers of two unfortunately surface somewhere no matter what conventions one chooses (perhaps one could work with a different cross ratio than \(\rho\) but that brings other disadvantages). To avoid interruption of the flow of ideas we defer proof to~\Cref{app:rel-cb-coeff-OPE-data}.
Using it for~\eqref{eq:phi-bulk-cb-exact}, the factor \(2^{-\Delta_{\phi}-\Delta_{\phi}}\) got absorbed as \(\mathcal{N}_{\phi}=2^{-\Delta_{\phi}-\Delta_{\phi}} \mathcal{N}_{\phi}^{\mathrm{FP}}\) such that the normalization factor \(\mathcal{N}_{\phi}\) factorized and canceled on both sides of the equation.

The large \(N\) expansion for all relevant quantities  starts at order \(N^0\)
\begin{align}
\label{eq:phi-data-largeN-exp}
&\Delta_{\phi}= \frac{d-1}{2}+ \mathcal{O} \left( \frac{1}{N} \right), &&\left\langle \phi\phi \right\rangle= \mathcal{N}_{\phi} \left[ \frac{\left( 1-\rho \right)^2}{\rho} \right]^{\frac{d-1}{2}} + \mathcal{O} \left( \frac{1}{N} \right) \notag \\
&\Delta_k=\Delta_k^{(0)}+ \mathcal{O} \left( \frac{1}{N} \right), &&A_k= \left[ 2^{\Delta_k}\frac{C_{\phi\phi\Oper_k}}{C_{\Oper_k\Oper_k}} \left\langle \Oper_k \right\rangle \right]^{(0)}+ \mathcal{O} \left( \frac{1}{N} \right) \eqend
\end{align}
Comparing with the expansion~\eqref{eq:bulk-cb-exp-phi} derived from bulk conformal partial wave decomposition leads to \(\Delta_k^{(0)}=2k,\,k\in\Z_{\geq 1}\) and \(\left[ 2^{\Delta_k}\frac{C_{\phi\phi\Oper_k}}{C_{\Oper_k\Oper_k}} \left\langle \Oper_k \right\rangle \right]^{(0)}=C \left[ \left\langle \phi\phi \right\rangle \right]_k^{(0)}\), explicitly known coefficients in~\eqref{eq:coeff-bulk-phi-explicit}. We will show below that the family of bulk primaries with these scaling dimensions consists of operators (in schematic form)
\begin{align}
\label{eq:bulk-primaries}
&\left\{ \sigma \right\}_{k=1},&& \left\{ \sigma^2 \right\}_{k=2},&&& \left\{ \sigma^3,\;\sigma\Box\sigma \right\}_{k=3},&&& \left\{ \sigma^4,\;\Box \sigma^3,\;\Box^2 \sigma^2 \right\}_{k=4},\;\ldots\eqend
\end{align}
At least naively, it seems to be the right choice as they have indeed the required scaling dimensions \(\Delta^{(0)}=2,\,4,\,6,\,8,\ldots\). Fortunately, a wealth of \CFT[] data for this tower of operators is summarized in the helpful reference~\cite{Henriksson:2022rnm}. This will allow us to
\begin{itemize}
\item Identify this family as the correct one contributing to the above bulk expansion by matching their subleading anomalous dimensions and some OPE coefficients
\item Determine a previously unknown bulk OPE coefficient \(C_{\sigma\sigma\sigma^{3}}\)
\end{itemize}
Plugging~\eqref{eq:phi-data-largeN-exp} into~\eqref{eq:phi-bulk-cb-exact} yields
\begin{align}
\label{eq:phi-bulk-cb-exp-explicit}
\left( 1-\rho \right)^{\frac{d-1}{2}}=1+\sum_{k\geq 1} \left[ 2^{\Delta_k}\frac{C_{\phi\phi\Oper_k}}{C_{\Oper_k\Oper_k}} \left\langle \Oper_k \right\rangle \right]^{(0)} \mathsf{G}_{2k}^B \eqcomma
\end{align}
with coefficients \(\left[ 2^{\Delta_k}\frac{C_{\phi\phi\Oper_k}}{C_{\Oper_k\Oper_k}} \left\langle \Oper_k \right\rangle \right]^{(0)}\) given in~\eqref{eq:coeff-bulk-phi-explicit}. In~\cref{app:bulk-phi-converg-ser} we will verify that the series on the right hand side indeed converges to the function on the left hand side. Let us remark, that first few coefficients of the above expansion could be easily computed without the technology of spectral functions, just by expanding both sides in a Taylor series and subsequent recursion. We can now test the right hand side term by term against the hypothesis~\eqref{eq:bulk-primaries}.

Before diving into the rather technical analysis, it will be essential to establish a conversion dictionary between OPE coefficients in our conventions and those in~\cite{Henriksson:2022rnm}. In general, the normalization invariant combination is (here we use \(\left\langle \ind \right\rangle_{\mathrm{FP}}\) just as a proxy for normalization of the correlators)
\begin{align}
\label{eq:norm-inv-ope-coeff}
  \frac{\left\langle \Oper_1\Oper_2\Oper_3 \right\rangle_{\mathrm{FP}}}{\sqrt{\left\langle \Oper_1\Oper_1 \right\rangle_{\mathrm{FP}}} \sqrt{\left\langle \Oper_2\Oper_2 \right\rangle_{\mathrm{FP}}} \sqrt{\left\langle \Oper_3\Oper_3 \right\rangle_{\mathrm{FP}}}} \eqcomma
\end{align}
and we apply it with \(\Oper_1=\Oper_2= \left\{ \phi,\,\sigma \right\}\) and \(\Oper_3\) belonging to the tower of primaries~\eqref{eq:bulk-primaries}. In~\cite{Henriksson:2022rnm}, from where we take \CFT{} data, all operators are unit normalized and only the square of OPE coefficients is available. Inspecting~\eqref{eq:bulk-cb-exp-phi}, one sees that normalization~\eqref{eq:phi-norm} of the \(\left\langle  \phi\phi\right\rangle\) correlator has been factorized, thus the bulk expansion coefficient on the right hand side was computed for a unit normalized \(\phi\). On the other hand, normalization of \(\sigma\) is as in~\eqref{eq:sigma-prop-norm}/\eqref{eq:sigma-prop-ope-lim}, hence we arrive at the dictionary \(\sigma_{\text{\textcolor{gray}{us}}}= \sqrt{\mathcal{N}_{\sigma}^{\mathrm{FP}}} \sigma_{\text{\cite{Henriksson:2022rnm}}}\). Taking this into account, the final conversion table reads
\begin{align}
&C_{\phi\phi\Oper}^{\text{\textcolor{gray}{us}}}\coloneq \left\langle \phi\phi\Oper \right\rangle_{\mathrm{FP}}=\sqrt{\left\langle \Oper\Oper \right\rangle_{\mathrm{FP}}}C_{\phi\phi\Oper\text{\cite{Henriksson:2022rnm}}} \eqcomma \label{eq:conversion-ope-coeff-phi}\\ &C^{\text{\textcolor{gray}{us}}}_{\sigma\sigma\Oper}\coloneq \left\langle \sigma\sigma\Oper \right\rangle_{\mathrm{FP}}=\sqrt{\left\langle \sigma\sigma \right\rangle_{\mathrm{FP}}}\sqrt{\left\langle \sigma\sigma \right\rangle_{\mathrm{FP}}}\sqrt{\left\langle \Oper\Oper \right\rangle_{\mathrm{FP}}}C_{\sigma\sigma\Oper\text{\cite{Henriksson:2022rnm}}} \label{eq:conversion-ope-coeff-sigma}\eqend
\end{align}

\paragraph{\(k=1\): \(\sigma\).} From the conformal block decomposition we know the coefficient of \(\mathsf{G}^B_{\sigma}\), given in~\eqref{eq:coeff-bulk-phi-explicit} evaluated at \(k=1\). We need to match it against the relevant term in the \(\phi\times\phi\) OPE contributing to this block
\begin{align}
\label{eq:ope-phi-phi-sigma}
&\phi\times\phi\supset \frac{\left\langle \phi\phi\sigma \right\rangle_{\mathrm{FP}}}{\left\langle \sigma\sigma \right\rangle_{\mathrm{FP}}} \sigma,\quad \text{with}&&\frac{\left\langle \phi\phi\sigma \right\rangle_{\mathrm{FP}}}{\left\langle \sigma\sigma \right\rangle_{\mathrm{FP}}}\overset{\text{\eqref{eq:conversion-ope-coeff-phi}}}{=} \frac{\sqrt{\mathcal{N}_{\sigma}^{\mathrm{FP}}}C_{\phi\phi\sigma\text{\cite{Henriksson:2022rnm}}}}{\mathcal{N}_{\sigma}^{\mathrm{FP}}} \eqend
\end{align}
Taking the expectation value of the above equation gives according to~\eqref{eq:rel-block-coeff-ope-data} the coefficient multiplying the conformal block \(\mathsf{G}^B_{\sigma}\) that should be equated to the known one
\begin{align}
\label{eq:phi-matching-sigma-coeff}
\left\langle  \phi\times\phi\right\rangle^{(0)} \supset \left[2^{\Delta_{\sigma}}\frac{\left\langle \phi\phi\sigma \right\rangle^{(-1/2)}_{\mathrm{FP}}}{\left\langle \sigma\sigma \right\rangle^{(0)}_{\mathrm{FP}}} \left\langle  \sigma\right\rangle^{(1/2)} \right]^{(0)} \mathsf{G}^B_{\sigma}=\text{\eqref{eq:coeff-bulk-phi-explicit}}\Big\vert_{k=1} \mathsf{G}^B_{\sigma}= \frac{1-d}{2} \mathsf{G}^B_{\sigma}\eqend
\end{align}
Using the dictionary leads to a prediction for the OPE coefficient in conventions of~\cite{Henriksson:2022rnm}
\begin{align}
\label{eq:ope-coeff-phi-phi-sigma}
C_{\phi\phi\sigma\text{\cite{Henriksson:2022rnm}}}^{(-1/2)} =\frac{\sqrt{\mathcal{N}_{\sigma}^{\mathrm{FP}}}}{2^{\Delta_{\sigma}}\left\langle \sigma \right\rangle^{(1/2)}} \frac{1-d}{2}= -\frac{2 \sqrt{\frac{(3-d) \Gamma (d-1)}{\Gamma \left(\frac{3-d}{2}\right)}}}{(d-3) \Gamma \left(\frac{d-1}{2}\right)^{3/2}}\eqcomma
\end{align}
with quantities on the right hand side defined in~\eqref{eq:sigma-prop-norm} and~\eqref{eq:sigma-saddle-point}.

Squaring this prediction to compare it with the known~\cite{Lang:1993ct} (even at subleading order that we don't need here) squared OPE coefficient from \texttt{ONdata.m} supplementing~\cite{Henriksson:2022rnm} (translated to our conventions \(d_{\text{\cite{Henriksson:2022rnm}}}=(d+1)_{\text{\textcolor{gray}{us}}}\))
\begin{align}
\label{eq:coeff-phi-phi-sigma}
\left( C^{(-1/2)}_{\phi\phi\sigma\text{\cite{Henriksson:2022rnm}}} \right)^2=\texttt{OpeN[Op[S,0,1]]}\Big\vert_{\frac{1}{N}}\coloneq\frac{2^d \cos \left(\frac{\pi  d}{2}\right) \Gamma \left(\frac{d}{2}\right)}{\pi ^{3/2} (d-3) \Gamma \left(\frac{d-1}{2}\right)} \eqcomma
\end{align}
verifies~\eqref{eq:ope-coeff-phi-phi-sigma}.

As expected, the lowest operator contributing to the \(\phi\times\phi\) OPE is \(\sigma\). Let us also point out that~\eqref{eq:phi-bulk-cb-exp-explicit} is sensitive directly to OPE coefficients and not just their squares, so we can fix the sign.

\paragraph{\(k=2\): \(\sigma^2\).} Having faced the normalization hell once in detail, we can now move faster. Writing the relevant term in the OPE and directly taking expectation value gives
\begin{align}
\label{eq:ope-phi-phi-sigma2}
&\left\langle \phi\phi \right\rangle^{(0)}\supset \frac{\left\langle \phi\phi\sigma^2 \right\rangle_{\mathrm{FP}}^{(-1)}}{\left\langle \sigma^2\sigma^2 \right\rangle^{(0)}_{\mathrm{FP}}} \left\langle \sigma^2 \right\rangle^{(1)} \eqcomma
\end{align}
where all quantities are computed at leading order of their respective large \(N\) expansions, such that we can use factorization
\begin{align}
\label{eq:k2-largeN-factor}
&\left\langle \sigma^2 \right\rangle^{(1)}=\lim_{x\to y}\left\langle \sigma(x) \sigma(y) \right\rangle\coloneq \left( \left\langle \sigma \right\rangle^{(1/2)} \right)^2+ \mathcal{O} \left( N^{0} \right) \eqcomma \notag \\
& \left\langle \sigma^{2} \sigma^2\right\rangle_{\mathrm{FP}}^{(0)}=2 \left[ \left\langle \sigma\sigma \right\rangle_{\mathrm{FP}}^{(0)} \right]^{2}+ \mathcal{O} \left( \frac{1}{N} \right) \eqend
\end{align}
Plugging it to~\eqref{eq:phi-bulk-cb-exp-explicit} for \(k=2\) and taking into account the conversion dictionaries~\eqref{eq:conversion-ope-coeff-phi}/\eqref{eq:rel-block-coeff-ope-data}, one gets
\begin{align}
\label{eq:phi-check-k2}
2^{\Delta_{\sigma^{2}}}\frac{\sqrt{2\left(\mathcal{N}_{\sigma}^{\mathrm{FP}}\right)^{2}}C^{(-1)}_{\phi\phi\sigma^2\text{\cite{Henriksson:2022rnm}}}}{2 \left[ \mathcal{N}_{\sigma}^{\mathrm{FP}} \right]^{2}} \left( \left\langle \sigma \right\rangle^{(1/2)} \right)^2=\eqref{eq:coeff-bulk-phi-explicit}\Big\vert_{k=2}= \frac{1}{8} \left(d^2+\frac{32}{d-5}+7\right) \eqcomma
\end{align}
which predicts the OPE coefficient (again, the relation \(d_{\text{\cite{Henriksson:2022rnm}}}=(d+1)_{\text{\textcolor{gray}{us}}}\) was used to convert to our conventions, as well as \(\mathcal{N}_{\sigma}^{\mathrm{FP}}\geq 0\) for \(d\in [0,5]\) inside the square root)
\begin{align}
\label{eq:ope-coeff-phi-phi-sigma2}
C^{(-1)}_{\phi\phi\sigma^2\text{\cite{Henriksson:2022rnm}}}= \frac{2^{-\Delta_{\sigma^{2}}}\sqrt{2} \mathcal{N}_{\sigma}^{\mathrm{FP}}}{\left( \left\langle \sigma \right\rangle^{(1/2)} \right)^{2}} \frac{1}{8} \left(d^2+\frac{32}{d-5}+7\right)= \frac{2^{d-\frac{1}{2}} \cos \left(\frac{\pi  d}{2}\right) \Gamma \left(\frac{d}{2}\right)}{\pi ^{3/2} (d-5) \Gamma \left(\frac{d-1}{2}\right)}\eqend
\end{align}
The value of its square computed in~\cite{Alday:2019clp} and cited in~\cite{Henriksson:2022rnm} precisely matches the above result
\begin{align}
\label{eq:coeff-sigma2-Hen}
\left( C^{(-1)}_{\phi\phi\sigma^2\text{\cite{Henriksson:2022rnm}}} \right)^2= \texttt{OpeN[Op[S,0,2]]}\Big\vert_{\frac{1}{N^2}}\eqend
\end{align}
Thus also the second lowest operator in the \(\phi\times\phi\) OPE is confirmed to be \(\sigma^2\), providing strong evidence for correctness of the hypothesis~\eqref{eq:bulk-primaries}.

For the next two members of this family, the \(k=3\) operators, we encounter the phenomenon of mixing that complicates matters significantly. Consequently, we won't be able to extract new OPE data from the \(\left\langle \phi\phi \right\rangle\) correlator, it serves just as a constraint. However, analysis of the \(\left\langle \sigma\sigma \right\rangle\) correlator for these operators in the next section, will allow us to make a prediction of a yet unknown OPE coefficient.

\paragraph{\(k=3\): \(\left\{ \sigma^3,\,\sigma\Box\sigma \right\}\).} Before even trying to reconstruct their contribution to the \(\left\langle  \phi\phi\right\rangle\) \(2\)-pt function we have to deal with \emph{mixing}. It was studied and partially solved in~\cite{Derkachov:1997gc,Derkachov:1998js}.

Discussion of mixing starts at the leading order \(N\to\infty\). There are two primaries of dimension \(6\), \(\sigma^{3}\) and the double twist primary
\begin{align}
\label{eq:sigma-box-sigma-basis}
\text{\enquote{\(\left( \sigma\Box\sigma \right)\)}}\coloneq \left[ \sigma\sigma \right]_{1,\,0}=X \left( \sigma\Box\sigma \right)+Y \left( \partial\sigma \right)^2 \eqend
\end{align}
It contains two terms with coefficients \(X,\,Y\) that can be fixed (up to overall normalization) by imposing the non-trivial primary condition \(\left[ \mathbf{K_{\mu}}, \left[ \sigma\sigma \right]_{1,\,0} \right]=0\), where \(\mathbf{K}_{\mu}\) is the generator of special conformal transformations. Result of this computation is presented for instance in~\cite[(14)]{Fitzpatrick:2011dm} and we use it in the following form
\begin{align}
\label{eq:sigma-box-sigma-explicit}
\left[ \sigma\sigma \right]_{1,\,0}=\sigma\Box\sigma- \frac{2\Delta_{\sigma}+2-(d+1)}{2\Delta_{\sigma}} \left( \partial\sigma \right)^2 \eqend
\end{align}
Below, we will need the normalization of its \(2\)-pt and \(1\)-pt functions at leading order. It is a straightforward application of the chain rule, so we report the final results
\begin{align}
  &\left\langle \left( \left[ \sigma\sigma \right]_{1,\,0} \right) \left( \left[ \sigma\sigma \right]_{1,\,0} \right) \right\rangle^{(0)}_{\mathrm{FP}}=2(d+1)(d-9)(d-5) \left( \mathcal{N}_{\sigma}^{\mathrm{FP}} \right)^2 \label{eq:sigma-box-sigma-2pt-norm} \\
  &\left\langle \left[ \sigma\sigma \right]_{1,\,0} \right\rangle^{(1)}=(1+d) \left( \left\langle \sigma \right\rangle^{(1/2)} \right)^2 \label{eq:sigma-box-sigma-1pt} \eqend
\end{align}

On top of the above primaries, there are also two descendants of dimension \(6\), \(\Box\sigma^2\) and \(\Box^2\sigma\).

Eigenstates of the dilatation operator at the next large \(N\) order correspond to particular linear combinations of these four operators. We are interested in the primaries with scaling dimensions
\begin{align}
\label{eq:scal-dim-k3-mix}
  \Delta_{\Oper_{\sigma^3}}=\Delta_{\sigma^3}^{(0)}+\gamma_{\sigma^3}^{(-1)}+\cdots \\
  \Delta_{\Oper_{[\sigma\sigma]_{1,0}}}=\Delta_{[\sigma\sigma]_{1,0}}^{(0)}+\gamma_{[\sigma\sigma]_{1,0}}^{(-1)}+\cdots \eqcomma
\end{align}
corresponding to linear combinations
\begin{align}
  \Oper_{\sigma^3}=\sigma^3+a_1^{(-1/2)}[\sigma\sigma]_{1,0}+a_2^{(-1/2)}\Box \sigma^2+a_3^{(-3/2)}\Box^2\sigma \label{eq:eigenstate-k3-mix-sig3}\\
  \Oper_{[\sigma\sigma]_{1,0}}=[\sigma\sigma]_{1,0}+b_1^{(-1/2)}\sigma^3+b_2^{(-3/2)}\Box\sigma^2+b_3^{(-3/2)}\Box^2\sigma \label{eq:eigenstate-k3-mix-SigBoxSig}\eqend
\end{align}
Coefficients \(a_1\) and \(b_1\) were computed in~\cite{Derkachov:1997gc,Derkachov:1998js}. With these preliminary results, we can now determine their contribution to the \(\left\langle \phi\phi \right\rangle\) \(2\)-pt function.

Recall that the correlator is computed at leading order, but it doesn't mean that we can discard mixing altogether. For \(\Oper_{\sigma^3}\) there is in fact no mixing as its leading \(1\)-pt function gets contribution just from the first term in~\eqref{eq:eigenstate-k3-mix-sig3}, \(\left\langle \Oper_{\sigma^3} \right\rangle^{(3/2)}= \left\langle \sigma^3 \right\rangle^{(3/2)}\). However, for \(\Oper_{[\sigma\sigma]_{1,0}}\) the first two terms in~\eqref{eq:eigenstate-k3-mix-SigBoxSig} contribute, \(\left\langle \Oper_{[\sigma\sigma]_{1,0}} \right\rangle^{(1)}= \left\langle [\sigma\sigma]_{1,0}+b_1^{(-1/2)}\sigma^3 \right\rangle^{(1)}\). Thus the term multiplying the conformal block of dimension \(\Delta=6\) corresponding to these two operators takes the form
\begin{align}
\label{eq:phi-k3-contrib}
\left\langle \phi\phi \right\rangle^{(0)}\supset 2^{\Delta_{k=3}}\left[ \frac{\left\langle  \phi\phi\sigma^{3}\right\rangle_{\mathrm{FP}}^{(-3/2)}}{\left\langle  \sigma^3\sigma^3\right\rangle_{\mathrm{\mathrm{FP}}}^{(0)}} \left\langle  \sigma^3\right\rangle^{(3/2)} + \frac{\left\langle \phi\phi [\sigma\sigma]_{1,0} \right\rangle_{\mathrm{FP}}^{(-1)}}{\left\langle [\sigma\sigma]_{1,0} [\sigma\sigma]_{1,0} \right\rangle_{\mathrm{FP}}^{(0)}} \left\langle [\sigma\sigma]_{1,0}+b_1^{(-1/2)}\sigma^3 \right\rangle^{(1)} \right] \mathsf{G}_6^B \eqend
\end{align}
According to~\cite{Henriksson:2022rnm}, neither of the OPE coefficients is known (in the large \(N\) expansion, result for that of \(\sigma^3\) is available in the \(\varepsilon\)-expansion). The \(1\)-pt functions, appearing at leading order, can be computed by large \(N\) factorization and \(b_1^{(-1/2)}\) is available in~\cite{Derkachov:1997gc,Derkachov:1998js}. Hence we get one equation for two unknown OPE coefficients
\begin{align}
\label{eq:phi-k3-eqn}
  &2^{\Delta_{k=3}}\left[ \frac{\left\langle  \phi\phi\sigma^{3}\right\rangle_{\mathrm{FP}}^{(-3/2)}}{\left\langle  \sigma^3\sigma^3\right\rangle_{\mathrm{\mathrm{FP}}}^{(0)}} \left\langle  \sigma^3\right\rangle^{(3/2)} + \frac{\left\langle \phi\phi [\sigma\sigma]_{1,0} \right\rangle_{\mathrm{FP}}^{(-1)}}{\left\langle [\sigma\sigma]_{1,0} [\sigma\sigma]_{1,0} \right\rangle_{\mathrm{FP}}^{(0)}} \left\langle [\sigma\sigma]_{1,0}+b_1^{(-1/2)}\sigma^3 \right\rangle^{(1)} \right]
\notag  \\ &= \text{\eqref{eq:coeff-bulk-phi-explicit}}\Big\vert_{k=3}=  -\frac{(d-5) \left[(d-1)(d-3)\right]^2}{48 (d-9) (d-7)} \eqend
\end{align}

The rest of the operator tower, \(k\geq 4\), imposes an infinite number of constraints on \CFT[B] data. However, ever increasing degeneracy of operators with a fixed \(k\) and lack of known \CFT{} data for them, makes it impossible to derive concrete predictions. For this reason, we don't list them here, though they can be used as crosschecks when this data is computed in the future by different means.
\subsection{\texorpdfstring{\(\left\langle \sigma\sigma \right\rangle\)}{<σσ>} bulk expansion}
As outlined above, we need to consider an exact bulk conformal block decomposition of the modified \(2\)-pt function~\eqref{eq:sigma-corr-mod}, and than expand it in large \(N\) to order \(N^0\), so that it can be matched against~\eqref{eq:bulk-cb-exp-sigma-lead}/\eqref{eq:bulk-cb-exp-sigma-sublead}
\begin{align}
\label{eq:sigma-bulk-cb-exp-exact}
\left( \frac{\rho}{1-\rho} \right)^{\Delta_{\sigma}} \left\langle \sigma\sigma \right\rangle=A_0 \mathsf{G}^B_{\mathds{1}}+\sum_{k\geq 1}A \left[ \left\langle \sigma\sigma \right\rangle \right]_{k} \mathsf{G}^B_{\Delta_k} \eqend
\end{align}
Large \(N\) expansions to the order required for our considerations of objects appearing in this formula are
\begin{align}
\label{eq:sigma-terms-largeN-exp}
  &\Delta_{\sigma}=2+\gamma_{\sigma}^{(-1)},&& \left\langle \sigma\sigma \right\rangle= \left( \left\langle\sigma \right\rangle^{(1/2)} \right)^2+ \left[ 2 \left\langle \sigma \right\rangle^{(1/2)} \left\langle \sigma \right\rangle^{(-1/2)}+ \left\langle \delta\sigma\delta\sigma \right\rangle^{(0)}\right] \notag \\
  &\Delta_k=\Delta_k^{(0)}+\gamma_k^{(-1)},&&A_k=A_k^{(1)}+A_k^{(0)} \eqend
\end{align}
Plugging them in and separating orders, one gets the desired relations among bulk expansion coefficients and \CFT[B] data.
\paragraph{Leading \(N^1\) order.} At this order, comparing with~\eqref{eq:bulk-cb-exp-sigma-lead}, one gets
\begin{align}
\label{eq:sigma-lead-bulk-exp-compar}
\left( \frac{\rho}{1-\rho} \right)^2 \left( \left\langle \sigma \right\rangle^{(1/2)} \right)^2=A_0^{(1)} \mathsf{G}^B_{\mathds{1}}+A_1^{(1)} \mathsf{G}^B_{\Delta_1^{(0)}}+\sum_{k\geq 2}A_k^{(1)} \mathsf{G}^B_{\Delta_k^{(0)}}\overset{\text{\eqref{eq:bulk-cb-exp-sigma-lead}}}{=}\sum_{k\geq 2}C_k^{(1)} \mathsf{G}^B_{2k} \eqcomma
\end{align}
which fixes the leading scaling dimensions and expansion coefficients
\begin{align}
\label{eq:sigma-lead-match-terms}
  &\Delta_k^{(0)}=2k,\,k\in\Z_{\geq 2},&& A_0^{(1)}=0, \notag \\
  &A_k^{(1)}= C_k^{(1)},\, k\in\Z_{\geq 2},&& A_1^{(1)}=0 \eqcomma
\end{align}
with the explicit coefficients \(C_k^{(1)}\) computed from the bulk spectral function provided in~\eqref{eq:sigma-lead-bulk-cb-coeff}.
\paragraph{Subleading \(N^0\) order.} Considering~\eqref{eq:sigma-bulk-cb-exp-exact} at this order gives
\begin{align}
\label{eq:sigma-bulk-cb-exp-sublead}
 &\text{\eqref{eq:sigma-corr-mod-sublead}}=A_0^{(0)} \mathsf{G}^B_{\mathds{1}}+A_1^{(0)} \mathsf{G}^B_{\Delta_1^{(0)}}+{\underbrace{A_1^{(1)}}_{0\text{ by~\eqref{eq:sigma-lead-match-terms}}}}\gamma_1^{(-1)}\partial_{\Delta} \mathsf{G}^B_{\Delta_1^{(0)}}+\sum_{k\geq 2} \left[ A_k^{(0)} \mathsf{G}^B_{\Delta_{k}^{(0)}}+A_k^{(1)}\gamma_k^{(-1)}\partial_{\Delta} \mathsf{G}^B_{\Delta_k^{(0)}}\right] \notag \\ &\overset{\text{\eqref{eq:bulk-cb-exp-sigma-sublead}}}{=}\mathcal{N}_{\sigma}^{\mathrm{FP}} \left\{\mathsf{G}^B_{\mathds{1}}-\frac{(d-1)(d-2)}{16} \mathsf{G}_2^B\right\}+\sum_{k\geq 2}\left\{C\left[\langle \sigma\sigma \rangle\right]^{(0)}_k\mathsf{G}^B_{2k} + D\left[\langle \sigma\sigma \rangle\right]^{(0)}_k\partial_{\Delta} \mathsf{G}^B_{2k}\right\} \eqcomma \end{align}
from where one concludes, by matching scaling dimensions of blocks on both sides and comparing coefficients in front of them
\begin{align}
\label{eq:sigma-sublead-match-terms}
  &\Delta_k^{(0)}=2k,\,k\in\Z_{\geq 2},&& A_k^{(0)}=C_{k}^{(0)},\,k\in\Z_{\geq 2} \notag \\
  &A_0^{(0)}= -\frac{16}{(d-1)(d-2)}A_1^{(0)}= \mathcal{N}_{\sigma}^{\mathrm{FP}},&& A_k^{(1)}\gamma_k^{(-1)}=D_k^{(0)},\,k\in\Z_{\geq 2} \eqend
\end{align}
The main outcome of the above analysis is that, like the \(\phi\times\phi\) OPE, also the \(\sigma\times\sigma\) OPE contains an infinite tower of operators of the schematic form~\eqref{eq:bulk-primaries} (apart from \(d=2\), where the coefficient of \(\sigma_2\) vanishes as apparent from~\eqref{eq:sigma-bulk-cb-exp-sublead}). It can be reinforced by matching also the anomalous dimensions \(\gamma_k^{(-1)}\) of this tower. Unfortunately, one doesn't get too far due to operator mixing, as we will now demonstrate.

\paragraph{\(k=1\): \(\sigma\).} This operator contributes just to the subleading term~\eqref{eq:sigma-bulk-cb-exp-sublead} (if \(d\neq 2\)) as its coefficient in the leading term~\eqref{eq:sigma-lead-bulk-exp-compar} vanishes due to~\eqref{eq:sigma-lead-match-terms}. Comparing its contribution via the bulk channel \(\sigma\times\sigma\) OPE with the coefficient of its block yields (recall~\eqref{eq:rel-block-coeff-ope-data} regarding factors of two)
\begin{align}
\label{eq:sigma-k1-sublead}
2^{-\Delta_{\sigma}^{(0)}-\Delta_{\sigma}^{(0)}+\Delta_{\sigma}^{(0)}}\left[ \frac{\left\langle \sigma\sigma\sigma \right\rangle^{(-1/2)}_{\mathrm{FP}}}{\left\langle \sigma\sigma \right\rangle^{(0)}_{\mathrm{FP}}} \left\langle \sigma \right\rangle^{(1/2)} \right]^{(0)}=- \mathcal{N}_{\sigma}^{\mathrm{FP}}\frac{(d-1)(d-2)}{16} \eqcomma
\end{align}
where we used that the leading order of the OPE coefficient is \(N^{- \tfrac{1}{2}}\). Thus all quantities are evaluated at leading order and we immediately get a prediction for it
\begin{align}
\label{eq:ope-coeff-sigma-sigma-sigma}
C_{\sigma\sigma\sigma\text{\cite{Henriksson:2022rnm}}}^{(-1/2)}=- 2^{\Delta_{\sigma}^{(0)}}\frac{\sqrt{\mathcal{N}_{\sigma}^{\mathrm{FP}}}}{\left\langle \sigma \right\rangle^{(1/2)}} \frac{(d-1)(d-2)}{16}= \frac{4(d-2)}{\Gamma\left(\frac{d-1}{2}\right)} \sqrt{ \frac{2^{d-3} \Gamma\left(\frac{d}{2}\right)}{\sqrt{\pi} \Gamma\left(\frac{5-d}{2}\right)} }\eqend
\end{align}
The square of this equation in fact reproduces the value of the squared OPE coefficient \(\texttt{OpeN[Op[S,0,1],Op[S,0,1],Op[S,0,1]]}\Big\vert_{\frac{1}{N}}\) cited in~\cite{Henriksson:2022rnm} and first computed in~\cite{Goykhman:2019kcj}, but \CFT[B] is sensitive also to its sign, which we fixed.

\paragraph{\(k=2\): \(\sigma^2\).} The \(\sigma^2\) operator participates both in the leading~\eqref{eq:sigma-lead-bulk-exp-compar} and subleading~\eqref{eq:sigma-bulk-cb-exp-sublead} equations, which we discuss in order.

At leading order, using the by now familiar strategy of matching the bulk OPE with the computed coefficient of the block, together with large \(N\) factorization (as all quantities are computed at leading order) and the conversion dictionaries~\eqref{eq:conversion-ope-coeff-sigma}/\eqref{eq:rel-block-coeff-ope-data}, gives
\begin{align}
\label{eq:sigma-k2-lead}
  \underbrace{2^{-\Delta_{\sigma}^{(0)}-\Delta_{\sigma}^{(0)}+\Delta_{\sigma^2}^{(0)}}}_{1}\left[ \frac{\left\langle \sigma\sigma\sigma^2 \right\rangle_{\mathrm{FP}}^{(0)}}{\left\langle \sigma^2\sigma^2 \right\rangle_{\mathrm{FP}}^{(0)} \left\langle \sigma^2 \right\rangle^{(1)}} \right]^{(1)}\coloneq \frac{\sqrt{\mathcal{N}_{\sigma}^{\mathrm{FP}}}\sqrt{\mathcal{N}_{\sigma}^{\mathrm{FP}}}\sqrt{2\left( \mathcal{N}_{\sigma}^{\mathrm{FP}} \right)^2}C_{\sigma\sigma\sigma^2\text{\cite{Henriksson:2022rnm}}}}{2 \left( \mathcal{N}_{\sigma}^{\mathrm{FP}} \right)^2 \left( \left\langle \sigma \right\rangle^{(1/2)} \right)^2}&=
  \text{\eqref{eq:sigma-lead-bulk-cb-coeff}}\Big\vert_{k=2} \notag \\
  &= \left( \left\langle \sigma \right\rangle^{(1/2)} \right)^2 \eqend
\end{align}
Thus the prediction for the leading OPE coefficient is
\begin{align}
\label{eq:ope-coeff-sigma-sigma-sigma2}
C_{\sigma\sigma\sigma^2\text{\cite{Henriksson:2022rnm}}}=\sqrt{2} \eqcomma
\end{align}
which precisely matches the squared value cited in~\cite{Henriksson:2022rnm}.

At subleading order, the computation splits into two parts -- matching the coefficients of blocks with/without derivatives. Those differentiated with respect to its scaling dimension encode anomalous dimensions multiplied by coefficients coming from the leading order we just discussed. So we start with them.

The anomalous dimension of \(\sigma^2\) is predicted by~\eqref{eq:sigma-lead-match-terms}/\eqref{eq:sigma-sublead-match-terms} as the ratio of the coefficients multiplying the differentiated block by the leading coefficient appearing in~\eqref{eq:sigma-k2-lead} on the right hand side
\begin{align}
\label{eq:k2-anom-dim-predict}
\gamma_2^{(-1)}= \frac{D \left[ \left\langle \sigma\sigma \right\rangle \right]_2^{(0)}}{A \left[ \left\langle \sigma\sigma \right\rangle \right]_2^{(1)}}=-\frac{2^d (d-3) d^2 \cos \left(\frac{\pi  d}{2}\right) \Gamma \left(\frac{d}{2}\right)}{\pi ^{3/2} \Gamma \left(\frac{d+3}{2}\right)} \eqend
\end{align}
It precisely agrees with the anomalous dimension of \(\sigma^2\), \(\texttt{DeltaN[Op[S,0,2]]}\Big\vert_{\frac{1}{N}}\), cited in~\cite{Henriksson:2022rnm} and computed in~\cite{Broadhurst:1996ur}.

\paragraph{\(k=3\): \( \left\{ \sigma^3,\,\sigma\Box\sigma \right\}\).}

These operators enter the leading part of the correlator~\eqref{eq:sigma-lead-match-terms}, as well as the subleading one~\eqref{eq:sigma-sublead-match-terms} -- both the terms with and without derivatives of conformal blocks. The subleading equation containing undifferentiated blocks involves \CFT{} data that needs to be evaluated at subleading order and is currently unknown. Hence it imposes a single constraint on four to be determined pieces of data. Consequently, it is not immediately useful to us, so we won't discuss it any further.

Instead, let us focus on a coupled system of equations originating from the leading part and the subleading part with differentiated blocks. It takes the form
\begin{align}
  &A_{\sigma^3}^{(1)}+A_{[\sigma\sigma]_{1,0}}^{(1)}= C\left[ \left\langle \sigma\sigma \right\rangle \right]_3^{(1)}=  \left( \left\langle \sigma \right\rangle^{(1/2)} \right)^2\frac{8}{9-d} \eqcomma \label{eq:sigma-k3-lead}\\
  &A_{\sigma^3}^{(1)}\gamma_{\sigma^3}^{(-1)}+A_{[\sigma\sigma]_{1,0}}^{(1)}\gamma_{[\sigma\sigma]_{1,0}}^{(-1)}
    \begin{aligned}[t]
  &= D\left[ \left\langle \sigma\sigma \right\rangle \right]_3^{(0)}   \\
  &= -\frac{4 \left( d^{5} - 13 d^{4} + 33d^{3} + 13d^{2} - 10d - 72 \right) \cos \left(\frac{\pi  d}{2}\right) \Gamma (d+1)}{3 \pi  (d-9) (d+1) \Gamma \left(\frac{d-3}{2}\right)^2}\eqend
     \end{aligned}
    \label{eq:sigma-k3-sublead-deriv}
\end{align}

As for \(\left\langle \phi\phi \right\rangle\), mixing of operators (previously discussed in~\eqref{eq:sigma-box-sigma-basis}--\eqref{eq:phi-k3-contrib}) enters scene at this stage. Recall that the conformal block expansion coefficients are a product of an OPE coefficient with a \(1\)-pt function. They essentially take the same form as in~\eqref{eq:phi-k3-contrib}, just with the replacement \(\phi\to\sigma\) and factors of two were inserted in compliance with~\eqref{eq:rel-block-coeff-ope-data}
\begin{align}
  &A_{\sigma^{3}}^{(1)}=\underbrace{2^{-\Delta_{\sigma}^{(0)}-\Delta_{\sigma}^{(0)}+\Delta_{\sigma^3}^{(0)}}}_{2^2=4} \left[ \frac{\left\langle \sigma\sigma\sigma^3 \right\rangle^{(-1/2)}_{\mathrm{FP}}}{\left\langle \sigma^3\sigma^3 \right\rangle^{(0)}_{\mathrm{FP}}} \left\langle \sigma^3 \right\rangle^{(3/2)} \right]^{(1)} \eqcomma \label{eq:A-coeff-form-sig3}\\
  &A_{[\sigma\sigma]_{1,0}}^{(1)}= \underbrace{2^{-\Delta_{\sigma}^{(0)}-\Delta_{\sigma}^{(0)}+\Delta_{[\sigma\sigma]_{1,0}}^{(0)}}}_{2^2=4}\left[ \frac{\left\langle \sigma\sigma[\sigma\sigma]_{1,0} \right\rangle^{(0)}_{\mathrm{FP}}}{\left\langle [\sigma\sigma]_{1,0}[\sigma\sigma]_{1,0} \right\rangle^{(0)}_{\mathrm{FP}}} \left\langle [\sigma\sigma]_{1,0} + b_1^{(-1/2)}\sigma^{3} \right\rangle^{(1)} \right]^{(1)}\eqend\label{eq:A-coeff-form-sigBoxsig}
\end{align}
The important point here is that all objects are to be evaluated at leading order, something within reach due to large \(N\) factorization.

Let us disentangle the known data from that to be computed. The normalization of the \(2\)-pt and \(1\)-pt function of \([\sigma\sigma]_{1,0}\) was computed in~\eqref{eq:sigma-box-sigma-2pt-norm} and~\eqref{eq:sigma-box-sigma-1pt}, respectively. The OPE coefficient \(\left\langle \sigma\sigma [\sigma\sigma]_{1,0} \right\rangle^{(0)}_{\mathrm{FP}}\) is known~\cite{Henriksson:2022rnm} and so are both the anomalous dimensions \(\gamma_{\sigma^3}^{(-1)},\,\gamma_{[\sigma\sigma]_{1,0}}^{(-1)}\)~\cite{Henriksson:2022rnm} (derived in~\cite{Vasiliev:1993ux,Derkachov:1997gc}). Therefore the only unknown that will be computed is the OPE coefficient \(\left\langle \sigma\sigma\sigma^3 \right\rangle^{(-1/2)}_{\mathrm{FP}}\) and for the moment we are treating the mixing coefficient \(b_1^{(-1/2)}\) defined in~\eqref{eq:eigenstate-k3-mix-SigBoxSig} also as unknown (in a moment we will compare with its value obtained in~\cite{Derkachov:1997gc,Derkachov:1998js}).

We are thus solving a system of two equations for two unknowns that has a straightforward solution
\begin{align}
&A_{\sigma^3}^{(1)}=\frac{D[\left\langle \sigma\sigma \right\rangle]_3^{(0)}-C[\left\langle \sigma\sigma \right\rangle]_3^{(1)}\gamma_{[\sigma\sigma]_{1,0}}^{(-1)}}{\gamma_{\sigma^3}^{(-1)}-\gamma_{[\sigma\sigma]_{1,0}}^{(-1)}} \eqcomma\label{eq:mix-A-sig3-sol} \\
&A_{[\sigma\sigma]_{1,0}}^{(1)}=\frac{D[\left\langle \sigma\sigma \right\rangle]_3^{(0)}-C[\left\langle \sigma\sigma \right\rangle]_3^{(1)}\gamma_{\sigma^{3}}^{(-1)}}{\gamma_{[\sigma\sigma]_{1,0}}^{(-1)}-\gamma_{\sigma^3}^{(-1)}} \eqcomma \label{eq:mix-A-sigBoxsig-sol}
\end{align}
with the non-differentiated/differentiated bulk block coefficients \(C_3/D_3\) explicitly given in~\eqref{eq:sigma-k3-lead}/\eqref{eq:sigma-k3-sublead-deriv}. We don't show the anomalous dimensions here since they can be easily accessed in~\cite{Henriksson:2022rnm}.

Plugging~\eqref{eq:A-coeff-form-sig3} into~\eqref{eq:mix-A-sig3-sol} and using the dictionary~\eqref{eq:conversion-ope-coeff-sigma} yields a prediction for the yet unknown (as far as we are aware) OPE coefficient
\begin{align}
\label{eq:OPE-sig-sig-sig3}
C_{\sigma\sigma\sigma^3\text{\cite{Henriksson:2022rnm}}}^{(-1/2)}=\frac{\sqrt{3! \mathcal{N}_{\sigma}^{\mathrm{FP}}}}{4 \left( \left\langle \sigma \right\rangle^3 \right)^{(3/2)}}A_{\sigma^3}^{(1)}=-\frac{16   \sqrt{6(d-3) \cos \left(\frac{\pi  d}{2}\right) \Gamma (d-1)}}{\sqrt{\pi}  \Gamma \left(\frac{d-1}{2}\right)} \frac{(d-2)}{\left( d^{2} - 32d + 63 \right)}\eqcomma
\end{align}
where the subscript~\cite{Henriksson:2022rnm} merely means it has been converted to those normalization conventions corresponding to unit normalized operators. However, \(d\) still represents dimension of the boundary. Based on correspondence with Johan Henriksson after the preprint of this paper appeared, (square of) the above OPE coefficient will be included in his database: \url{https://github.com/johhen1/ON-model}.

Turning now to equation \eqref{eq:A-coeff-form-sigBoxsig}, which contains a single unknown---the mixing coefficient \(b_{1}\)---we can express this coefficient in a straightforward way, yielding
\begin{align}
\label{eq:mix-coeff-sol}
  b_1^{(-1/2)}&=\frac{1}{\left( \left\langle \sigma \right\rangle^3 \right)^{(3/2)}} \left[ A_{[\sigma\sigma]_{1,0}}^{(1)} \frac{\sqrt{\left\langle [\sigma\sigma]_{1,0}[\sigma\sigma]_{1,0} \right\rangle}}{4 \mathcal{N}_{\sigma}^{\mathrm{FP}} C_{\sigma\sigma[\sigma\sigma]_{1,0}}^{(0)}}-\left\langle [\sigma\sigma]_{1,0} \right\rangle^{(1)} \right]\notag\\
  &=-\frac{2 (d-9) (d-2) (d+1)}{d^{2}-32d+63} \eqend
\end{align}
Comparing with the expression obtained in~\cite{Derkachov:1997gc,Derkachov:1998js}, we get a perfect match upon accounting for differences in normalization conventions.\footnote{The overall normalization for the \(\sigma\) operator in~\cite{Derkachov:1997gc,Derkachov:1998js} (see also~\cite{Derkachov:1997ch} compares with ours as \(\sigma^{\text{DM}} = - \frac{1}{\sqrt{N}}\sigma^{\text{us}}\), while instead of the double twist operator \(\left[ \sigma \sigma  \right]_{1,0}\) as we define in \eqref{eq:sigma-box-sigma-explicit}, Derkachov and Manashov use a rescaled one denoted \(O_{3}^{(2)}\) and related to ours by \(O_{3}^{(2)} = \frac{1}{N}\frac{4}{9-d} \left[ \sigma \sigma  \right]_{1,0}^{\text{us}}\).} This strongly supports our method of extracting the coefficients and also provides more solid ground to the correctness of the new OPE coefficient \eqref{eq:OPE-sig-sig-sig3}.

Before closing this section, let us comment on the structure of results we obtained. Paragraphs below are based on observations of an anonymous referee, whom we thank for suggestions leading to an improved presentation of conclusions:

\paragraph{Structure of poles and zeros in \(d\).}
The advantage of the large \(N\) expansion is that it outputs expressions exact in the dimension \(d\). Many of them display a peculiar structure of poles and zeros. We anticipate three main mechanisms -- \emph{(i)} null states; \emph{(ii)} the fact that vertices in the effective action~\eqref{eq:Z-conformal-general} with an odd number of \(\sigma\)-fields (in particular the cubic vertex) vanish in \(D=d+1=3\); \emph{(iii)} level crossing requiring replacement of non-degenerate perturbation theory with a degenerate one -- however certain aspects are technical and necessitate a detailed discussion below. It is useful to start with vanishing of the norm of the double-twist operator \([\sigma\sigma]_{1,0}\) as it affects other formulas.
\begin{itemize}
\item \eqref{eq:sigma-box-sigma-2pt-norm}: The vanishing of the norm of this operator in \(D=d+1=6,\,10\) happens due to null states. In \(D=6\), the second piece in~\eqref{eq:sigma-box-sigma-explicit} vanishes and hence the operator becomes \(\sigma\Box\sigma\). But \(\sigma\) is a free field in \(D=6\) since \(\Delta_{\sigma}=\frac{D-2}{2}\), which implies that \(\Box\sigma\) is null. A similar mechanism works for \(D=10\), where the two pieces in~\eqref{eq:sigma-box-sigma-explicit} combine to \(\sigma\Box\sigma+\left(\partial\sigma\right)^2=\frac{1}{2}\Box (\sigma)^2\). The same logic implies that \(\sigma^2\) is a free field in \(D=10\) as \(\Delta_{\sigma^2}=\frac{D-2}{2}\). Hence the double-twist operator is again a null state with a vanishing norm. Finally, the zero at \(D=d+1=0\) is kinematic due to a trace over space in the definition of the double-twist operator under consideration. When space is just a point, the trace and thus also the operator just vanish.
\item \eqref{eq:phi-k3-eqn}: Coefficient of the block \(\mathsf{G}_6^{B}\) in the \(\left\langle \phi\phi \right\rangle\) correlator has zeros at \(d=1,\,3,\,5\) and poles at \(d=7,\,9\). Zeros are easy to explain, so let's start from them. At odd \(d\), the correlator is just a polynomial, for the listed dimensions under consideration its form is \(1,\,1-\rho,\,(1-\rho)^2\). Since expansion of the blocks starts with \(\rho^{\tfrac{\Delta}{2}}\) it is clear that only a finite number of blocks can contribute: for \(d=1\) only \(\mathds{1}\) contributes, for \(d=3\) \(\{\mathds{1},\,\sigma\}\) contribute and finally for \(d=5\) \(\{\mathds{1},\,\sigma,\,\sigma^2\}\) contribute. In these dimensions, coefficients of all other blocks must develop zeros which we indeed observe for the block \(\mathsf{G}_6^B\) corresponding to the operators \(\sigma^3,[\sigma\sigma]_{{1,0}}\). On the other hand, the reason for poles at \(d=7,\,9\) is trickier. It is based on a bulk shadow relation between blocks \(\left(\mathsf{G}_{\Delta_L}^B,\, \mathsf{G}_{\Delta_H}^B\right)\) contributing to the \(2\)-pt function and the observation that the block of the dual pair with lower scaling dimension \(\Delta_L\) develops a singularity (in \(d\)) whose \(\rho\)-expansion starts at order \(\tfrac{\Delta_L}{2}-\left(\Delta_{L}+\tfrac{1-d}{2}\right)+1\). Equating it with the \(\rho\)-expansion of the higher (non-singular) block starting at \(\tfrac{\Delta_H}{2}\), one concludes the shadow relation \(\Delta_L+\Delta_{H}=D\). Then the coefficient of the higher block must develop a pole (in \(d\)) in order to cancel the singularity of the lower block as their sum contributing to the correlator is perfectly regular. Let us sketch the argument only for \(D=d+1=8\) as for the other pole the logic is the same. In this case the bulk shadow dual pair of blocks is \(\left( \mathsf{G}_6^B,\, \mathsf{G}_2^B \right)\) corresponding to the operators \({\sigma^3,\,[\sigma\sigma]_{1,0}}\) and \(\sigma\), respectively. The last argument of the hypergeometric function (Pochhammer in the denominator) in the definition~\eqref{eq:bulk-CB} of the lower block \(\mathsf{G}_2^B\) becomes \(\Delta+\tfrac{1-d}{2}=-1\) and thus the block develops a singularity, whose \(\rho\)-expansion starts as \(\rho^3\) which coincides with the starting power of the higher block \(\mathsf{G}_6^B\). Its coefficient must thus produce a pole at \(D=d+1=8\) to cancel this singularity since the \(2\)-pt function is equal to \(\left( 1-\rho \right)^3\), hence regular. For \(D=10\), the bulk shadow dual pair is \(\left(\mathsf{G}_{6}^{B},\, \mathsf{G}_{4}^B\right)\) and the argument proceeds verbatim. To summarize, we have \(\left\langle \phi\phi \right\rangle \supset c_{L} \mathsf{G}_{\Delta_L}^{B}+c_H \mathsf{G}_{\Delta_H}^B\): at \(d=7,\,9\) the correlator is regular, the block \(\mathsf{G}_{\Delta_L}^B\) develops a pole in \(d\) while its coefficient \(c_{L}\) stays finite and on the other hand the block \(\mathsf{G}_{\Delta_H}^B\) remains also regular, therefore its coefficient \(c_{H}\) must develop a compensating pole in \(d\).
\item \eqref{eq:OPE-sig-sig-sig3}: The zero at \(d=2\) comes from vanishing of the cubic \(\sigma\)-vertex in~\eqref{eq:ope-coeff-sigma-sigma-sigma} since construction of the \(3\)-pt function \(\left\langle \sigma\sigma\sigma^3 \right\rangle\) requires insertion of at least one cubic vertex even at leading large \(N\). Zeros at \(d=1,\,3\) correspond to the lower and upper critical dimension and originate exclusively from the normalization of the \(\sigma\)-field \(\mathcal{N}_{\sigma}^{\mathrm{FP}}\) (for instance at the upper critical dimension the theory becomes free, hence \(\sigma\sim\phi^2\), but \(\left\langle \phi^2\phi^2:\phi^6: \right\rangle\) indeed vanishes by Wick contractions). Now, let's discuss the more interesting mechanism \emph{(iii)} for generating poles. They are roots of the denominator \(d^2-32d+63=0\), \(d_{\pm}=16\pm\sqrt{193}\). At first sight this looks suspicious, but it can be checked that this condition follows from vanishing of the denominator in~\eqref{eq:mix-A-sig3-sol}, namely degeneracy of the primaries \(\sigma^3\) and \([\sigma\sigma]_{{1,0}}\) even at subleading large \(N\) order (at generic dimensions \(d\), degeneracy at subleading order is of course lifted). So these poles warn us that at these peculiar dimensions non-degenerate conformal perturbation theory encoded in~\eqref{eq:eigenstate-k3-mix-sig3}/\eqref{eq:eigenstate-k3-mix-SigBoxSig} breaks down and would have to be replaced by degenerate perturbation theory. It is worth emphasizing that the correlator \(\left\langle \delta\sigma\delta\sigma \right\rangle\) is insensitive to this issue as it sees only the weighted averages~\eqref{eq:sigma-k3-lead}/\eqref{eq:sigma-k3-sublead-deriv} which are regular at \(d_{\pm}\) (each of the two terms corresponding to individual eigenoperators diverges, but their residues cancel).
\item \eqref{eq:mix-coeff-sol}: The poles are precisely those just described above. By (first order) non-degenerate conformal perturbation theory the mixing angle \(b_1\) between \([\sigma\sigma]_{1,0}\) and \(\sigma^3\) carries an energy denominator \(\tfrac{1}{\gamma_{[\sigma\sigma]_{1,0}}-\gamma_{\sigma^3}}\sim\tfrac{1}{E_{[\sigma\sigma]_{1,0}}-E_{\sigma^3}}\), where \(E\) are energies when the theory is quantized on the (generalized) cylinder by a Weyl map. However, at \(d_{\pm}\) this energy difference vanishes as explained above and thus non-degenerate perturbation theory breaks down and becomes degenerate. Next, we analyze the zeros. Mixing of the operators with two and three \(\sigma\)-fields needs a vertex from the effective action with an odd number of \(\sigma\)-fields. But by \emph{(ii)} those vanish in \(d=2\), so for this dimension there is no mixing and the entire contribution to the \(\mathsf{G}_6^B\) block comes from the double-twist operator (since \(A_{\sigma^3}\eqdim{2} 0\)). Finally, the zero at \(d=9\) is slightly tricky. As we have shown at \(d=9\) the double-twist operator becomes null, so it is plausible that in \(\mathsf{O}_{[\sigma\sigma]_{1,0}}=[\sigma\sigma]_{1,0}+b_1 \sigma^3\) the mixing coefficient \(b_1\) vanishes to prevent the \(\sigma^3\) admixture become the dominant contribution. However, the double-twist operator is also null in \(d=5\) and there is no zero in \(b_1\), so this argument is not watertight.
\end{itemize}

As we have learned the structure of poles and zeros following from conditions \emph{(i), (ii), (iii)} is rather constraining. It is remarkable that the result for the unknown OPE coefficient \(C_{\sigma\sigma\sigma^3}\) could have been (almost) guessed. The denominator of the last term in~\eqref{eq:OPE-sig-sig-sig3} comes from the known anomalous dimensions \(\gamma_{[\sigma\sigma]_{1,0}},\,\gamma_{\sigma^3}\) and carefully thinking about degenerate conformal perturbation theory formulated as condition \emph{(iii)}. Its numerator follows from the fact that even at leading large \(N\) a cubic \(\sigma\)-vertex is required, which by \emph{(ii)} vanishes at \(d=2\). The first term is just a canonical normalization \(\sqrt{3! \mathcal{N}_{\sigma}^{\mathrm{FP}}}\) that could have been computed by Wick contractions at the leading large \(N\) order. What remains to be determined is just a constant which takes the value \((-1)\).

\paragraph{Resolving mixing for \(\left\langle \phi\phi \right\rangle\).}
To disentangle mixing of the operators \([\sigma\sigma]_{1,0}\) and \(\sigma^3\) in the \(\left\langle \phi\phi \right\rangle\) correlator, one would have to supply~\eqref{eq:phi-k3-eqn} providing the coefficient of the \(\mathsf{G}_6^B\) block with an equation giving the coefficient of the differentiated block \(\partial_{\Delta} \mathsf{G}_6^B\) (as was done for the \(\sigma\)-correlator). There are two ways how to provision this data: \emph(i) compute \(\left\langle \phi\phi \right\rangle\) at subleading \(1/N\) order, which was done in~\cite{OhnoOkabe1983a,OhnoOkabe1983b,OhnoOkabe1983c,OhnoOkabe1984} but only in a very implicit form (see comments in the next section) not suitable for the task; \emph{(ii)} compute \(\left\langle \phi\phi \right\rangle\) at \(\epsilon^3\)-order in the \(\epsilon\)-expansion, which to the best of our knowledge is not available either as the state of the art result in this regard is the order \(\epsilon^2\) computation in~\cite{Bissi:2018mcq}. So at this point full unmixing is unfortunately not possible. However, if one contents with a weaker result,~\eqref{eq:phi-k3-eqn} can be solved for the unknown OPE coefficient \(C_{\phi\phi[\sigma\sigma]_{1,0}}\) directly in a double \(\left(\tfrac{1}{N},\epsilon\right)\) expansion. The equation contains just two OPE coefficients not known in the large \(N\) expansion, \(C_{\phi\phi[\sigma\sigma]_{1,0}}\) and \(C_{\phi\phi\sigma^3}\), but the latter one is known in the \(\epsilon\)-expansion. So one expands the equation in \(\epsilon\) to prepare it for the input of this latter OPE coefficient and at the same time one expands the tabulated result for \(C_{\phi\phi\sigma^3}\) to leading large \(N\) order to make it consistent with the equation. Then one just solves for the only remaining unknown \(C_{\phi\phi[\sigma\sigma]_{1,0}}\) (the tabulated result for \(C_{\phi\phi\sigma^3}\) gives only the square and we considered the positive branch for its square root needed; the choice matters, but by our analysis we unambiguously fixed the signs of the two previous OPE coefficients \(C_{\phi\phi\sigma},\,C_{\phi\phi\sigma^2}\) to be positive, hence we believe that this is the correct option) yielding at leading \(1/N\) order (the definition of \(\epsilon\) in~\cite{Henriksson:2022rnm} is \(d=4-\epsilon\))
\begin{align}
\label{eq:ope-phi-phi-dt-double-exp}
C_{\phi\phi[\sigma\sigma]_{1,0}}=\frac{\epsilon^2}{12 \sqrt{6} N}+\cdots \eqend
\end{align}
Future derivations, either working at large \(N\) and thus outputting exact functions of \(d\) or operating within the \(\epsilon\)-expansion providing coefficients exact in \(N\), should match this double-expansion Taylor coefficient as a crosscheck. It was computed from the input of the squared OPE coefficient \(C_{\phi\phi\sigma^3}\) known to order \(\epsilon^4\)~\cite{Henriksson:2022rnm}. Once higher orders are computed it immediately produces higher \(\epsilon\)-orders in the above result via~\eqref{eq:phi-k3-eqn}.

\paragraph{Alternatives to methods used.}
We used a formalism based on orthogonality of conformal partial waves developed in~\cite{Hogervorst:2017sfd} and adapted to boundaries in~\cite{Hogervorst:2017kbj} (it belongs to a category known as Euclidean inversion formulas). There are also Lorentzian inversion formulas that have been constructed for \CFT[D] over the past years, both for the bulk and defect/boundary channel~\cite{Mazac:2018biw,Lemos:2017vnx,Liendo:2019jpu}. However, they all carry some limitations.~\cite{Mazac:2018biw} covers both channels, nevertheless explicit expressions for the inversion kernels are known only for the case when the difference of scaling dimensions of the two bulk operators in the \(2\)-pt function is odd (which is not applicable to our case as we consider identical operators, thus the difference is zero). Both~\cite{Lemos:2017vnx} and~\cite{Liendo:2019jpu} are derived for \CFT[D] with defects of co-dimension strictly bigger than one. It seems possible to take a limit to \CFT[B], but it is singular and requires non-trivial analysis. Even when these technical difficulties are overcome, for the particular problem at hand, we believe that the method used was the most efficient one. We started from known position space correlators, so there was no need to take certain discontinuities. Also note that when deriving the spectral function for the \(\sigma\)-correlator, in practice the integral inversion transform was applied to \(\left\langle \phi^2\phi^2 \right\rangle\) and such integral has much nicer convergence properties compared to the case when the transform would have been directly applied to \(\left\langle \delta\sigma\delta\sigma \right\rangle\). This was possible thanks to the nice property of the formalism used -- it serves as a generalized Fourier transform on hyperbolic spaces and thus converts convolution to multiplication. Hence the spectral function for \(\sigma\) could be computed by a simple algebraic inverse (avoiding an integral with a slow fall-off).

Having said that, Lorentzian inversion formulas could be useful at subleading large \(N\) order where position space correlators are not known and computing their discontinuities might be easier. It is also worth mentioning that one of the main powers in the original sense~\cite{Caron-Huot:2017vep}, namely analyticity in spin, is lost in \CFT[B] since spectral functions corresponding to \(2\)-pt functions of scalar operators don't depend on spin at all (neither in the bulk nor in the boundary channel).

Finally, let us mention that a shadow formalism for \CFT[B] for spectral functions (essentially a harmonic analysis generalization of~\cite{Karateev:2018oml} to theories with a boundary) is work in progress. It could provide also an independent derivation of a ``Lorentzian inversion formula'' at least in the sense that the inversion integral runs over a region of the cross ratio line corresponding to Lorentzian configurations.
 \section{Outlook}
\label{sec:outlook}
In this paper we studied the \(\OO(N)\) model in the large \(N\) approximation at the critical point, with boundary conditions chosen to correspond to the \emph{ordinary} universality class. To get a firm handle on \CFT[B] data, as summed up in~\Cref{sec:summary-results}, techniques of bulk/boundary conformal partial wave expansions were used.

There are a couple of future directions worth investigating, ranked by their (anticipated) difficulty:
\begin{enumerate}
\item\label{item:1} For simplicity, we focused entirely on the ordinary transition. But all methods used are general and should be extendable to the special and extraordinary/normal transition without major difficulties. Minor technical complications might be encountered, however the expectation is that the extension to the other two boundary universality classes should be fairly straightforward.
\item\label{item:2} In order to get additional constraints for \CFT[B] data, one could try to run the logic of conformal perturbation theory in reverse (typically, knowledge about the critical point is assumed and used to extend it to its vicinity). While working in the hyperbolic Weyl frame, the non-critical theory is still a theory in \EAdS. In particular, it supports a well defined (even though non-local) \CFT{} at its boundary. Scaling dimensions of certain primary operators (double twist families) in this boundary \CFT{} were already obtained in~\cite{Carmi:2018qzm,Dujava:2025php} (see also~\cite{Dujava:2025nqp} for a detailed introduction to this topic) and their OPE coefficients can in principle be computed as well. Linking this known data to the critical theory (the \CFT[B] that was the subject of this paper) via conformal perturbation theory provides an additional handle on \CFT[B] data of interest.
\item\label{item:3} One could try to extend the computation of the simplest \(\left\langle \phi\phi \right\rangle\) correlator to the next large \(N\) order (that is \(1/N\) in our conventions). Partial results were published a long time ago in a series of papers~\cite{OhnoOkabe1983a,OhnoOkabe1983b,OhnoOkabe1983c,OhnoOkabe1984,Ohno:2025qfv}. The \(2\)-pt function was computed in position space and expressed implicitly via an integral representation with the single cross ratio as the upper bound of the integral. Reconstructing BOE of \(\phi\) translates to a computation of asymptotic expansion of this integral. In contrast to the leading large \(N\) order, where only a single boundary primary \(\widehat{\phi}\) contributes, at the subleading order there is an infinite tower. To our knowledge, only the leading term in the asymptotic expansion was derived, allowing to fix the scaling dimension of \(\widehat{\phi}\). Results for the rest of the tower don't exist.

  We propose that instead of attempting the calculation in position space it might be simpler to go after the associated boundary spectral function (that anyhow packages BOE data more efficiently than a position space representation). As we saw in~\Cref{app:bdry-spec-bootstrap}, it is highly constraint by basic principles. Thus one might hope for an efficient mixed analytic/bootstrap approach. It remains to be seen if it turns out as a viable option.
\end{enumerate}


\acknowledgments{It is our pleasure to thank Ji\v{r}\'{i} Novotn\'{y} and
  Tom\'{a}\v{s} Tuleja for discussions during our group seminars on \CFT[B]. We
  are especially grateful to Ji\v{r}\'{i} for reading a draft of this paper.
  Finally, we are also indebted to Jon\'{a}\v{s} Dujava for useful \LaTeX{}
  scripts from a previous project with PV as well as several typographical suggestions.

  The work of PV was supported by the Grant Agency of the Czech Republic under
  the grant GACR 24-11722S, and in its final stages also by the Czech Ministry
  of Education, Youth and Sports, project No. FORTE —
  CZ.02.01.01/00/22\_008/0004632, co-funded by the European Union.}

\appendix
\crefalias{section}{appendix} \section{Relation between bulk/boundary block coefficients and \texorpdfstring{\CFT[B]}{BCFT} data}
\label{app:rel-cb-coeff-OPE-data}
In this appendix we derive the relation between the coefficient of a bulk/boundary conformal block and \CFT[B] data for the general case of a \(2\)-pt function \(\left\langle \Oper_1\Oper_2 \right\rangle\). For the bulk expansion, formula~\eqref{eq:rel-block-coeff-ope-data} will be then a special instance upon setting \(\Oper_1=\Oper_2\).
\paragraph{Bulk expansion.} Before we start, it is useful to recall that the bulk OPE limit \(x\to y\) in particular implies \(x_{\perp}=y_{\perp}+ \mathcal{O} \left( y_{\perp}^2 \right)\), where the correction terms are associated with descendants. For establishing the desired relation between the data, they can be safely dropped, which will be taken advantage of in the steps below.

Let's consider a \(2\)-pt function \(\left\langle \Oper_1\Oper_2 \right\rangle\) and an operator \(\Oper_3\) contributing to its bulk conformal block decomposition via the OPE. We prefer to isolate the correlator on the left hand side and attach the factor \(\tfrac{\rho}{1-\rho}\) (linking \(2\)-pt functions with their bulk block expansions) to the block on the right hand side as
\begin{align}
\label{eq:2pt-fn-bulk-block-contrib}
\left\langle \Oper_1(x)\Oper_2(y) \right\rangle\supset A \left( \frac{1-\rho}{\rho} \right)^{\frac{\Delta_1+\Delta_2}{2}} \mathsf{G}_{\Oper_3}^{B}(\rho)\eqend
\end{align}
The task is to output a formula for the coefficient \(A\) in terms of \CFT[B] data. To take the OPE limit on the right hand side is straightforward. From~\eqref{eq:cross-ratios}, the OPE limit of \(\rho\) gives
\begin{align}
\label{eq:rho-ope-lim}
\rho\underset{x\to y}{\sim} \frac{\left( x-y \right)^2}{\left( 2x_{\perp} \right) \left( 2y_{\perp} \right)}\sim \frac{\left( x-y \right)^2}{\left( 2y_{\perp} \right)^2}\eqend
\end{align}
Combining it with the leading OPE behavior of the bulk block \(\mathsf{G}_{\Oper_3}^B(\rho)\sim \rho^{\frac{\Delta_3}{2}}\), one gets for the limit on the right hand side
\begin{align}
\label{eq:ope-lim-rhs}
\mathrm{RHS}\sim A \rho^{- \frac{\Delta_1+\Delta_2-\Delta_3}{2}}\sim A\left[ \frac{\left( 2y_{\perp} \right)^2}{\left( x-y \right)^{2}} \right]^{\frac{\Delta_1+\Delta_2-\Delta_3}{2}}\eqend
\end{align}

Next, let's work out the left hand side from the OPE
\begin{align}
\label{eq:ope-generic}
  \Oper_1(x)\times \Oper_2(y)\supset \left[ \left( x-y \right)^2 \right]^{-\frac{\Delta_1+\Delta_2-\Delta_3}{2}} \frac{C_{\Oper_1\Oper_2\Oper_3}}{C_{\Oper_3\Oper_3}}\Oper_3(y)\eqcomma
\end{align}
where we introduced the \(2\)-pt function normalization \(C_{\Oper_3\Oper_3}\) since we don't work with unit normalized operators. Taking expectation value of the above equation on flat half plane \(\left\langle \ind \right\rangle_{\mathrm{FHP}}\) and using an expression for flat half space correlators in terms of those on the hyperboloid via a Weyl transform, one arrives at
\begin{align}
\label{eq:lhs-by-ope}
\frac{1}{\left( x_{\perp} \right)^{\Delta_1} \left( y_{\perp} \right)^{\Delta_2}} \left\langle \Oper_1(x)\Oper_2(y) \right\rangle= \left\langle \Oper_1(x)\Oper_2(y) \right\rangle_{\mathrm{FHP}}\sim \left[ \left( x-y \right)^2 \right]^{-\frac{\Delta_1+\Delta_2-\Delta_3}{2}} \frac{C_{\Oper_1\Oper_2\Oper_3}}{C_{\Oper_3\Oper_3}} \underbrace{\left\langle \Oper_3(y) \right\rangle_{\mathrm{FHP}}}_{\frac{\left\langle \Oper_3 \right\rangle}{\left( y_{\perp} \right)^{\Delta_3}}}\eqend
\end{align}
Simply collecting factors and using \(x_{\perp}\sim y_{\perp}\) yields the bulk limit of the \(2\)-pt function on \(\mathsf{H}_{d+1}\)
\begin{align}
\label{eq:2pt-by-ope}
\left\langle \Oper_1(x)\Oper_2(y) \right\rangle\sim \frac{y_{\perp}^{\Delta_1+\Delta_2-\Delta_3}}{\left[ \left( x-y \right)^2 \right]^{\frac{\Delta_1+\Delta_2-\Delta_3}{2}}} \frac{C_{\Oper_1\Oper_2\Oper_3}}{C_{\Oper_3\Oper_3}} \left\langle \Oper_3 \right\rangle\eqcomma
\end{align}
that should be compared to~\eqref{eq:ope-lim-rhs}, leading to the desired expression of the bulk block coefficient \(A\) in terms of \CFT[B] data
\begin{align}
\label{eq:bulk-cb-coeff-bcft-data-gen}
A=2^{-\Delta_1-\Delta_2+\Delta_3} \frac{C_{\Oper_1\Oper_2\Oper_3}}{C_{\Oper_3\Oper_3}} \left\langle \Oper_3 \right\rangle \eqcomma
\end{align}
which was applied in~\eqref{eq:rel-block-coeff-ope-data} for \(\Oper_1=\Oper_2\equiv\Oper\).

\paragraph{Boundary expansion.} A similar justification can be made for the case of the BOE.
Here, the relevant limit is the boundary limit \(y_{\perp} \to 0\) or, in terms of the cross-ratio, \(\rho  \to 1\).
Assuming we decompose the two-point function \(\langle  \mathsf{O}_{1} \mathsf{O}_{2} \rangle\) into boundary conformal blocks and thus know the coefficient \(B\) in front of a particular conformal block,
\begin{align}
    \label{eq:2pt-fn-bdy-block-contrib}
    \langle  \mathsf{O}_{1}(x) \mathsf{O}_{2}(y) \rangle \supset B \mathsf{G}_{\widehat{\mathsf{O}}}^{\partial} (\rho ) \eqcomma
\end{align}
we can give an interpretation to \(B\) by comparing the leading behavior of the conformal block in this limit with that expected from the BOE.
The conformal block behaves like \( \left( 1-\rho \right)^{\widehat{\Delta}}\), which in coordinates means
\begin{align}
    \label{eq:boe-lim-rhs}
    \text{RHS} \sim B\left( 1-\rho  \right)^{\widehat{\Delta }} \underset{y_{\perp} \to 0}{\sim} B\left( \frac{4 x_{\perp}y_{\perp}}{x_{\perp}^{2} + \left( \mathbf{x}-\mathbf{y} \right)^{2}} \right)^{\widehat{\Delta}}\eqend
\end{align}
Now if we work on the flat half plane, the leading behavior of the BOE is
\begin{align}
    \label{eq:boe-generic}
    \mathsf{O}_{2}(y) \supset  \frac{b_{\mathsf{O}_{2}\mathsf{\widehat{O}}}}{y_{\perp}^{\Delta_{2}-\widehat{\Delta}}} \widehat{\mathsf{O}}(\mathbf{y}) \eqend
\end{align}
Plugging this into the 2-pt function \(\langle  \mathsf{O}_{1} \mathsf{O}_{2} \rangle_{\text{FHP}}\) gives
\begin{align}
    \langle  \mathsf{O}_{1}(x) \mathsf{O}_{2}(y) \rangle_{\text{FHP}} \supset \frac{b_{\mathsf{O}_{2}\widehat{\mathsf{O}}}}{y_{\perp}^{\Delta_{2}-\widehat{\Delta}}} \underbrace{\langle \mathsf{O}_{1}(x) \widehat{\mathsf{O}}(\mathbf{y})\rangle_{\text{FHP}}}_{\frac{b_{\mathsf{O}_{1}\widehat{\mathsf{O}}}}{x_{\perp}^{\Delta_{1}-\widehat{\Delta}} \left( x_{\perp}^{2}+\left( \mathbf{x}-\mathbf{y} \right)^{2} \right)^{\widehat{\Delta}}}}\eqcomma
\end{align}
and comparing this with \eqref{eq:boe-lim-rhs} allows us to identify the coefficient \(B\) to be
\begin{align}
    B = 2^{-2 \widehat{\Delta}} b_{\mathsf{O}_{1}\widehat{\mathsf{O}}} b_{\mathsf{O}_{2} \widehat{\mathsf{O}}}\eqend
\end{align}

\section{Bootstrap of the boundary spectral function for \texorpdfstring{\(\left\langle \delta\sigma \delta\sigma \right\rangle\)}{<δσδσ>}}
\label{app:bdry-spec-bootstrap}

We will ``derive'' here the boundary spectral function~\eqref{eq:spec-bdy-d2} associated with the
connected \(2\)-pt function \(\left\langle \delta\sigma \delta\sigma \right\rangle\) by bootstrap arguments (by which we mean merely a certain set of minimal assumptions). For concreteness, let us consider the case \(d=2\) of a \( \CFT[B]_3 \). This section is not completely rigorous, however some ideas might be worth presenting nevertheless.

Strategy for determining the boundary spectral function will be to fix positions of its poles and zeroes. To do so, we take advantage of the relation~\eqref{eq:B-inverse-from-boundary-spec} specialized to \(d=2\)
\begin{align}
\label{eq:sigma-bubble-inverse-2d}
\mathsf{Spec}^{\partial}_{\left\langle \delta\sigma\delta\sigma \right\rangle}=- \frac{1}{\left( 8\pi \right)^2} \frac{1}{\mathsf{Spec}^{\partial}_{\left\langle \phi^2\phi^2 \right\rangle}} \eqcomma
\end{align}
where the (to be determined) functions read
\begin{align}
\label{eq:spec-bdry-sigma-bubble-2d}
& \mathsf{Spec}^{\partial}_{\left\langle\phi^2\phi^2 \right\rangle}=- \frac{1}{64 \pi} \frac{\cot \left( \frac{\pi}{2} \widehat{\Delta} \right)}{\widehat{\Delta}-1},&& \mathsf{Spec}^{\partial}_{\left\langle \delta\sigma\delta\sigma \right\rangle}=\frac{1}{\pi} \left( \widehat{\Delta}-1 \right)\tan \left( \frac{\pi}{2} \widehat{\Delta}\right) \eqend
\end{align}
The inverse relation implies that the set of zeroes of \(\mathsf{Spec}^{\partial}_{\left\langle \delta\sigma\delta\sigma \right\rangle}\) is given by the set of poles of \(\mathsf{Spec}^{\partial}_{\left\langle \phi^2\phi^2 \right\rangle}\), that is by the boundary spectrum of the bubble function \(\left\langle \phi^2\phi^2 \right\rangle\).

At leading large \(N\), the \(\phi\)-field is free and its BOE contains a single boundary operator, \(\phi_{\frac{1}{2}}\underset{\mathrm{BOE}}{\sim} \widehat{\phi}_1\) of dimension \(\widehat{\Delta}=1\) (for the ordinary transition). Thus it is natural to assume for the boundary expansion of \(\phi^2\) the following form
\begin{align}
\label{eq:BOE-phi2-2d}
\text{\textbf{Assumption 1}:}\quad \left( \phi^2 \right)_1\underset{\mathrm{BOE}}{\sim} \bigoplus_{k=1}^{\infty} \left( \widehat{\phi}^{2k} \right)_{2k} \eqcomma
\end{align}
implying the boundary spectrum \(\widehat{\Delta}_k=2k,\,k\in\Z_{\geq 1}\). Boundary spectral functions are boundary shadow symmetric, therefore we just concluded that the set of poles of \(\mathsf{Spec}^{\partial}_{\left\langle \phi^2\phi^2 \right\rangle}\) is \(2\Z\). At the same time, by~\eqref{eq:sigma-bubble-inverse-2d}, this set fixes the set of zeroes of the function \(\mathsf{Spec}^{\partial}_{\left\langle \delta\sigma\delta\sigma \right\rangle}\) we are trying to bootstrap.

\begin{figure}[ht]
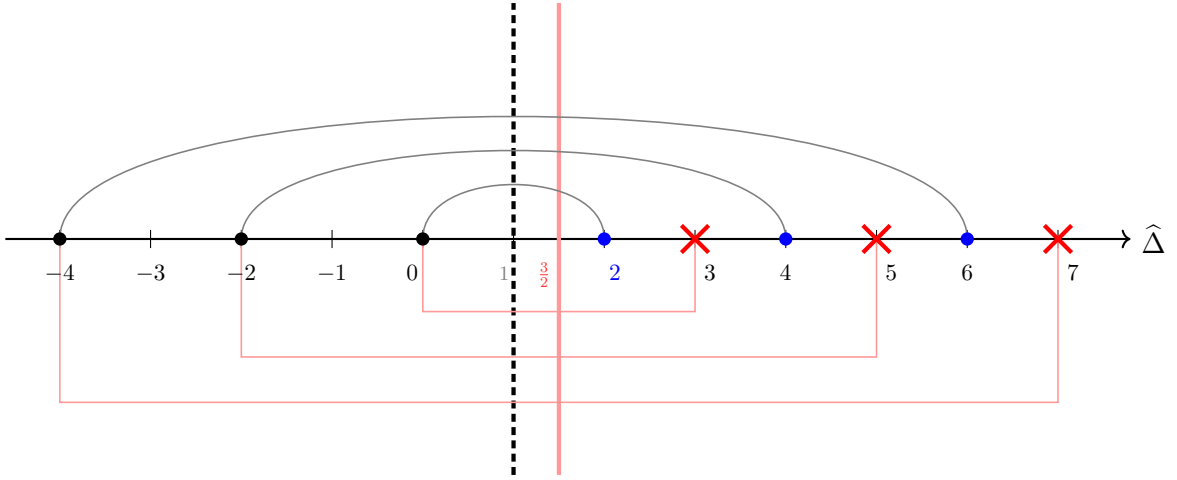

    \centering
    \ShadowSpec
    \caption{Boundary shadow symmetry of boundary spectral functions together with bulk shadow symmetry between \(\Phi^2\) and \(\sigma\) allows to fix the boundary spectrum of \(\sigma\). Black/red vertical lines represent boundary/bulk shadow symmetry mirrors (principal series). Blue dots denote boundary spectrum of \(\phi^2\), while black dots are their boundary shadow symmetry partners. By applying bulk shadow symmetry on black dots, one gets red crosses that represent boundary spectrum of \(\sigma\).}
    \label{fig:bdry-bulk-shadow}
\end{figure}

Having fixed its zeroes, next we should determine its poles. At this point comes in our second crucial assumption -- bulk shadow symmetry relating \(\phi^2\) and \(\sigma\) -- that will allow us to do that
\begin{align}
\label{eq:bulk-shadow-symm-sig-phi2}
\text{\textbf{Assumption 2}:}\quad \left( \phi^2 \right)_1 \xleftrightarrow[\text{bulk shadow}]{\Delta\leftrightarrow 3-\Delta} \left( \sigma \right)_{3-1=2}\eqend
\end{align}
Let's take it seriously and extend its action even to BOE spectra, in a sense of~\Cref{fig:bdry-bulk-shadow}.
One starts with the natural assumption for the boundary spectrum of \(\phi^2\)~\eqref{eq:BOE-phi2-2d} (blue dots). Then one considers as an auxiliary step their boundary shadow symmetry partners (black dots). Now, on these the bulk shadow symmetry assumption~\eqref{eq:bulk-shadow-symm-sig-phi2} is applied to get the red crosses that represent the boundary spectrum of \(\sigma\). By boundary shadow symmetry, they are extended to the set of poles \(\left( 2\Z+1 \right)\setminus \left\{ 1 \right\}\) of \(\mathsf{Spec}^{\partial}_{\left\langle \delta\sigma\delta\sigma \right\rangle}\). Finally, we assume they are simple poles
\begin{align}
\label{eq:assum-simple-poles}
\text{\textbf{Assumption 3}: }\text{poles of } \mathsf{Spec}^{\partial}_{\left\langle \delta\sigma\delta\sigma \right\rangle} \text{ at } \left( 2\Z+1 \right)\setminus \left\{ 1 \right\}\;: \text{ \emph{simple}}
\end{align}
This assumption about poles leads to a natural ansatz for the spectral function (for a moment ignoring the missing pole at \(\widehat{\Delta}=1\))
\begin{align}
\label{eq:bdry-spec-2d-ansatz}
\mathsf{Spec}^{\partial}_{\left\langle \delta\sigma\delta\sigma \right\rangle}=\sum_{k\in\Z}\frac{a_k}{\widehat{\Delta}-\left( 2k+1 \right)}\eqend
\end{align}
The requirement~\eqref{eq:BOE-phi2-2d} of simple zeroes at \(2\Z\) imposes infinitely many constraints on the coefficients \(a_k\). Under mild conditions (convergence of sum and boundedness for \(\abs{\Im\widehat{\Delta}}\to\infty\)), the only solution is that all coefficients are equal, \(a_k=A,\,\forall k\). Therefore, the spectral function is fixed up to a constant
\begin{align}
\label{eq:bdry-spec-sigma-bootstrap-const}
\mathsf{Spec}^{\partial}_{\left\langle \delta\sigma\delta\sigma \right\rangle}=A\sum_{k\in\Z}\frac{1}{\widehat{\Delta}-\left( 2k+1 \right)}=A \frac{\pi}{2} \tan \left( \frac{\pi}{2} \widehat{\Delta} \right)\eqcomma
\end{align}
which can be pinned down by determining a single residue, that is a single squared BOE coefficient. The obvious candidate is of course the boundary operator of scaling dimension \(\widehat{\Delta}=3\) -- the displacement operator -- whose BOE coefficient is linked via the Ward identity~\eqref{eq:Ward-id} to the \(1\)-pt function \(\left\langle \sigma \right\rangle\). Plugging in this known residue gives the constant \(A=\tfrac{2}{\pi^2}\). The final step is to remove the unwanted pole in~\eqref{eq:assum-simple-poles} at \(\widehat{\Delta}=1\) by multiplying the spectral function by a factor \(\left( \widehat{\Delta}-1 \right)\), which determines it completely
\begin{align}
\label{eq:bdry-spec-sigma-bootrap}
\mathsf{Spec}^{\partial}_{\left\langle \delta\sigma\delta\sigma \right\rangle}=\frac{1}{\pi} \left( \widehat{\Delta}-1 \right) \tan \left( \frac{\pi}{2} \widehat{\Delta} \right)\eqend
\end{align}
This finishes the proof of~\eqref{eq:spec-bdry-sigma-bubble-2d} we set out to achieve in this appendix.

 \section{Ward identity for BOE coefficient of displacement operator}
\label{app:Ward-id}
The boundary limit of the connected \(\sigma\)-propagator \(\left\langle \delta\sigma\delta\sigma \right\rangle\) is dominated by the boundary operator of lowest scaling dimension in the BOE of \(\delta\sigma\) -- the displacement operator. So to get a handle on its BOE coefficient, one should compare this limit of a known \(2\)-pt function with its prediction from BOE.

We first compute the boundary limit. In \(\left\langle \delta\sigma(x)\delta\sigma(y) \right\rangle\), the bulk point \(y\) is held fixed, while \(x\) approaches the boundary. Recalling~\eqref{eq:sigma-prop-connected} and~\eqref{eq:cross-ratios}, one gets
\begin{align}
\label{eq:sigma-conn-bdry-lim}
\left\langle \delta\sigma(x)\delta\sigma(y) \right\rangle\xrightarrow[x_{\perp}\to 0]{} \mathcal{N}_{\sigma} \left( 1-\rho \right)^{d+1}= \mathcal{N}_{\sigma} \left[ \frac{\left( 2x_{\perp} \right) \left( 2y_{\perp} \right)}{\left( \mathbf{x}-\mathbf{y} \right)^2+y_{\perp}^{2}} \right]^{d+1}\eqend
\end{align}
In order to compare it with the BOE computation, we need to Weyl transform this expression back to the flat half plane
\begin{align}
\label{eq:sigma-conn-bdry-lim-FHP}
  \left\langle \delta\sigma(x)\delta\sigma(y) \right\rangle_{\mathrm{FHP}}&\coloneq \frac{1}{x_{\perp}^2}\frac{1}{y_{\perp}^2} \left\langle \delta\sigma(x)\delta\sigma(y) \right\rangle \notag \\
  &\xrightarrow[x_{\perp}\to 0]{} \mathcal{N}_{\sigma}2^{2(d+1)} \frac{1}{x_{\perp}^{2-(d+1)}} \frac{1}{y_{\perp}^{2-(d+1)}} \frac{1}{\left[ \left( \mathbf{x}-\mathbf{y} \right)^2 \right]^{d+1}}\Bigg(1+ \mathcal{O} \left( y_{\perp}^2 \right)\Bigg)\eqend
\end{align}

Next, we compute this limit from the BOE of \(\delta\sigma\)
\begin{align}
\label{eq:delta-sig-BOE}
\delta\sigma(x) \underset{\text{BOE}}{\sim} \frac{1}{x_{\perp}^{2-(d+1)}} \frac{B_{\sigma \widehat{D}}}{C_{\widehat{D} \widehat{D}}} \left( \widehat{D}(\mathbf{x})+\text{desc.} \right)+\cdots \eqcomma
\end{align}
where \(C_{\widehat{D} \widehat{D}}\) is the norm of the \(2\)-pt function of the displacement operator and the BOE coefficient was deliberately denoted by a different symbol than \(b_{\sigma\widehat{D}} \) in~\eqref{eq:sigma-BOE} as it is normalized differently. The above definition is chosen such that \(\left\langle \sigma \widehat{D} \right\rangle\sim B_{\sigma \widehat{D}}\), keeping in mind that the displacement operator is not unit normalized (as opposed to all other boundary operators in our conventions). Using it to compute the leading contribution to the boundary limit, we arrive at
\begin{align}
\label{eq:sigma-conn-bdry-lim-BOE}
\left\langle \delta\sigma(x)\delta\sigma(y) \right\rangle_{\mathrm{FHP}}\xrightarrow[x_{\perp}\to 0]{} \frac{1}{x_{\perp}^{2-(d+1)}}\frac{1}{y_{\perp}^{2-(d+1)}} \left( \frac{B_{\sigma\widehat{D}} }{C_{\widehat{D} \widehat{D}}} \right)^2 \Big[ \underbrace{\left\langle \widehat{D}(\mathbf{x}) \widehat{D}(\mathbf{y}) \right\rangle}_{\frac{C_{\widehat{D} \widehat{D}}}{\left[ \left( \mathbf{x}-\mathbf{y} \right)^2 \right]^{(d+1)}}} +\underbrace{\text{desc.}}_{\mathcal{O} \left( y_{\perp}^2 \right)}\Big] +\cdots \eqend
\end{align}
Matching~\eqref{eq:sigma-conn-bdry-lim-FHP} with~\eqref{eq:sigma-conn-bdry-lim-BOE} yields
\begin{align}
\label{eq:BOE-displ-relation}
\mathcal{N}_{\sigma}2^{2(d+1)}=\frac{\left( B_{\sigma \widehat{D}} \right)^2}{C_{\widehat{D} \widehat{D}}}\eqend
\end{align}
Supplementing it with the Ward identity~\cite[(5.30)]{Billo:2016cpy} for the BOE coefficient \(B_{\sigma \widehat{D}}\) as defined in~\eqref{eq:delta-sig-BOE}
\begin{align}
\label{eq:Ward-id}
B_{\sigma \widehat{D}}=\frac{2^{d+1}}{\mathrm{vol}S^d}\Delta_{\sigma} \left\langle \sigma \right\rangle
\end{align}
allows to make a prediction for the norm of the displacement operator \(2\)-pt function
\begin{align}
\label{eq:norm-displ-2pt}
C_{\widehat{D} \widehat{D}}= \left( \frac{\Delta_{\sigma} \left\langle \sigma \right\rangle}{\sqrt{\mathcal{N}_{\sigma}} \mathrm{vol}S^d} \right)^{2}= \frac{8^{-d} (d-3) (d-1)^2 \pi ^{\frac{3}{2}-d}  \Gamma \left(\frac{d+1}{2}\right) \Gamma (2 d-2)}{\cos \left(\frac{\pi  d}{2}\right)\Gamma \left(\frac{d}{2}+1\right) \Gamma \left(\frac{d}{2}\right)^2} \eqcomma
\end{align}
which precisely matches~\cite{McAvity:1995zd} and~\cite[(C.9)]{Giombi:2020rmc}.

Finally, let us remark that the squared BOE coefficient \(\left( b_{\sigma\widehat{D}}  \right)^2=\mathcal{N}_{\sigma}2^{2(d+1)}\) as computed in~\eqref{eq:sigma-BOE-coeff} is related to the one in~\eqref{eq:delta-sig-BOE} via
\begin{align}
\label{eq:BOE-displ-rel}
\left( b_{\sigma\widehat{D}}  \right)^2= \left( \frac{B_{\sigma\widehat{D}} }{\sqrt{C_{\widehat{D} \widehat{D}}}} \right)^{2} \eqend
\end{align}


\section{Summing bulk conformal block expansion for \texorpdfstring{\(\phi\)}{ϕ}}
\label{app:bulk-phi-converg-ser}

Our goal is to verify the bulk expansion of the modified correlator \(\left( \tfrac{\rho}{1-\rho} \right)^{\Delta_{\phi}} \left\langle \phi\phi \right\rangle\) given in~\eqref{eq:phi-bulk-cb-exp-explicit} with coefficients~\eqref{eq:coeff-bulk-phi-explicit}. For convenience, let us repeat here the explicit formula to be proven
\begin{align}
\label{eq:phi-bulk-to-prove}
\left( 1-\rho \right)^{\frac{d-1}{2}}=1+\sum_{k\geq 1}\underbrace{\frac{\Gamma \left(\frac{d+1}{2}\right)^2 \Gamma \left(k-\frac{d+1}{2}\right)}{k! \Gamma \left(\frac{d+1}{2}-k\right)^2 \Gamma \left(2 k-\frac{d+1}{2}\right)}}_{\text{\eqref{eq:coeff-bulk-phi-explicit}}} \underbrace{\rho^k\HypGeo{k,\,k+\frac{1-d}{2}}{2k+\frac{1-d}{2}}[\rho]}_{\mathsf{G}^B_{2k}}\eqend
\end{align}
Strategy of the proof is straightforward. We write the left hand side as a Taylor series
\begin{align}
\label{eq:phi-mod-Taylor}
(1-\rho)^{\frac{d-1}{2}} = \sum_{n=0}^{\infty} \left[ \binom{\frac{d-1}{2}}{n}(-1)^n \right] \rho^n
\end{align}
within its radius of convergence \(\rho\leq 1\) (the boundary value \(\rho=1\) included for \(d>1\)). On the right hand side, we use Taylor series for the hypergeometric function within its radius of convergence \(\rho< 1\), reorganize sums to collect the coefficient in front of a power of \(\rho\) and finally compare it with the one above.

Explicitly, the double sum on the right hand side takes the form
\begin{align}
\label{eq:phi-cb-exp-Taylor}
1+\Gamma\left(\frac{d+1}{2}\right)^2\sum_{\substack{k\geq 1 \\ l\geq 0}}\left[ \frac{\Gamma \left(k-\frac{d+1}{2}\right) \left( k \right)_l \left( k+\frac{1-d}{2} \right)_l}{k! l! \Gamma \left(\frac{d+1}{2}-k\right)^2 \Gamma \left(2 k-\frac{d+1}{2}\right) \left( 2k+\frac{1-d}{2} \right)_{l}} \right]\rho^{k+l} \eqcomma
\end{align}
where the summation runs over a rectangular lattice \(\Z_{\geq 0}\times\Z_{\geq 1}\). To separate the power \(\rho^{k+l}\) let us reorganize the sum. Instead of summing horizontally/vertically we sum diagonally, introducing thus a change of variables
\begin{align}
\label{eq:change-sum-var}
&n=k+l,\; n\in\Z_{\geq 1} \notag \\
  &m=k-l,\;m\in \mathcal{I}_n\coloneq \left\{ -n+2,-n+4,\ldots,n-2,n \right\} \eqend
\end{align}
After this change of variables, the right hand side becomes
\begin{align}
\label{eq:phi-cb-exp-sum-fin}
1+\sum_{n\geq 1}\rho^n\left[\Gamma\left(\frac{d+1}{2}\right)^2\sum_{m\in \mathcal{I}_n} \frac{\Gamma \left(\frac{n+m}{2}-\frac{d+1}{2}\right) \left( \frac{n+m}{2} \right)_{\frac{n-m}{2}} \left( \frac{n+m}{2}+\frac{1-d}{2} \right)_{\frac{n-m}{2}}}{\left( \frac{n+m}{2} \right)! \left( \frac{n-m}{2} \right)! \Gamma \left(\frac{d+1}{2}-\frac{n+m}{2}\right)^2 \Gamma \left(n+m-\frac{d+1}{2}\right) \left( n+m+\frac{1-d}{2} \right)_{\frac{n-m}{2}}} \right] \eqend
\end{align}
The expression inside the brackets \([\ldots]\) is a finite sum over the set \(\mathcal{I}_n\)~\eqref{eq:change-sum-var}. It could be further simplified but it is not even needed. With the help of \wmathematica{} it can be immediately evaluated and shown to be equal to the coefficient in front of the power of \(\rho\) on the left hand side~\eqref{eq:phi-mod-Taylor}. This finishes the proof of~\eqref{eq:phi-bulk-cb-exp-explicit}. The same strategy can be used to verify other bulk expansions.


\bibliography{ref} \bibliographystyle{JHEP}

@article{Giombi:2020rmc,
    author = "Giombi, Simone and Khanchandani, Himanshu",
    title = "{CFT in AdS and boundary RG flows}",
    eprint = "2007.04955",
    archivePrefix = "arXiv",
    primaryClass = "hep-th",
    doi = "10.1007/JHEP11(2020)118",
    journal = "JHEP",
    volume = "11",
    pages = "118",
    year = "2020"
}

@article{Dujava:2025php,
    author = "Dujava, Jon{\'a}{\v{s}} and Va{\v{s}}ko, Petr",
    title = "{Finite-coupling spectrum of O(N) model in AdS}",
    eprint = "2503.16345",
    archivePrefix = "arXiv",
    primaryClass = "hep-th",
    doi = "10.1007/JHEP12(2025)036",
    journal = "JHEP",
    volume = "12",
    pages = "036",
    year = "2025"
}

@article{Carmi:2018qzm,
    author = "Carmi, Dean and Di Pietro, Lorenzo and Komatsu, Shota",
    title = "{A Study of Quantum Field Theories in AdS at Finite Coupling}",
    eprint = "1810.04185",
    archivePrefix = "arXiv",
    primaryClass = "hep-th",
    doi = "10.1007/JHEP01(2019)200",
    journal = "JHEP",
    volume = "01",
    pages = "200",
    year = "2019"
}

@article{McAvity:1995zd,
    author = "McAvity, D. M. and Osborn, H.",
    title = "{Conformal field theories near a boundary in general dimensions}",
    eprint = "cond-mat/9505127",
    archivePrefix = "arXiv",
    reportNumber = "DAMTP-95-1, UBC-TP-95-002",
    doi = "10.1016/0550-3213(95)00476-9",
    journal = "Nucl. Phys. B",
    volume = "455",
    pages = "522--576",
    year = "1995"
}

@article{Liendo:2012hy,
    author = "Liendo, Pedro and Rastelli, Leonardo and van Rees, Balt C.",
    title = "{The Bootstrap Program for Boundary CFT$_d$}",
    eprint = "1210.4258",
    archivePrefix = "arXiv",
    primaryClass = "hep-th",
    reportNumber = "YITP-SB-12-37",
    doi = "10.1007/JHEP07(2013)113",
    journal = "JHEP",
    volume = "07",
    pages = "113",
    year = "2013"
}

@article{Metlitski:2020cqy,
    author = "Metlitski, Max A.",
    title = "{Boundary criticality of the O(N) model in d = 3 critically revisited}",
    eprint = "2009.05119",
    archivePrefix = "arXiv",
    primaryClass = "cond-mat.str-el",
    doi = "10.21468/SciPostPhys.12.4.131",
    journal = "SciPost Phys.",
    volume = "12",
    number = "4",
    pages = "131",
    year = "2022"
}

@article{Cuomo:2021kfm,
    author = "Cuomo, Gabriel and Komargodski, Zohar and Mezei, M\'ark",
    title = "{Localized magnetic field in the O(N) model}",
    eprint = "2112.10634",
    archivePrefix = "arXiv",
    primaryClass = "hep-th",
    doi = "10.1007/JHEP02(2022)134",
    journal = "JHEP",
    volume = "02",
    pages = "134",
    year = "2022"
}

@article{Bissi:2018mcq,
    author = {Bissi, Agnese and Hansen, Tobias and S{\"o}derberg, Alexander},
    title = "{Analytic Bootstrap for Boundary CFT}",
    eprint = "1808.08155",
    archivePrefix = "arXiv",
    primaryClass = "hep-th",
    doi = "10.1007/JHEP01(2019)010",
    journal = "JHEP",
    volume = "01",
    pages = "010",
    year = "2019"
}

@article{Bissi:2022mrs,
    author = "Bissi, Agnese and Sinha, Aninda and Zhou, Xinan",
    title = "{Selected topics in analytic conformal bootstrap: A guided journey}",
    eprint = "2202.08475",
    archivePrefix = "arXiv",
    primaryClass = "hep-th",
    doi = "10.1016/j.physrep.2022.09.004",
    journal = "Phys. Rept.",
    volume = "991",
    pages = "1--89",
    year = "2022"
}

@article{Hogervorst:2017kbj,
    author = "Hogervorst, Matthijs",
    title = "{Crossing Kernels for Boundary and Crosscap CFTs}",
    eprint = "1703.08159",
    archivePrefix = "arXiv",
    primaryClass = "hep-th",
    reportNumber = "YITP-SB-17-8",
    month = "3",
    year = "2017"
}

@article{Diatlyk:2024ngd,
    author = "Diatlyk, Oleksandr and Sun, Zimo and Wang, Yifan",
    title = "{Surprises in the ordinary: O(N) invariant surface defect in the {\ensuremath{\epsilon}}-expansion}",
    eprint = "2411.16522",
    archivePrefix = "arXiv",
    primaryClass = "hep-th",
    reportNumber = "PUPT-2655",
    doi = "10.1007/JHEP06(2025)131",
    journal = "JHEP",
    volume = "06",
    pages = "131",
    year = "2025"
}

@article{Padayasi:2021sik,
    author = "Padayasi, Jaychandran and Krishnan, Abijith and Metlitski, Max A. and Gruzberg, Ilya A. and Meineri, Marco",
    title = "{The extraordinary boundary transition in the 3d O(N) model via conformal bootstrap}",
    eprint = "2111.03071",
    archivePrefix = "arXiv",
    primaryClass = "cond-mat.stat-mech",
    doi = "10.21468/SciPostPhys.12.6.190",
    journal = "SciPost Phys.",
    volume = "12",
    number = "6",
    pages = "190",
    year = "2022"
}

@article{Liendo:2019jpu,
    author = "Liendo, Pedro and Linke, Yannick and Schomerus, Volker",
    title = "{A Lorentzian inversion formula for defect CFT}",
    eprint = "1903.05222",
    archivePrefix = "arXiv",
    primaryClass = "hep-th",
    reportNumber = "DESY-19-039",
    doi = "10.1007/JHEP08(2020)163",
    journal = "JHEP",
    volume = "08",
    pages = "163",
    year = "2020"
}

@article{Lauria:2017wav,
    author = "Lauria, Edoardo and Meineri, Marco and Trevisani, Emilio",
    title = "{Radial coordinates for defect CFTs}",
    eprint = "1712.07668",
    archivePrefix = "arXiv",
    primaryClass = "hep-th",
    doi = "10.1007/JHEP11(2018)148",
    journal = "JHEP",
    volume = "11",
    pages = "148",
    year = "2018"
}

@article{Lemos:2017vnx,
    author = "Lemos, Madalena and Liendo, Pedro and Meineri, Marco and Sarkar, Sourav",
    title = "{Universality at large transverse spin in defect CFT}",
    eprint = "1712.08185",
    archivePrefix = "arXiv",
    primaryClass = "hep-th",
    reportNumber = "DESY 17-239, HU-EP-17/31, DESY-17-239, HU-EP-17-31",
    doi = "10.1007/JHEP09(2018)091",
    journal = "JHEP",
    volume = "09",
    pages = "091",
    year = "2018"
}

@article{Ohno:2025qfv,
    author = "Ohno, Kaoru and Okabe, Yutaka",
    title = "{Comment on CFT in AdS and boundary RG flows: O(1/N) Result}",
    eprint = "2511.06577",
    archivePrefix = "arXiv",
    primaryClass = "hep-th",
    month = "11",
    year = "2025"
}

@article{Henriksson:2022rnm,
    author = "Henriksson, Johan",
    title = "{The critical O(N) CFT: Methods and conformal data}",
    eprint = "2201.09520",
    archivePrefix = "arXiv",
    primaryClass = "hep-th",
    doi = "10.1016/j.physrep.2022.12.002",
    journal = "Phys. Rept.",
    volume = "1002",
    pages = "1--72",
    year = "2023"
}

@article{Alday:2019clp,
    author = "Alday, Luis F. and Henriksson, Johan and van Loon, Mark",
    title = "{An alternative to diagrams for the critical O(N) model: dimensions and structure constants to order 1/N$^{2}$}",
    eprint = "1907.02445",
    archivePrefix = "arXiv",
    primaryClass = "hep-th",
    doi = "10.1007/JHEP01(2020)063",
    journal = "JHEP",
    volume = "01",
    pages = "063",
    year = "2020"
}

@article{Diehl:2020rfx,
    author = "Diehl, H. W.",
    title = "{Why boundary conditions do not generally determine the universality class for boundary critical behavior}",
    eprint = "2006.15425",
    archivePrefix = "arXiv",
    primaryClass = "hep-th",
    doi = "10.1140/epjb/e2020-10422-9",
    journal = "Eur. Phys. J. B",
    volume = "93",
    number = "10",
    pages = "195",
    year = "2020"
}

@article{Rastelli:2017ecj,
    author = "Rastelli, Leonardo and Zhou, Xinan",
    title = "{The Mellin Formalism for Boundary CFT$_d$}",
    eprint = "1705.05362",
    archivePrefix = "arXiv",
    primaryClass = "hep-th",
    reportNumber = "YITP-SB-2017-18",
    doi = "10.1007/JHEP10(2017)146",
    journal = "JHEP",
    volume = "10",
    pages = "146",
    year = "2017"
}

@article{Mazac:2018biw,
    author = "Maz{\'a}{\v{c}}, Dalimil and Rastelli, Leonardo and Zhou, Xinan",
    title = "{An analytic approach to BCFT$_{d}$}",
    eprint = "1812.09314",
    archivePrefix = "arXiv",
    primaryClass = "hep-th",
    reportNumber = "PUPT-2580; YITP-44",
    doi = "10.1007/JHEP12(2019)004",
    journal = "JHEP",
    volume = "12",
    pages = "004",
    year = "2019"
}

@article{Billo:2016cpy,
    author = "Bill{\`o}, Marco and Gon{\c{c}}alves, Vasco and Lauria, Edoardo and Meineri, Marco",
    title = "{Defects in conformal field theory}",
    eprint = "1601.02883",
    archivePrefix = "arXiv",
    primaryClass = "hep-th",
    doi = "10.1007/JHEP04(2016)091",
    journal = "JHEP",
    volume = "04",
    pages = "091",
    year = "2016"
}

@article{Mukhametzhanov:2018zja,
    author = "Mukhametzhanov, Baur and Zhiboedov, Alexander",
    title = "{Analytic Euclidean Bootstrap}",
    eprint = "1808.03212",
    archivePrefix = "arXiv",
    primaryClass = "hep-th",
    doi = "10.1007/JHEP10(2019)270",
    journal = "JHEP",
    volume = "10",
    pages = "270",
    year = "2019"
}

@article{Pappadopulo:2012jk,
    author = "Pappadopulo, Duccio and Rychkov, Slava and Espin, Johnny and Rattazzi, Riccardo",
    title = "{OPE Convergence in Conformal Field Theory}",
    eprint = "1208.6449",
    archivePrefix = "arXiv",
    primaryClass = "hep-th",
    reportNumber = "LPTENS-12-31",
    doi = "10.1103/PhysRevD.86.105043",
    journal = "Phys. Rev. D",
    volume = "86",
    pages = "105043",
    year = "2012"
}

@article{Qiao:2017xif,
    author = "Qiao, Jiaxin and Rychkov, Slava",
    title = "{A tauberian theorem for the conformal bootstrap}",
    eprint = "1709.00008",
    archivePrefix = "arXiv",
    primaryClass = "hep-th",
    reportNumber = "CERN-TH-2017-176",
    doi = "10.1007/JHEP12(2017)119",
    journal = "JHEP",
    volume = "12",
    pages = "119",
    year = "2017"
}

@article{Rychkov:2015lca,
    author = "Rychkov, Slava and Yvernay, Pierre",
    title = "{Remarks on the Convergence Properties of the Conformal Block Expansion}",
    eprint = "1510.08486",
    archivePrefix = "arXiv",
    primaryClass = "hep-th",
    reportNumber = "CERN-PH-TH-2015-253",
    doi = "10.1016/j.physletb.2016.01.004",
    journal = "Phys. Lett. B",
    volume = "753",
    pages = "682--686",
    year = "2016"
}

@article{Lang:1993ct,
    author = "Lang, K. and Ruhl, W.",
    title = "{Critical nonlinear O(N) sigma models at 2 {\ensuremath{<}} d {\ensuremath{<}} 4: The Degeneracy of quasiprimary fields and it resolution}",
    reportNumber = "KL-TH-93-7",
    doi = "10.1007/BF01413189",
    journal = "Z. Phys. C",
    volume = "61",
    pages = "495--510",
    year = "1994"
}

@article{Vasiliev:1981dg,
    author = "Vasiliev, A. N. and Pismak, Yu. M. and Khonkonen, Yu. R.",
    title = "{1/$N$ Expansion: Calculation of the Exponents $\eta$ and Nu in the Order 1/$N^2$ for Arbitrary Number of Dimensions}",
    doi = "10.1007/BF01019296",
    journal = "Theor. Math. Phys.",
    volume = "47",
    pages = "465--475",
    year = "1981"
}

@article{Goykhman:2019kcj,
    author = "Goykhman, Mikhail and Smolkin, Michael",
    title = "{Vector model in various dimensions}",
    eprint = "1911.08298",
    archivePrefix = "arXiv",
    primaryClass = "hep-th",
    doi = "10.1103/PhysRevD.102.025003",
    journal = "Phys. Rev. D",
    volume = "102",
    number = "2",
    pages = "025003",
    year = "2020"
}

@article{Broadhurst:1996ur,
    author = "Broadhurst, David J. and Gracey, J. A. and Kreimer, D.",
    title = "{Beyond the triangle and uniqueness relations: Nonzeta counterterms at large N from positive knots}",
    eprint = "hep-th/9607174",
    archivePrefix = "arXiv",
    reportNumber = "OUT-4102-46, LTH-360, MZ-TH-95-28",
    doi = "10.1007/s002880050500",
    journal = "Z. Phys. C",
    volume = "75",
    pages = "559--574",
    year = "1997"
}

@article{Derkachov:1997gc,
    author = "Derkachov, Sergey E. and Manashov, A. N.",
    title = "{On the stability problem in the O(N) nonlinear sigma model}",
    eprint = "hep-th/9705020",
    archivePrefix = "arXiv",
    reportNumber = "SPBU-IP-97-08",
    doi = "10.1103/PhysRevLett.79.1423",
    journal = "Phys. Rev. Lett.",
    volume = "79",
    pages = "1423--1427",
    year = "1997"
}

@article{Derkachov:1998js,
    author = "Derkachov, Sergey E. and Manashov, A. N.",
    title = "{Critical dimensions of composite operators in the nonlinear sigma model}",
    doi = "10.1007/BF02557145",
    journal = "Theor. Math. Phys.",
    volume = "116",
    pages = "1034--1049",
    year = "1998"
}

@article{Derkachov:1997ch,
    author = "Derkachov, Sergey E. and Manashov, A. N.",
    title = "{The Simple scheme for the calculation of the anomalous dimensions of composite operators in the 1/N expansion}",
    eprint = "hep-th/9710015",
    archivePrefix = "arXiv",
    reportNumber = "SPBU-IP-97-17, NORDITA-97-69",
    doi = "10.1016/S0550-3213(98)00103-5",
    journal = "Nucl. Phys. B",
    volume = "522",
    pages = "301--320",
    year = "1998"
}

@article{Vasiliev:1993ux,
    author = "Vasiliev, A. N. and Stepanenko, A. S.",
    title = "{A Method of calculating the critical dimensions of composite operators in the massless nonlinear sigma model}",
    doi = "10.1007/BF01015903",
    journal = "Theor. Math. Phys.",
    volume = "94",
    pages = "471--481",
    year = "1993"
}

@article{OhnoOkabe1983a,
    author = "Ohno, K. and Okabe, Y.",
    title = "{THE 1/N EXPANSION FOR THE N VECTOR MODEL IN THE SEMIINFINITE SPACE}",
    doi = "10.1143/PTP.70.1226",
    journal = "Prog. Theor. Phys.",
    volume = "70",
    pages = "1226--1239",
    year = "1983"
}

@article{OhnoOkabe1983b,
title = {Surface critical exponents in the 1/n expansion},
journal = {Physics Letters A},
volume = {95},
number = {1},
pages = {41-43},
year = {1983},
issn = {0375-9601},
doi = {https://doi.org/10.1016/0375-9601(83)90777-6},
url = {https://www.sciencedirect.com/science/article/pii/0375960183907776},
author = {Kaoru Ohno and Yutaka Okabe},
}

@article{OhnoOkabe1983c,
title = {The 1/n expansion for the special transition in semi-infinite systems},
journal = {Physics Letters A},
volume = {99},
number = {1},
pages = {54-57},
year = {1983},
issn = {0375-9601},
doi = {https://doi.org/10.1016/0375-9601(83)90064-6},
url = {https://www.sciencedirect.com/science/article/pii/0375960183900646},
author = {K. Ohno and Y. Okabe},
}

@article{OhnoOkabe1984,
    author = {Ohno, Kaoru and Okabe, Yutaka},
    title = {The 1/n Expansion for the Extraordinary Transition of Semi-Infinite System},
    journal = {Progress of Theoretical Physics},
    volume = {72},
    number = {4},
    pages = {736-745},
    year = {1984},
    doi = {10.1143/PTP.72.736},
}

@article{LubenskyRubin1975a,
  title = {Critical phenomena in semi-infinite systems. I. $\ensuremath{\epsilon}$ expansion for positive extrapolation length},
  author = {Lubensky, T. C. and Rubin, Morton H.},
  journal = {Phys. Rev. B},
  volume = {11},
  issue = {11},
  pages = {4533--4546},
  year = {1975},
  doi = {10.1103/PhysRevB.11.4533},
}

@article{LubenskyRubin1975b,
    author = "Lubensky, T. C. and Rubin, Morton H.",
    title = "{Critical phenomena in semi-infinite systems. 2. Mean-field theory}",
    doi = "10.1103/PhysRevB.12.3885",
    journal = "Phys. Rev. B",
    volume = "12",
    pages = "3885--3901",
    year = "1975"
}

@article{BrayMoore1977,
    author = "Bray, A. J. and Moore, M. A.",
    title = "{Critical behaviour of semi-infinite systems}",
    doi = "10.1088/0305-4470/10/11/021",
    journal = "J. Phys. A",
    volume = "10",
    number = "11",
    pages = "1927",
    year = "1977"
}

@article{Mills1971,
  title = {Surface Effects in Magnetic Crystals near the Ordering Temperature},
  author = {Mills, D. L.},
  journal = {Phys. Rev. B},
  volume = {3},
  issue = {11},
  pages = {3887--3895},
  year = {1971},
  doi = {10.1103/PhysRevB.3.3887},
}

@article{BinderHohenberg1972,
  title = {Phase Transitions and Static Spin Correlations in Ising Models with Free Surfaces},
  author = {Binder, K. and Hohenberg, P. C.},
  journal = {Phys. Rev. B},
  volume = {6},
  issue = {9},
  pages = {3461--3487},
  year = {1972},
  doi = {10.1103/PhysRevB.6.3461},
}

@article{BinderHohenberg1974,
  title = {Surface effects on magnetic phase transitions},
  author = {Binder, K. and Hohenberg, P. C.},
  journal = {Phys. Rev. B},
  volume = {9},
  issue = {5},
  pages = {2194--2211},
  year = {1974},
  doi = {10.1103/PhysRevB.9.2194},
}

@article{DiehlDietrich1980,
  title = {Multicritical behaviour at surfaces},
  author = {Diehl, H. W. and Dietrich, S.},
  journal = {Z. Phys. B Condens. Matter},
  volume = {34},
  issue = {4},
  pages = {405--419},
  year = {1980},
  doi = {10.1007/BF01304094},
}

@article{DiehlDietrich1981,
  title = {Field-theoretical approach to static critical phenomena in semi-infinite systems},
  author = {Diehl, H. W. and Dietrich, S.},
  journal = {Z. Phys. B Condens. Matter},
  volume = {42},
  issue = {1},
  pages = {65--86},
  year = {1981},
  doi = {10.1007/BF01298293},
}

@incollection{Diehl1986,
  title = {Field-Theoretical Approach to Critical Behavior at Surfaces},
  author = {Diehl, H. W.},
  booktitle = {Phase Transitions and Critical Phenomena},
  editor = {Domb, C. and Lebowitz, J. L.},
  volume = {10},
  pages = {75--267},
  year = {1986},
  publisher = {Academic Press},
}

@article{Wilson1971a,
  title = {Renormalization Group and Critical Phenomena. I. Renormalization Group and the Kadanoff Scaling Picture},
  author = {Wilson, Kenneth G.},
  journal = {Phys. Rev. B},
  volume = {4},
  issue = {9},
  pages = {3174--3183},
  year = {1971},
  month = {Nov},
  publisher = {American Physical Society},
  doi = {10.1103/PhysRevB.4.3174},
  url = {https://link.aps.org/doi/10.1103/PhysRevB.4.3174}
}

@article{Wilson1971b,
  title = {Renormalization Group and Critical Phenomena. II. Phase-Space Cell Analysis of Critical Behavior},
  author = {Wilson, Kenneth G.},
  journal = {Phys. Rev. B},
  volume = {4},
  issue = {9},
  pages = {3184--3205},
  year = {1971},
  month = {Nov},
  publisher = {American Physical Society},
  doi = {10.1103/PhysRevB.4.3184},
  url = {https://link.aps.org/doi/10.1103/PhysRevB.4.3184}
}

@article{WilsonFisher1972,
  title = {Critical Exponents in 3.99 Dimensions},
  author = {Wilson, Kenneth G. and Fisher, Michael E.},
  journal = {Phys. Rev. Lett.},
  volume = {28},
  issue = {4},
  pages = {240--243},
  year = {1972},
  month = {Jan},
  publisher = {American Physical Society},
  doi = {10.1103/PhysRevLett.28.240},
  url = {aps.org}
}

@article{Wilson1975,
  title = {The renormalization group: Critical phenomena and the Kondo problem},
  author = {Wilson, Kenneth G.},
  journal = {Rev. Mod. Phys.},
  volume = {47},
  issue = {4},
  pages = {773--840},
  year = {1975},
  month = {Oct},
  publisher = {American Physical Society},
  doi = {10.1103/RevModPhys.47.773},
  url = {https://link.aps.org/doi/10.1103/RevModPhys.47.773}
}

@article{Cardy:1984bb,
    author = "Cardy, John L.",
    title = "{Conformal Invariance and Surface Critical Behavior}",
    doi = "10.1016/0550-3213(84)90241-4",
    journal = "Nucl. Phys. B",
    volume = "240",
    pages = "514--532",
    year = "1984"
}

@article{Cardy:1989ir,
    author = "Cardy, John L.",
    title = "{Boundary Conditions, Fusion Rules and the Verlinde Formula}",
    reportNumber = "UCSB-TH-89-06",
    doi = "10.1016/0550-3213(89)90521-X",
    journal = "Nucl. Phys. B",
    volume = "324",
    pages = "581--596",
    year = "1989"
}

@article{Diehl:1996kd,
    author = "Diehl, H. W.",
    title = "{The Theory of boundary critical phenomena}",
    eprint = "cond-mat/9610143",
    archivePrefix = "arXiv",
    doi = "10.1142/S0217979297001751",
    journal = "Int. J. Mod. Phys. B",
    volume = "11",
    pages = "3503--3523",
    year = "1997"
}

@article{McAvity:1993ue,
    author = "McAvity, D. M. and Osborn, H.",
    title = "{Energy momentum tensor in conformal field theories near a boundary}",
    eprint = "hep-th/9302068",
    archivePrefix = "arXiv",
    reportNumber = "DAMTP-93-01",
    doi = "10.1016/0550-3213(93)90005-A",
    journal = "Nucl. Phys. B",
    volume = "406",
    pages = "655--680",
    year = "1993"
}

@article{Rattazzi:2008pe,
    author = "Rattazzi, Riccardo and Rychkov, Vyacheslav S. and Tonni, Erik and Vichi, Alessandro",
    title = "{Bounding scalar operator dimensions in 4D CFT}",
    eprint = "0807.0004",
    archivePrefix = "arXiv",
    primaryClass = "hep-th",
    doi = "10.1088/1126-6708/2008/12/031",
    journal = "JHEP",
    volume = "12",
    pages = "031",
    year = "2008"
}

@article{El-Showk:2012cjh,
    author = "El-Showk, Sheer and Paulos, Miguel F. and Poland, David and Rychkov, Slava and Simmons-Duffin, David and Vichi, Alessandro",
    title = "{Solving the 3D Ising Model with the Conformal Bootstrap}",
    eprint = "1203.6064",
    archivePrefix = "arXiv",
    primaryClass = "hep-th",
    reportNumber = "LPTENS-12-07",
    doi = "10.1103/PhysRevD.86.025022",
    journal = "Phys. Rev. D",
    volume = "86",
    pages = "025022",
    year = "2012"
}

@article{Lauchli:2025fii,
    author = {L{\"a}uchli, Andreas M. and Herviou, Lo{\"\i}c and Wilhelm, Patrick H. and Rychkov, Slava},
    title = "{Exact diagonalization, matrix product states and conformal perturbation theory study of a 3D Ising fuzzy sphere model}",
    eprint = "2504.00842",
    archivePrefix = "arXiv",
    primaryClass = "cond-mat.stat-mech",
    doi = "10.21468/SciPostPhys.19.3.076",
    journal = "SciPost Phys.",
    volume = "19",
    number = "3",
    pages = "076",
    year = "2025"
}

@article{Dey:2026cso,
    author = {Dey, Arjun and Herviou, Loic and Mudry, Christopher and Rychkov, Slava and L{\"a}uchli, Andreas Martin},
    title = "{Conformal Data for the $O(2)$ Wilson-Fisher CFT in $(2+1)$-Dimensional Spacetime from Exact Diagonalization and Matrix Product States on the Fuzzy Sphere}",
    eprint = "2604.18705",
    archivePrefix = "arXiv",
    primaryClass = "cond-mat.str-el",
    month = "4",
    year = "2026"
}

@article{Podo:2026hfh,
    author = "Podo, Alessandro and Rychkov, Slava",
    title = "{Direct Experimental Test of Conformal Invariance via Grazing Scattering: A Proposal for X-ray and Neutron Experiments}",
    eprint = "2605.06773",
    archivePrefix = "arXiv",
    primaryClass = "cond-mat.stat-mech",
    month = "5",
    year = "2026"
}

@article{Zhu:2022gjc,
    author = "Zhu, Wei and Han, Chao and Huffman, Emilie and Hofmann, Johannes S. and He, Yin-Chen",
    title = "{Uncovering Conformal Symmetry in the 3D Ising Transition: State-Operator Correspondence from a Quantum Fuzzy Sphere Regularization}",
    eprint = "2210.13482",
    archivePrefix = "arXiv",
    primaryClass = "cond-mat.stat-mech",
    doi = "10.1103/PhysRevX.13.021009",
    journal = "Phys. Rev. X",
    volume = "13",
    number = "2",
    pages = "021009",
    year = "2023"
}

@article{Zhou:2024dbt,
    author = "Zhou, Zheng and Zou, Yijian",
    title = "{Studying the 3d Ising surface CFTs on the fuzzy sphere}",
    eprint = "2407.15914",
    archivePrefix = "arXiv",
    primaryClass = "hep-th",
    doi = "10.21468/SciPostPhys.18.1.031",
    journal = "SciPost Phys.",
    volume = "18",
    number = "1",
    pages = "031",
    year = "2025"
}

@article{Dedushenko:2024nwi,
    author = "Dedushenko, Mykola",
    title = "{Ising BCFT from Fuzzy Hemisphere}",
    eprint = "2407.15948",
    archivePrefix = "arXiv",
    primaryClass = "hep-th",
    month = "7",
    year = "2024"
}

@article{Feng:2026iii,
    author = "Feng, Jiechao and Wang, Taige",
    title = "{Studying 3D O(N) Surface CFT on the Fuzzy Sphere}",
    eprint = "2604.21091",
    archivePrefix = "arXiv",
    primaryClass = "cond-mat.str-el",
    month = "4",
    year = "2026"
}

@article{Toldin:2021kun,
    author = "Toldin, Francesco Parisen and Metlitski, Max A.",
    title = "{Boundary Criticality of the 3D O(N) Model: From Normal to Extraordinary}",
    eprint = "2111.03613",
    archivePrefix = "arXiv",
    primaryClass = "cond-mat.stat-mech",
    doi = "10.1103/PhysRevLett.128.215701",
    journal = "Phys. Rev. Lett.",
    volume = "128",
    number = "21",
    pages = "215701",
    year = "2022"
}

@mastersthesis{Dujava:2025nqp,
    author = "Dujava, Jon{\'a}{\v{s}}",
    title = "{Strongly Coupled Quantum Field Theory in Anti-de Sitter Spacetime}",
    eprint = "2507.07111",
    archivePrefix = "arXiv",
    primaryClass = "hep-th",
    school = "Charles U., Prague (main)",
    year = "2025"
}

@article{Giombi:2025pxx,
    author = "Giombi, Simone and Sun, Zimo",
    title = "{Higher loops in AdS: applications to boundary CFT}",
    eprint = "2506.14699",
    archivePrefix = "arXiv",
    primaryClass = "hep-th",
    doi = "10.1007/JHEP12(2025)011",
    journal = "JHEP",
    volume = "12",
    pages = "011",
    year = "2025"
}

@article{Hu:2025yrs,
    author = "Hu, Runzhe and Li, Wenliang",
    title = "{Accurate boundary bootstrap for the three-dimensional O($N$) normal universality class}",
    eprint = "2508.20854",
    archivePrefix = "arXiv",
    primaryClass = "hep-th",
    month = "8",
    year = "2025"
}

@article{Drukker:2026nvv,
    author = "Drukker, Nadav and Komatsu, Shota and Wallberg, Anders",
    title = "{Crosscap Defects}",
    eprint = "2604.19868",
    archivePrefix = "arXiv",
    primaryClass = "hep-th",
    reportNumber = "CERN-TH-2026-095",
    month = "4",
    year = "2026"
}

@article{Sun:2026mib,
    author = "Sun, Xinyu and Jian, Shao-Kai and Yao, Hong",
    title = "{Analytic Bootstrap for $O(N)$ Boundary Conformal Field Theories with Interacting Boundaries}",
    eprint = "2605.28933",
    archivePrefix = "arXiv",
    primaryClass = "hep-th",
    month = "5",
    year = "2026"
}

@article{Diatlyk:2026eta,
    author = "Diatlyk, Oleksandr and Giombi, Simone and Sun, Zimo",
    title = "{Boundary criticality in the Gross-Neveu-Yukawa model at higher orders}",
    eprint = "2606.07510",
    archivePrefix = "arXiv",
    primaryClass = "hep-th",
    month = "6",
    year = "2026"
}

@article{Hogervorst:2017sfd,
    author = "Hogervorst, Matthijs and van Rees, Balt C.",
    title = "{Crossing symmetry in alpha space}",
    eprint = "1702.08471",
    archivePrefix = "arXiv",
    primaryClass = "hep-th",
    reportNumber = "YITP-SB-17-7",
    doi = "10.1007/JHEP11(2017)193",
    journal = "JHEP",
    volume = "11",
    pages = "193",
    year = "2017"
}

@article{Caron-Huot:2017vep,
    author = "Caron-Huot, Simon",
    title = "{Analyticity in Spin in Conformal Theories}",
    eprint = "1703.00278",
    archivePrefix = "arXiv",
    primaryClass = "hep-th",
    doi = "10.1007/JHEP09(2017)078",
    journal = "JHEP",
    volume = "09",
    pages = "078",
    year = "2017"
}

@article{Karateev:2018oml,
    author = "Karateev, Denis and Kravchuk, Petr and Simmons-Duffin, David",
    title = "{Harmonic Analysis and Mean Field Theory}",
    eprint = "1809.05111",
    archivePrefix = "arXiv",
    primaryClass = "hep-th",
    reportNumber = "CALT-TH 2018-036",
    doi = "10.1007/JHEP10(2019)217",
    journal = "JHEP",
    volume = "10",
    pages = "217",
    year = "2019"
}

@article{Fitzpatrick:2011dm,
    author = "Fitzpatrick, A. Liam and Kaplan, Jared",
    title = "{Unitarity and the Holographic S-Matrix}",
    eprint = "1112.4845",
    archivePrefix = "arXiv",
    primaryClass = "hep-th",
    reportNumber = "SLAC-PUB-14979",
    doi = "10.1007/JHEP10(2012)032",
    journal = "JHEP",
    volume = "10",
    pages = "032",
    year = "2012"
}

@article{Cornalba:2007fs,
    author = "Cornalba, Lorenzo",
    title = "{Eikonal methods in AdS/CFT: Regge theory and multi-reggeon exchange}",
    eprint = "0710.5480",
    archivePrefix = "arXiv",
    primaryClass = "hep-th",
    month = "10",
    year = "2007"
}

@article{Moshe:2003xn,
    author = "Moshe, Moshe and Zinn-Justin, Jean",
    title = "{Quantum field theory in the large N limit: A Review}",
    eprint = "hep-th/0306133",
    archivePrefix = "arXiv",
    doi = "10.1016/S0370-1573(03)00263-1",
    journal = "Phys. Rept.",
    volume = "385",
    pages = "69--228",
    year = "2003"
}

@article{Penedones:2010ue,
    author = "Penedones, Joao",
    title = "{Writing CFT correlation functions as AdS scattering amplitudes}",
    eprint = "1011.1485",
    archivePrefix = "arXiv",
    primaryClass = "hep-th",
    doi = "10.1007/JHEP03(2011)025",
    journal = "JHEP",
    volume = "03",
    pages = "025",
    year = "2011"
}

@phdthesis{Penedones:2007ns,
    author = "Penedones, Joao",
    title = "{High Energy Scattering in the AdS/CFT Correspondence}",
    eprint = "0712.0802",
    archivePrefix = "arXiv",
    primaryClass = "hep-th",
    school = "University of Porto",
    month = "12",
    year = "2007"
}

@article{Fei:2014yja,
    author = "Fei, Lin and Giombi, Simone and Klebanov, Igor R.",
    title = "{Critical $O(N)$ models in $6-\epsilon$ dimensions}",
    eprint = "1404.1094",
    archivePrefix = "arXiv",
    primaryClass = "hep-th",
    reportNumber = "PUPT-2463",
    doi = "10.1103/PhysRevD.90.025018",
    journal = "Phys. Rev. D",
    volume = "90",
    number = "2",
    pages = "025018",
    year = "2014"
}

@Inbook{Koornwinder1984,
    author="Koornwinder, Tom H.",
    editor="Askey, R. A.
    and Koornwinder, T. H.
    and Schempp, W.",
    title="Jacobi Functions and Analysis on Noncompact Semisimple Lie Groups",
    bookTitle="Special Functions: Group Theoretical Aspects and Applications",
    year="1984",
    publisher="Springer Netherlands",
    address="Dordrecht",
    pages="1--85",
    isbn="978-94-010-9787-1",
    doi="10.1007/978-94-010-9787-1_1",
    url="https://doi.org/10.1007/978-94-010-9787-1_1"
}

\end{document}